\documentclass[conference]{IEEEtran}
\usepackage{cite}
\usepackage{amsmath,amssymb,amsfonts}
\usepackage{algorithmic}
\usepackage{graphicx}
\usepackage{textcomp}
\usepackage{xcolor}
\usepackage{booktabs}
\def\BibTeX{{\rm B\kern-.05em{\sc i\kern-.025em b}\kern-.08em
    T\kern-.1667em\lower.7ex\hbox{E}\kern-.125emX}}
\usepackage{kotex}
\usepackage{multirow}
\usepackage{epsfig}
\usepackage{paralist}
\usepackage[center]{subfigure}
\usepackage{enumitem}
\usepackage{color}
\usepackage{xspace}

\usepackage{bbold}
\usepackage{mathtools}
\usepackage{comment}
\usepackage{microtype}
\usepackage{algorithmic}
\usepackage[ruled,vlined,linesnumbered,resetcount]{algorithm2e}
\usepackage{amsthm}
\usepackage{mathrsfs}
\usepackage[hidelinks]{hyperref}

\usepackage{alphalph}

\include{pythonlisting}
\SetKwInput{KwInput}{Input}                % Set the Input
\SetKwInput{KwOutput}{Output}              % set the Output
\newcommand\Comment[1]{\footnotesize\blue}
\SetKwComment{Comment}{$\triangleright$\ }{}
\newcommand\CommentNT[1]{\footnotesize\blue}
\SetKwComment{CommentNT}{\ }{}

\SetCommentSty{mycommfont}

\let\oldnl\nl
\newcommand{\noLineNumber}{\renewcommand{\nl}{\let\nl\oldnl}}

\DeclareMathOperator*{\argmax}{argmax}

\setlist[itemize]{leftmargin=*}

\newtheorem{problem}{Problem}
\newtheorem{theorem}{Theorem}
\newtheorem{lemma}{Lemma}
\newtheorem{example}{Example}

\newcommand{\InputGraph}{G}
\newcommand{\SummaryGraph}{\overline{G}}
\newcommand{\ReconstructedGraph}{\hat{G}}
\newcommand{\AdjacencyMatrix}{A^{(G)}}
\newcommand{\ReconstructedAdjacencyMatrix}{A^{(\hat{G})}}
\newcommand{\TargetNodeSet}{T}
\newcommand{\SizeBudget}{k}

\newcommand{\PersonalizedError}{RE^{(\TargetNodeSet)}}
\newcommand{\NonPersonalizedError}{RE^{(V)}}

\newcommand{\Weight}{W^{(\TargetNodeSet)}}

\newcommand{\PIS}{\Pi_{S}}

\newcommand{\SuperIdxSubnodei}{S_{u}}
\newcommand{\SuperIdxSubnodej}{S_{v}}

\newcommand{\CostFunction}{Cost^{(\TargetNodeSet)}}

\newcommand{\CostReductionAB}{\Delta Cost^{(\TargetNodeSet)}(A,B;\SummaryGraph)}
\newcommand{\RelReductionAB}{\Delta \overline{Cost}^{(\TargetNodeSet)}(A,B;\SummaryGraph)}

\newcommand{\MergeAB}{merge(A,B;\SummaryGraph)}

\newcommand{\MaxIterations}{t_{max}}
\newcommand{\threshold}{\theta}

\newcommand{\DistanceWeightConstant}{Z}

\newtheoremstyle{problemstyle}  % <name>
{3pt}                                               % <space above>
{3pt}                                               % <space below>
{\normalfont}                               % <body font>
{}                                                  % <indent amount}
{\bfseries\itshape}                 % <theorem head font>
{\normalfont\bfseries:}         % <punctuation after theorem head>
{.5em}                                          % <space after theorem head>
{}                                                  % <theorem head spec (can be left empty, meaning `normal')>
\theoremstyle{problemstyle}
\newcommand\blue[1]{\textcolor{blue}{#1}}
\newcommand{\OurModel}{\textsc{PeGaSus}\xspace}
\newcommand{\method}{\OurModel}

\newcommand{\SSumM}{\textsc{SSumM}\xspace}
\newcommand{\SAAGs}{\textsc{SAAGs}\xspace}
\newcommand{\kGrass}{\textsc{k-Grass}\xspace}
\newcommand{\SL}{\textsc{S2L}\xspace}

\newcommand{\answerMeasure}{x}
\newcommand{\approxMeasure}{\hat{x}}
\newcommand{\SMAPE}{\textsc{SMAPE}\xspace}
\newcommand{\Spearman}{\textsc{SC}\xspace}
\newcommand{\RWR}{\textsc{RWR}\xspace}
\newcommand{\HOP}{\textsc{HOP}\xspace}
\newcommand{\PHP}{\textsc{PHP}\xspace}

\newcommand{\smallsection}[1]{{\vspace{0.02in} \noindent {\bf{\underline{\smash{#1}}}}}}

\AtBeginDocument{%
  \providecommand\BibTeX{{%
    \normalfont B\kern-0.5em{\scshape i\kern-0.25em b}\kern-0.8em\TeX}}}

\begin{document}

\title{Personalized Graph Summarization: Formulation, Scalable Algorithms, and Applications}

\author{
\IEEEauthorblockN{Shinhwan Kang,\textsuperscript{1} Kyuhan Lee,\textsuperscript{1} and Kijung Shin\textsuperscript{1,2}}
\IEEEauthorblockA{\textit{\textsuperscript{1}Kim Jaechul Graduate School of AI and \textsuperscript{2}School of Electrical Engineering, KAIST, Seoul, South Korea} \\
\{shinhwan.kang, kyuhan.lee, kijungs\}@kaist.ac.kr} 
}

\maketitle

\noindent{\small \textit{``Everything is related to everything else, but near things are more related than distant things''} - \textbf{1st Law of Geography} (Tobler 1970).}

\vspace{2mm}

\begin{abstract}
    \textit{Are users of an online social network interested equally in all connections in the network? If not, how can we obtain a summary of the network personalized to specific users?
Can we use the summary for approximate query answering?}

As massive graphs (e.g., online social networks, hyperlink networks, and road networks) have become pervasive, graph compression has gained importance for the efficient processing of such graphs with limited  resources. 
Graph summarization is an extensively-studied lossy compression method. It provides a summary graph where nodes with similar connectivity are merged into supernodes, and a variety of graph queries can be answered approximately from the summary graph.

In this work, we introduce a new problem, namely \textit{personalized graph summarization}, where the objective is to obtain a summary graph where more emphasis is put on connections closer to a given set of target nodes. Then, we propose \OurModel, a linear-time algorithm for the problem.
Through experiments on six real-world graphs, we demonstrate that \OurModel is (a) \textit{Effective:} node-similarity queries for target nodes can be answered significantly more accurately from personalized summary graphs than from  non-personalized ones of similar size, (b) \textit{Scalable:} it summarizes graphs with up to one billion edges, and (c) \textit{Applicable to distributed multi-query answering}: it successfully replaces graph partitioning for communication-free multi-query processing.    
\end{abstract}


\section{Introduction}
\label{sec:intro}

A graph is an abstract data structure, and it naturally represents a wide range of data, including hyperlink networks \cite{boldi2004webgraph,page1999pagerank}, online social networks \cite{dhulipala2016compressing,shin2019sweg}, collaboration networks \cite{ramasco2004self}, and co-purchasing networks \cite{leskovec2007dynamics}.
Such real-world graphs grow rapidly as data modeled by them are accumulated at an unprecedented pace.

As a result, many real-world graphs are too large to fit in main memory, while real-time processing of various complex graph queries requires them to be resident in main memory of a single machine. Real-time answering of complex queries on graphs often requires fast random access into memory, and thus, if graphs are disk-resident and/or distributed across multiple machines, answering such queries incurs significant I/O overhead, preventing real-time processing. 

As a promising approach to address the above challenge, 
\textit{graph summarization} \cite{shin2019sweg,lee2020ssumm,khan2015set,beg2018scalable,riondato2017graph,lefevre2010grass,ko2020incremental} has received much attention among many graph-compression techniques \cite{dhulipala2016compressing,boldi2004webgraph,chierichetti2009compressing,buehrer2008scalable,fan2012query,henecka2015lossy,tsalouchidou2018scalable}.
%\red{[KJ: Add a sentence.. while graph summarization means]}
Its objective is to compress a given graph $\InputGraph$ in a lossy way while satisfying a given space budget, which is typically the amount of main memory in a machine.
The output, which we call \textit{summary graph}, is in a form of a graph where each node indicates a group of nodes in $\InputGraph$ and each edge between two groups indicates the presence of edges between a large fraction of pairs of nodes in the two groups. 
Note that we use the term ``graph summarization'' to refer to this  specific way of compression throughout this paper, while the term has been used to refer to a number of related but different concepts \cite{liu2018graph}.

\begin{figure}[t]
    \centering
    \subfigure[Input graph]{
        \includegraphics[width=0.33\textwidth]{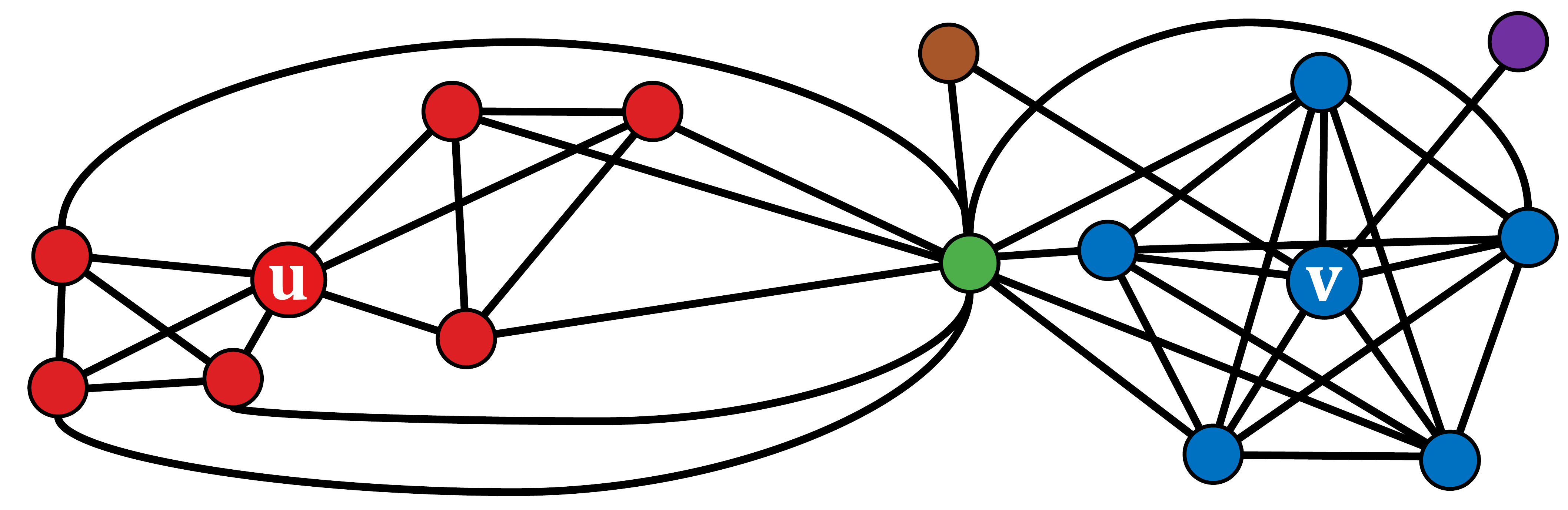}
    } \\
    \vspace{-3mm}
    \subfigure[Summary graph personalized to the node $u$]{
        \includegraphics[width=0.215\textwidth]{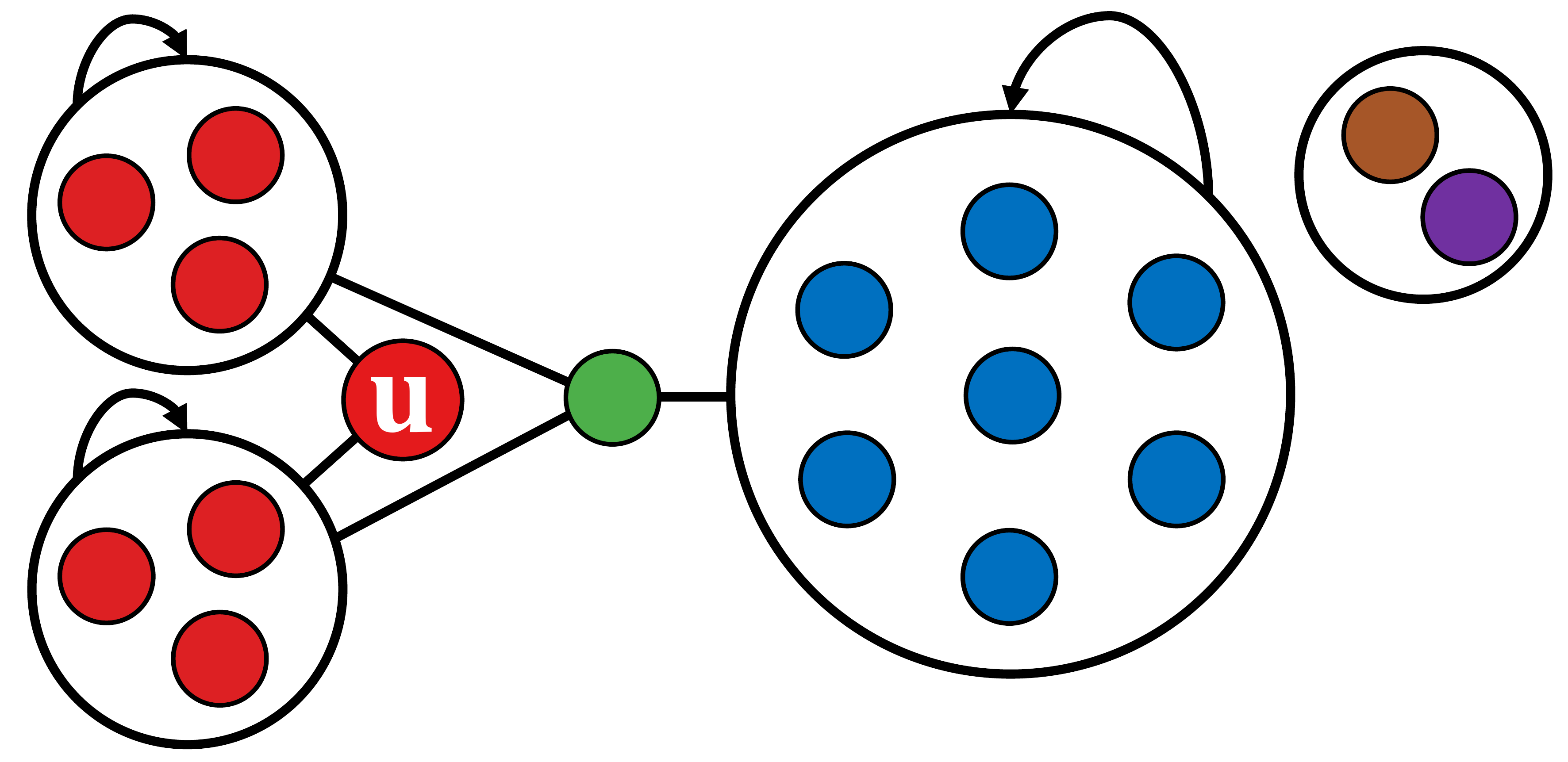}
    }
    \subfigure[Summary graph personalized to the node $v$]{
        \includegraphics[width=0.215\textwidth]{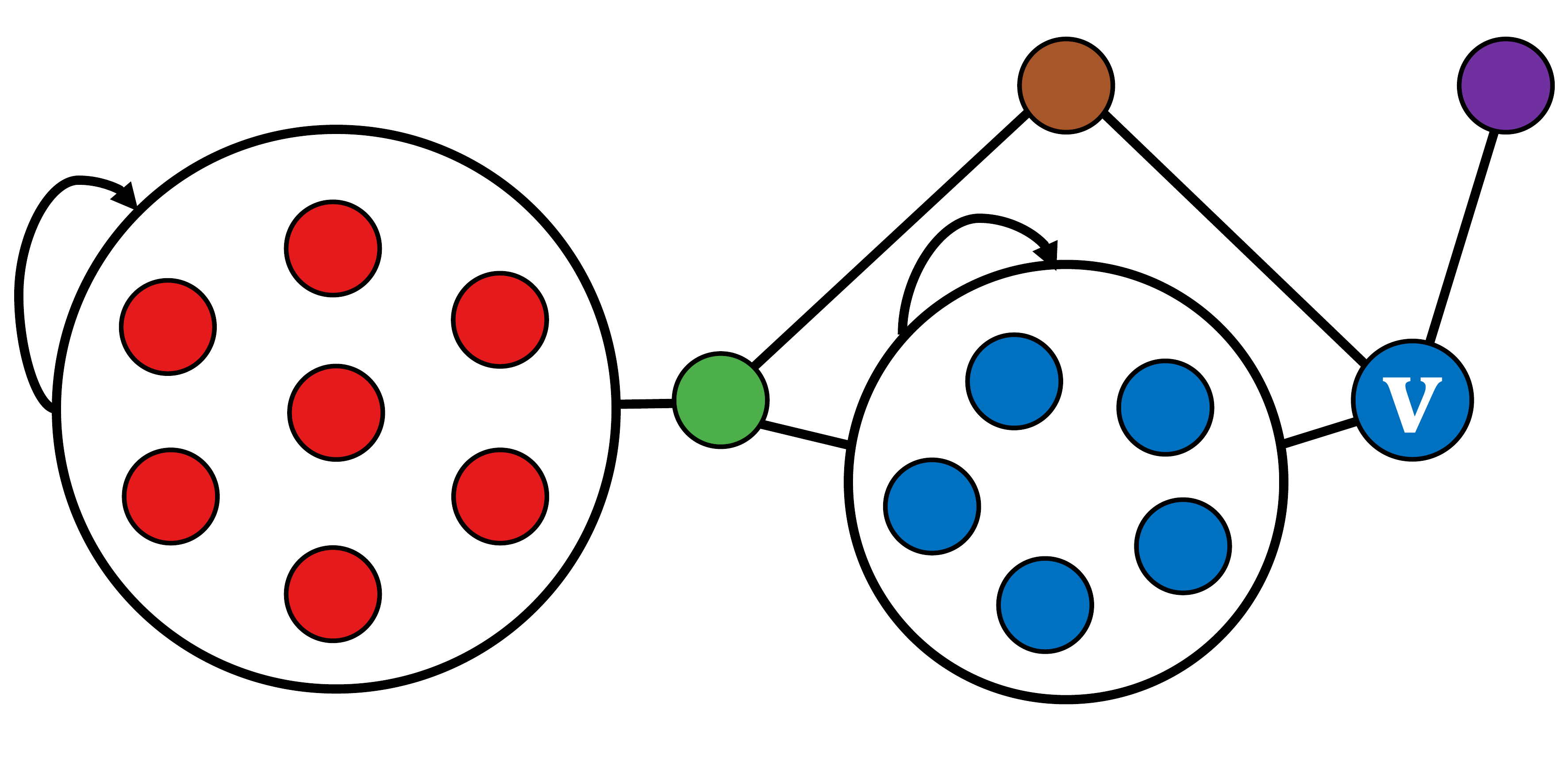}
    } \\
    \vspace{-2mm}
    \caption{\textbf{An illustrating example of personalized graph summarization.}
    From the input graph (top), two summary graphs are obtained. Supernodes group nodes, and each superedge between two supernodes indicates dense connections across the nodes in them. Similarly, the self-loop of each supernode indicates dense connections within the supernode.
    Note that one summary graph (middle) is personalized to the the node $u$, accurately preserving the edges near $u$.
    The other (bottom) is personalized to the the node $v$, which focus relatively more on the edges near $v$.
    \label{fig:IdeaFigure}}
\end{figure}

A key benefit of graph summarization is that a variety of graph queries can be directly be approximately answered from the output, in addition to many other benefits, including the interpretability of its output, the scalability its solvers, and its combinability with other graph-compression techniques.\footnote{Since an output is in the form of a graph, it can be further compressed using any graph-compression techniques.}
This is because, the neighborhood query (i.e., retrieving the neighbors of a given node) can be approximately answered directly from a summary graph, and many graphs algorithms, including node degrees \cite{riondato2017graph}, clustering coefficients \cite{riondato2017graph}, eigenvector centrality \cite{lefevre2010grass}, hops between nodes, and random walk with restart, access graphs only by the neighborhood query (see Appendix~\ref{sec:appendix:query} for examples).

\begin{figure}[t]
    \vspace{-4mm}
    \centering
    \subfigure[Effective]{
        \label{fig:crown:pe}
        \includegraphics[width=0.133\textwidth]{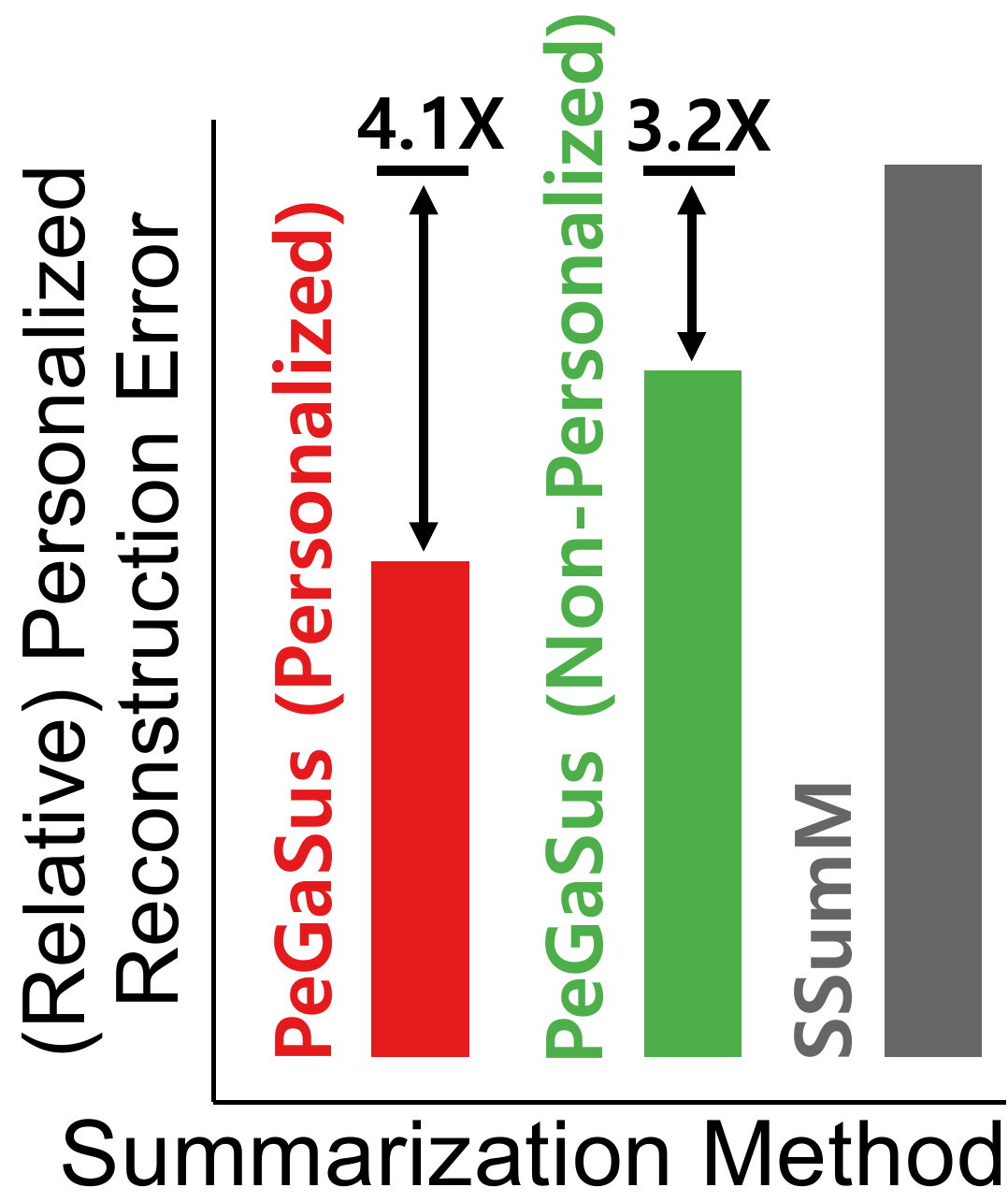}
    }
    \subfigure[Scalable]{
        \label{fig:crown:scalable}
        \includegraphics[width=0.154\textwidth]{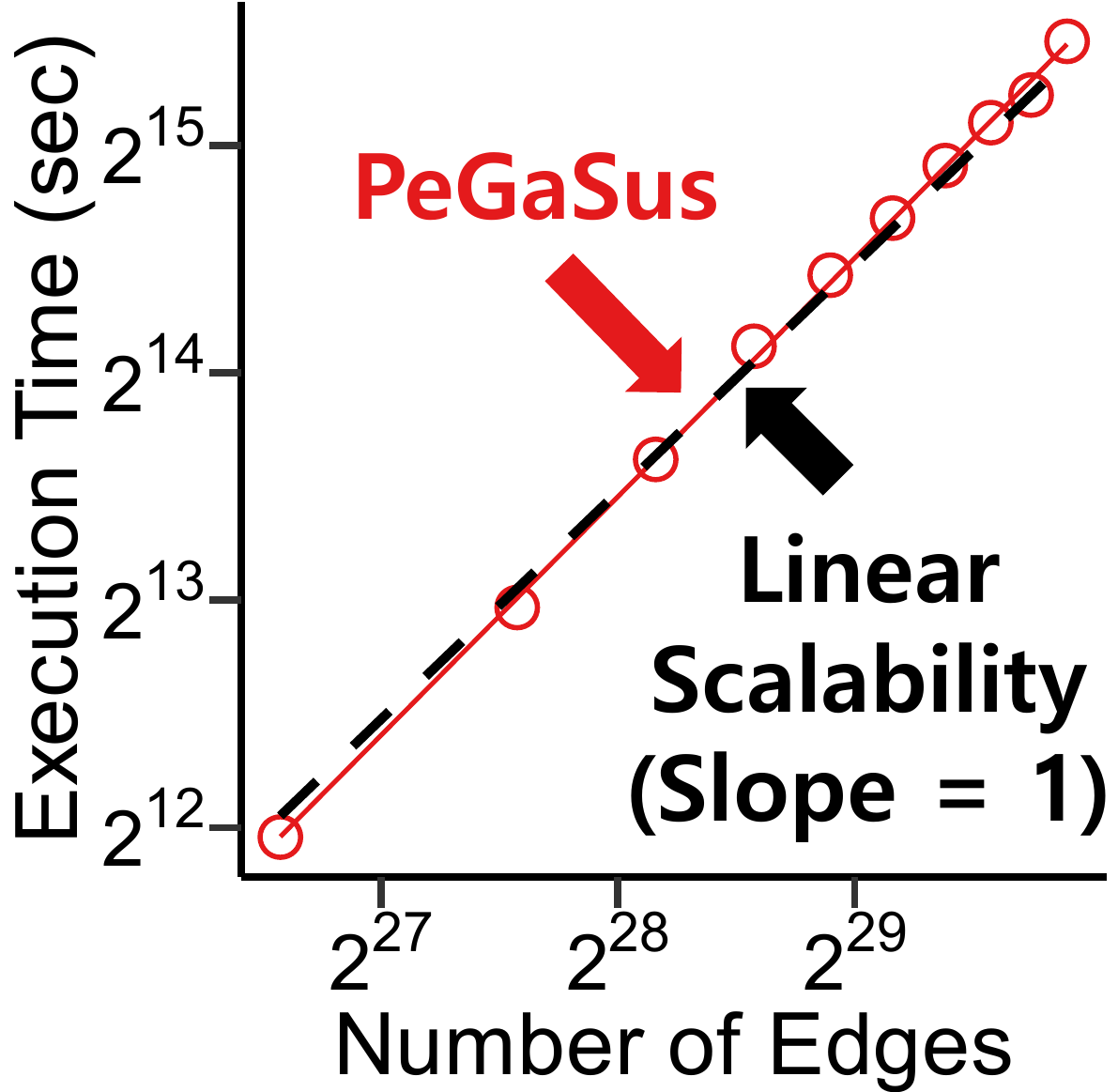}
    }
    \subfigure[Applicable]{
        \label{fig:crown:app}
        \includegraphics[width=0.1485\textwidth]{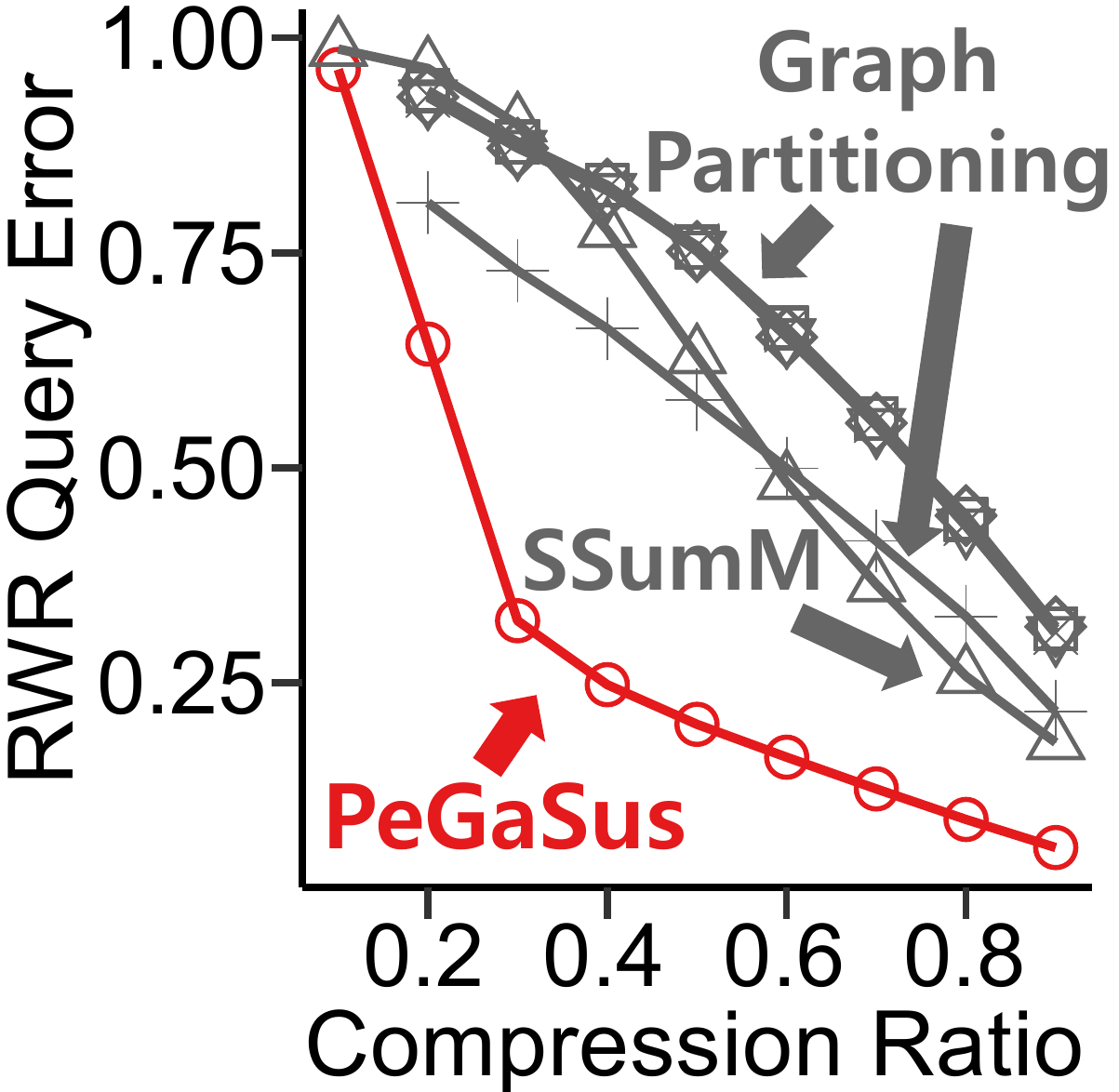}
    } \\ \vspace{-2mm}
    \caption{\label{fig:crown} \textbf{\OurModel is effective, linearly scalable, and applicable to distributed multi-query answering.} (a) \OurModel successfully personalizes summary graphs, (b) \OurModel is linear in the number of edges, scaling to one billion edges.
    (c) \OurModel answers RWR queries accurately in a distributed but ``communication-free'' manner. See Sect.~\ref{sec:experiments} for details. 
    }
\end{figure}

In this work, we introduce a new problem called personalized graph summarization.
It is motivated by the fact that people often have different levels of interest in different parts of a graph, following Tobler's first law of geography \cite{tobler1970computer}: \textit{``everything is related to everything else, but near things are more related than distant things.''}
For example, users in online social networks are more interested in connections of their close friends than in those of strangers.
Moreover, travelers navigating a road network are more interested in the roads near them than in those far from them.

Given a graph $G$, a set of target nodes $T$, and a space budget $k$, \textit{personalized graph summarization} is to find a summary graph of $G$ that is personalized to $T$ while satisfying the budget $k$.
We formulate it as an optimization problem whose objective, namely \textit{personalized error}, weighs error differently depending on the distance between the location of error and $T$. Specifically, the weight of error at closer locations is larger, requiring the output summary graph to be focused more on parts closer to target nodes $T$ (see Fig.~\ref{fig:IdeaFigure} for an example).

Our algorithmic contribution is to design \OurModel (\textbf{Pe}rsonalized \textbf{G}r\textbf{a}ph \textbf{Su}mmarization with \textbf{S}calability), a linear-time algorithm for the personalized summarization problem.
It is largely based on \SSumM, which is the state-of-the-art algorithm  \cite{lee2020ssumm} for non-personalized graph summarization, with an improved balance between exploration and exploitation. %, and (b) binarization of summary graphs for further compression.
Through extensive experiments, we demonstrate that \OurModel provides personalized summary graphs (see Fig.~\ref{fig:crown}(a)) from which three types of node-similarity queries for target nodes are answered significantly more accurately than from non-personalized summary graphs of similar size obtained by previous solvers \cite{lee2020ssumm,lefevre2010grass,riondato2017graph,beg2018scalable}.

\smallsection{Application to Distributed Multi-Query Answering.} Much effort has been made to minimize the communication between machines when answering queries on graphs distributed over multiple machines \cite{shao2013trinity, sarwat2013horton+, ngo2018worst, khan2018smart}. 
We demonstrate that \OurModel can be useful for  ``communication-free'' distributed multi-query answering.
Specifically, we assume approximate multi-query answering by multiple machines that act independently without I/O overhead over the network.
To this end, using \OurModel, we produce multiple personalized summary graphs with focuses on different regions of the input graph with a mapping function from nodes to summary graphs. 
The summary graphs are loaded in different machines, and we assign each query to machines on which the queries can be processed accurately, based on the mapping function.
We demonstrate through experiments that three types of node-similarity queries are answered more accurately without communication from multiple summary graphs than from partitions of graphs (see Fig.~\ref{fig:crown}(c)).

We summarize our contribution as follows:
\begin{itemize}
    \item \textbf{Problem Formulation:} We introduce a new problem, personalized graph summarization, and demonstrate the usefulness of personalizing summary graphs (see Fig.~\ref{fig:crown}(a)). 
    \item \textbf{Algorithm Design:} We propose \OurModel, a linear-time algorithm for the problem. We show empirically that it scales to a graph with one billion edges (see Fig.~\ref{fig:crown}(b)).
    \item \textbf{Extensive Experiments:} Using six real-world graphs, we exhibit the effectiveness of \OurModel and its applicability to distributed multi-query answering (see Fig.~\ref{fig:crown}(c)).
\end{itemize}
The source code and the datasets are available at \cite{appendixurl}. 

In Sect.~\ref{sec:preliminaries}, we introduce some concepts and formally define the personalized graph summarization problem. 
In Sect.~\ref{sec:methods}, we present \OurModel.
In Sect.~\ref{sec:application}, we discuss its application to distributed multi-query answering.   
In Sect.~\ref{sec:experiments}, we evaluate \OurModel. 
After reviewing related studies in Sect.~\ref{sec:relatedworks}, we draw a conclusion in Sect.~\ref{sec:conclusion}.

\vspace{-1mm}
\section{Concepts \& Problem Definition}
\label{sec:preliminaries}

\begin{table}[]
	\vspace{-3mm}
	\small
	\begin{center}
		\caption{Frequently-used symbols and their definitions.}
		\vspace{-2mm}
		\scalebox{0.88}{
		    \label{tab:symbol}
			\begin{tabular}{l|l}
				\toprule 
				\textbf{Symbol}  & \textbf{Definition}\\
				\midrule
				$\InputGraph = (V,E)$ & input graph with nodes $V$ and edges $E$\\
				$\{u,v\}\in E$ & edge between nodes $u\in V$ and $v\in V$\\
				$\AdjacencyMatrix$ & adjacency matrix of $\InputGraph$\\
				\midrule
				$\SummaryGraph = (S,P)$ & summary graph with supernodes $S$, superedges $P$\\
				$\SuperIdxSubnodei \in S$ & supernode containing the node $u\in V$ \\
				$\{A,B\}\in P$ & superedge between supernodes $A\in S$ and $B\in S$\\
				$\PIS$ & set of possible unordered pairs of supernodes\\
				${A \choose 2}$ & set of size-2 subsets of $A$ \\
				\midrule
				$\ReconstructedGraph = (V,\hat{E})$ &  reconstructed graph with nodes $V$, edges $\hat{E}$\\
				$\ReconstructedAdjacencyMatrix$  &  reconstructed adjacency matrix of $\ReconstructedGraph$\\		
				\midrule
				$\TargetNodeSet\subseteq V$ & target node set \\
				$\alpha$ & degree of personalization \\
				$\SizeBudget$  &  desired size in bits of the output summary graph\\ 
				\midrule
				$A\cup B$ & supernode into which supernodes $A$ and $B$ are merged \\
				$\theta$ & threshold for adaptive thresholding \\
				$L$ & list for adaptive thresholding \\
				$\beta$ & parameter for adaptive thresholding  \\
				$\{C_1,\dots,C_q\}$ & candidate groups \\
				$\MaxIterations$ & maximum number of iterations \\
				\bottomrule
			\end{tabular}
		}
	\end{center}
	   \vspace{-2mm}
\end{table}

In this section, we introduce basic concepts and formalize our new problem, namely the personalized graph summarization problem.
The frequently-used notations are listed in Table~\ref{tab:symbol}.

\vspace{-1mm}
\subsection{Basic Concepts}
\label{sec:notations}
\vspace{-1mm}

\smallsection{Input Graph}. An input graph $\InputGraph = (V,E)$ consists of a set of \textit{nodes} $V = \{1,2, ..., |V|\}$ and a set of \textit{edges} $E\subseteq{ V \choose 2 }$.
We assume that $\InputGraph$ is undirected without self-loops for simplicity.
The \textit{adjacency matrix} $\AdjacencyMatrix \in \mathbb{R}^{|V|\times|V|}$ of $\InputGraph$ encodes the connectivity in $\InputGraph$ as follows:
\vspace{-2mm}
\begin{equation*}
\AdjacencyMatrix_{uv} := 
    \begin{cases}
        1, & \text{if $\{u, v\} \in E$,} \\
        0, & \text{otherwise,}
    \end{cases} \ \forall u\in V, \forall v\in V.
    \label{eq:adjmatrix}
\end{equation*}
\vspace{-2mm}

\smallsection{Summary Graph}. A \textit{summary graph} $\SummaryGraph = (S,P)$ of $\InputGraph = (V,E)$ consists of a set of \textit{supernodes} $S$ and a set of \textit{superedges} $P$, and the set $S$ is a partition of $V$.
That is, supernodes are disjoint sets whose union is $V$, and equivalently, each node belongs to exactly one supernode.
We use $\SuperIdxSubnodei \in S$ to denote the supernode containing each node $u \in V$.
The summary graph $\SummaryGraph$ is undirected and it may have self-loops.
Thus, $P\subseteq \PIS$ holds where $\PIS:={S \choose 2}\cup\{\{A,A\}: A\in S\}$ denotes the set of all possible unordered pairs of supernodes.
We use
$\{A,B\} \in P$ to denote the superedge that joins $A \in S$ and $B \in S$.

\begin{figure}[t]
    \centering
    \subfigure[Example with exact reconstruction]{
        \label{fig:minhashing:top}
        \includegraphics[width=0.48\textwidth]{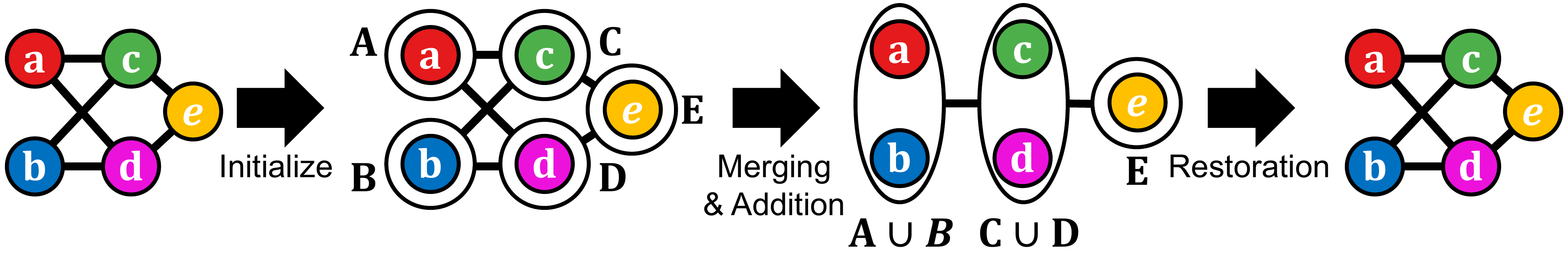}
    } \\
    \vspace{-1mm}
    \subfigure[Example with reconstruction error]{
        \label{fig:minhashing:bottom}
        \includegraphics[width=0.48\textwidth]{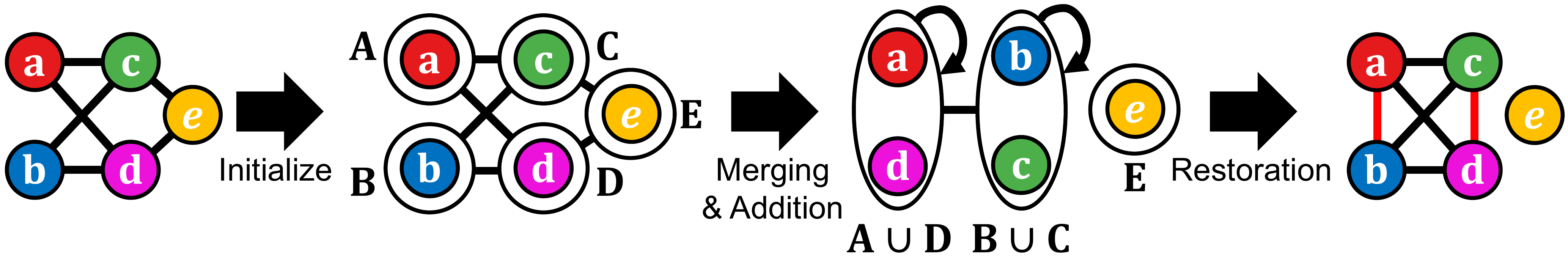}
    } \\
    \vspace{-2mm}
    \caption{\label{fig:minhashing} 
    \textbf{Two illustrating examples of graph summarization.}
    Note that merging supernodes with similar connectivity (e.g., $A$ and $B$) yields a more concise and accurate summary graph than merging those with dissimilar connectivity (e.g., $A$ and $D$).
   For restoration in (b), the superedge between $A\cup D$ and $B\cup C$ is interpreted as $\{a,b\}$, $\{a,c\}$, $\{b,d\}$, and $\{c,d\}$, The self-loops on $A\cup D$ and $B\cup C$ are interpreted as $\{a,d\}$ and $\{b,c\}$, resp.}
%   \vspace{-4mm}
\end{figure}

\smallsection{Reconstructed Graph}. 
From a summary graph $\SummaryGraph= (S,P)$, we can reconstruct a graph $\ReconstructedGraph = (V,\hat{E})$ with the set of nodes $V$ and the set of reconstructed edges $\hat{E} \subseteq{ V \choose 2 }$. 
In the reconstructed graph $\ReconstructedGraph$, there exists an edge between nodes $u$ and $v$ (i.e., $\{u,v\}\in \hat{E}$) if and only if the supernodes containing them (i.e., $\SuperIdxSubnodei$ and $\SuperIdxSubnodej$) are adjacent in $\SummaryGraph$ (see Fig.~\ref{fig:minhashing} for an example).
That is, the \textit{reconstructed graph} $\ReconstructedGraph$ from $\SummaryGraph$ is defined as the graph whose adjacency matrix satisfies
\begin{equation*} 
\ReconstructedAdjacencyMatrix_{uv} = 
    \begin{cases}
        1, & \text{if $u\neq v$ and $\{ \SuperIdxSubnodei, \SuperIdxSubnodej \} \in P$,} \\
        0, & \text{otherwise,}
    \end{cases} \forall u\in V, \forall v\in V.
    %\label{eq:reconadjmatrix}
\end{equation*}
Note that, a self-loop $\{A,A\}\in P$ indicates that each pair of nodes in $A$ (i.e., ${A \choose 2}$) is adjacent in $\ReconstructedGraph$.

\smallsection{(Non-personalized) Graph Summarization}.
Given (a) a graph $\InputGraph$ and (b) a budget on the number of supernodes \cite{lefevre2010grass} or the number of bits to encode $\SummaryGraph$ \cite{lee2020ssumm},
\textit{graph summarization} aims to find the summary graph $\SummaryGraph$ to minimize the difference  between $\InputGraph$ and the reconstructed graph $\ReconstructedGraph$ (e.g., Manhattan distance between their adjacency matrices), while satisfying the budget. A variety of graph queries can be answered directly from $\SummaryGraph$ without reconstructing $\ReconstructedGraph$ entirely (see Appendix~\ref{sec:appendix:query}).

\vspace{-1mm}
\subsection{Problem Formulation}
\label{sec:formulation}
\vspace{-1mm}

It is easy to imagine cases where people (e.g., users in online social networks and travelers navigating a road network) have different levels  of interest in different parts of a graph.
With this intuition, we generalize graph summarization \cite{lefevre2010grass,lee2020ssumm} to \textit{personalized graph summarziation}, where we aim to find a summary graph personalized to given target nodes, as formalized in Problem~\ref{prob:personalizedgs} (see Fig.~\ref{fig:IdeaFigure} for an example).

\vspace{0.5mm}
\noindent\fbox{%
        \parbox{0.98\columnwidth}{%
        \vspace{-2mm}
\begin{problem}[Personalized Graph Summarization]~
    \label{prob:personalizedgs}
    \begin{itemize}[leftmargin=*]
    	\item \textbf{Given:}
    	    \begin{itemize}
    	        \item a graph $\InputGraph = (V,E)$
    	        \item  a set of target nodes $\TargetNodeSet(\subseteq V)$
    	        \item a size budget $\SizeBudget >0$
    	    \end{itemize}
    	\item \textbf{Find:} a summary graph $\SummaryGraph = (S,P)$
    	\item \textbf{to Minimize:} personalized error $\PersonalizedError(\SummaryGraph)$
    	\item \textbf{Subject to:} $Size(\SummaryGraph) \leq $ \SizeBudget.
    \end{itemize}
\end{problem}
        \vspace{-2mm}
        }%
    }
\vspace{0.5mm}

\noindent Below, we describe $\PersonalizedError(\SummaryGraph)$ and $Size(\SummaryGraph)$.

\smallsection{Personalized Error}. 
We design \textit{peronsalized error} as a weighted reconstruction error in the form of
\begin{equation}
    \label{eq:personalizederror}
    \PersonalizedError(\SummaryGraph) := \sum_{u=1}^{|V|}\sum_{v=1}^{|V|} \Weight_{uv}|\AdjacencyMatrix_{uv} - \ReconstructedAdjacencyMatrix_{uv}|.
\end{equation}
Based on the first law of geography \cite{tobler1970computer}, i.e., \textit{``everything is related to everything else, but near things are more related than distant things,''}
we design the personalized weight $\Weight_{uv}$ on each node pair $\{u,v\}$ so that it depends on the number of hops between them and the target node set $T$ as follows:

\begin{equation}
\label{eq:distanceweightsubnodeij}
\Weight_{uv}:=\frac{\alpha^{-(D(u, \TargetNodeSet)+D(v, \TargetNodeSet))}}{\DistanceWeightConstant},
\end{equation}
where $D(u, \TargetNodeSet):=\min_{t \in T}\#hops(u,t)$ is the minimum number of hops between $u$ and any target node, 
$\DistanceWeightConstant$ is the constant that makes the average weight $1$,\footnote{$Z:= \frac{\left(\sum_{u=1}^{V} \alpha^{-D(u,T)}\right)^{2} - \left(\sum_{v=1}^{V} \alpha^{-2\times D(v,T)}\right)}{|V|(|V|-1)}$ is the average personalized weight over all possible pairs of nodes.}
and $\alpha\geq 1$ is a constant that controls the degree of personalization. 
That is, the closer an edge is to target nodes, the larger its weight is.
In other words, we aim to find a summary graph of $\InputGraph$ with greater focuses on parts closer to target nodes.

\smallsection{Graph Size}. For the size $Size(\SummaryGraph)$ of a summary graph $\SummaryGraph$, we use the number of bits to encode $\SummaryGraph = (S,P)$ as in \cite{lee2020ssumm}.
We assume an encoding method that requires 
$2\cdot \log_{2}|S|$ bits to encode each superedge (i.e., $\log_{2}|S|$ bits per incident supernode) and $\log_{2}|S|$ bits to encode the supernode membership of each node $v\in V$.
That is,
\begin{equation}
    Size(\SummaryGraph):= 2|P|\log_{2}|S| + |V|\log_{2}|S|.
    \label{eq:size:summarygraph}
\end{equation}
Similarly, the size of an input graph $\InputGraph = (V,E)$ in bits is
\begin{equation}
    Size(\InputGraph):= 2|E|\log_{2}|V|. 
    \label{eq:size:inputgraph}
\end{equation}

\begin{figure*}[!t]
    \centering
    \vspace{-6mm}
    \includegraphics[width=\textwidth]{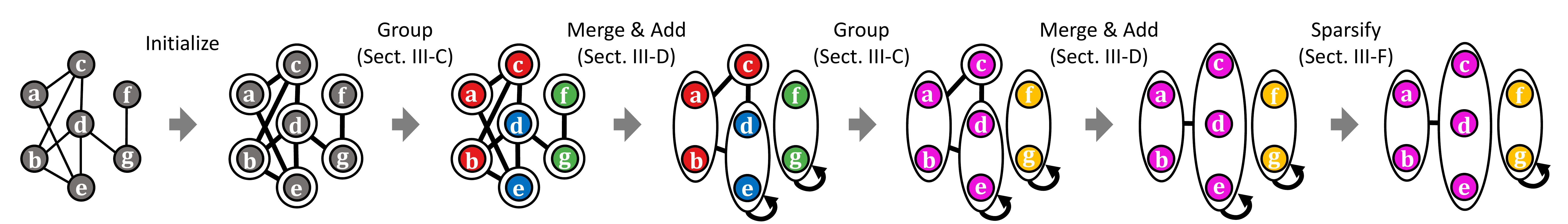} \\
    \vspace{-2mm}
    \caption{\label{fig:overview} \textbf{A pictorial description of \OurModel with an illustrative toy example.} Supernodes are colored according the groups that they belong to (i.e., shingles) at each iteration.
    Note that supernodes are merged only among those in the same group.
    }
\end{figure*}

\vspace{-1mm}
\section{Proposed Methods: \OurModel}
\label{sec:methods}

In this section, we present \OurModel (\textbf{Pe}rsonalized \textbf{G}r\textbf{a}ph \textbf{Su}mmarization with \textbf{S}calability), our proposed algorithm for Problem~\ref{prob:personalizedgs}.
We first outline \OurModel. Then, we present the cost function based on which \OurModel performs greedy search. After that, we describe the detailed procedures of greedy search. After making a comparison with \SSumM \cite{lee2020ssumm}, which \OurModel is largely based on, we analyze the time and space complexity.
It should be noted that \OurModel is a heuristic without approximation guarantees.
\footnote{Most graph-summarization algorithms \cite{lefevre2010grass,beg2018scalable,lee2020ssumm} are heuristic without approximation guarantees. While S2L~\cite{riondato2017graph} guarantees a constant approximation ratio, empirically, \SSumM outperforms S2L \cite{lee2020ssumm}.}

\subsection{Overview (Alg.~\ref{alg:overview})}
\label{sec:overview}
We provide the pseudocode of \OurModel in Alg.~\ref{alg:overview} and an illustrative example in Fig.~\ref{fig:overview}.
Given an input graph $\InputGraph=(V,E)$, a size budget $\SizeBudget$, a target node set $\TargetNodeSet$, the degree of personalization $\alpha$, a parameter for adaptive thresholding $\beta$, and the maximum number of iterations $\MaxIterations$, \OurModel produces a summary graph $\SummaryGraph=(S,P)$ personalized for $\TargetNodeSet$.
First, it initializes the supernode set $S$ so that each node $u\in V$ composes a singleton supernode $\SuperIdxSubnodei$ (line~\ref{alg:line:intialize}).
For each edge $\{u,v\} \in E$, we create a superedge between supernodes $\SuperIdxSubnodei$ and $\SuperIdxSubnodej$ (see Fig.~\ref{fig:minhashing} for examples).
Then, \OurModel updates $\SummaryGraph$ by repeating the following steps until its size meets the budget or the maximum number of iterations is reached (line~\ref{alg:line:termination}).
\begin{itemize}
    \item \textbf{Candidate Generation (line~\ref{alg:line:candiGenerate1}, Sect.~\ref{sec:candidategeneration})}: \OurModel divides $S$ into groups of supernodes with similar connectivity whose merger is likely to reduce the personalized cost (i.e., our objective function described in Sect.~\ref{sec:costfunction}).
    The groups are updated in every iteration.
    \item \textbf{Merging and Addition (line~\ref{alg:line:merge1}-\ref{alg:line:merge2}, Sect.~\ref{sec:mergingandencoding})}: Within each group, \OurModel repeats merging supernodes and selectively adding superedges incident to the merged supernode. 
\end{itemize}
During the above process, \OurModel balances exploitation and exploration using the threshold $\theta$, and $\theta$ is updated adaptively (line~\ref{alg:line:changingtheta}), as described in detail in Sect.~\ref{sec:thresholding}.
If the size of $\SummaryGraph$ still exceeds the budget $\SizeBudget$, \OurModel sparsifies $\SummaryGraph$ until the budget is met (lines~\ref{alg:line:notmerge1}-\ref{alg:line:notmerge2}), as described in detail in Sect.~\ref{sec:furthersparsification}.

\begin{algorithm}[t]
    \small
    \caption{Overview of \OurModel \label{alg:overview}}
    \DontPrintSemicolon
    \KwInput{(1) input graph $\InputGraph = (V, E)$, (2) size budget $\SizeBudget$,\\
    \hspace{9.8mm}(3) target node set $\TargetNodeSet$, (4) degree of personalization $\alpha$,\\
    \hspace{9.8mm}(5) parameter for adaptive thresholding $\beta$, \\
    \hspace{9.8mm}(6) max. number of iterations $\MaxIterations$,\\
    }
    \KwOutput{summary graph $\SummaryGraph = (S, P)$}
    \SetAlgoLined
        $S \leftarrow \{\{u\} : u \in V\}$; $P \leftarrow \{\{\{u\}, \{v\}\} : \{u,v\}\in E\}$ \label{alg:line:intialize}\\
        $t \leftarrow 1$; $\threshold \leftarrow 0.5$; $L \leftarrow$ an empty list \label{alg:line:initializeList}\\ 
        \While{$t \leq \MaxIterations$ \normalfont{and} $Size(\SummaryGraph) > \SizeBudget$ \label{alg:line:termination}}{  
            $C \leftarrow$ generate candidate groups  \Comment*[r]{Sect.~\ref{sec:candidategeneration}} \label{alg:line:candiGenerate1}
            \For{\normalfont{\textbf{each} group}\ $C_{i} \in C$}{ \label{alg:line:greedilyMandE1}
                 greedily merge nodes in $C_{i}$ with the threshold $\theta$ (update $S$, $P$, and $L$) \Comment*[r]{Sect.~\ref{sec:mergingandencoding}} \label{alg:line:greedilyMandE2}
            }
            $\threshold \leftarrow$ $\lfloor \beta\times|L|\rfloor$-th largest entry in $L$
            \Comment*[r]{Sect.~\ref{sec:thresholding}}\label{alg:line:changingtheta}
            $L \leftarrow$ an empty list; $t \leftarrow t+1$ \label{alg:line:resetList}
        }
        \If{$Size(\SummaryGraph) > \SizeBudget$}{ \label{alg:line:further1}
            sparsify $\SummaryGraph$ further \Comment*[r]{Sect.~\ref{sec:furthersparsification}} \label{alg:line:furthersparsification} 
        }\label{alg:line:further3}
        \textbf{return} $\SummaryGraph = (S, P)$ 
\end{algorithm}

\subsection{Personalized Cost Function}
\label{sec:costfunction}

In this subsection, we introduce the cost function that \OurModel uses while performing greedy search. 

\smallsection{Total Cost.}
In order to decide a pair of supernodes to be merged, both the size (i.e., Eq.~\eqref{eq:size:summarygraph}) of summary graphs after mergers and the personalized error (i.e., Eq.~\eqref{eq:personalizederror}) needs to be taken into consideration. We define the \textit{personalized cost} of each summary graph $\SummaryGraph$ as
\begin{equation}
    \label{eq:costfunc}
    \CostFunction(\SummaryGraph) := Size(\SummaryGraph) + \log|V| \cdot \PersonalizedError(\SummaryGraph).
\end{equation}
Note that $\log|V|\cdot\NonPersonalizedError(\SummaryGraph)$ equals the number of bits required to encode the difference between the input graph $\InputGraph$ and the graph $\ReconstructedGraph$ reconstructed from $\SummaryGraph$; and  $\log|V|\cdot\PersonalizedError(\SummaryGraph)$ can be regarded as its personalized version.\footnote{\label{footnote:encode}$\sum_{u=1}^{|V|}\sum_{v=1}^{|V|}|\AdjacencyMatrix_{uv} - \ReconstructedAdjacencyMatrix_{uv}|$ entries in the adjacency matrices are flipped, and due to symmetry, only a half of them needs to be specified.
Encoding the location (i.e., row and column) of each of them requires $2\cdot \log |V|$ bits.}
Since both the size and the personalized error are interpreted as the number of bits, aiming to minimize their sum in Eq.~\eqref{eq:costfunc} aligns with the minimum description length (MDL) principle \cite{grunwald2007minimum}.

\smallsection{Cost Decomposition.}
We define the cost for each unordered pair of supernodes $\{A,B\}\in \PIS$ (see Sect.~\ref{sec:notations} for $\PIS$) as
\begin{multline}
\label{eq:CostAB}
    \CostFunction_{AB}(\SummaryGraph):=  \\ 2\log_{2}|S| \cdot \mathbb{1}_{P}(\{A,B\})  + \log_{2}|V|\cdot \PersonalizedError_{AB}(\SummaryGraph),
\end{multline}
%수식 확인, 추가
where $\mathbb{1}_{P}(\{A,B\})=1$ if $\{A,B\} \in P$ and $0$ otherwise; and $\PersonalizedError_{AB}$ is the personalized error between $A$ and $B$, i.e., 
\begin{equation}
    \PersonalizedError_{AB}(\SummaryGraph):= \sum_{u\in A}\sum_{v\in B} \Weight_{uv}|\AdjacencyMatrix_{uv} - \ReconstructedAdjacencyMatrix_{uv}|.
\end{equation}
Recall that, by definition of $\PIS$, $A$ may equal $B$, and in such a case, $\PersonalizedError_{AB}(\SummaryGraph)= \sum_{
u \neq v\in A}\Weight_{uv}|\AdjacencyMatrix_{uv} - \ReconstructedAdjacencyMatrix_{uv}|$.

Then, the personalized cost in Eq.~\eqref{eq:costfunc} is decomposed into the cost for each supernode pair as follows: 
\begin{equation}
    \CostFunction(\SummaryGraph) =  |V|\log_{2}|S| + \sum_{\{A,B\}\in \PIS} \CostFunction_{AB}(\SummaryGraph).
\end{equation}
Additionally, we define the cost for each supernode $A\in S$ as
\begin{equation}
    \CostFunction_{A}(\SummaryGraph):=  \sum_{B\in S} \CostFunction_{AB}(\SummaryGraph). \label{eq:nodecost}
\end{equation}

\smallsection{Cost Reduction.}
Let $\MergeAB$ be the summary graph obtained if supernodes $A\in S$ and $B\in S$ are merged in the summary graph $\SummaryGraph=(S,P)$ (see Sect.~\ref{sec:mergingandencoding} for details),
Then, the reduction in the personalized cost in Eq.~\eqref{eq:costfunc} is
\begin{multline}
   \CostReductionAB  :=  \CostFunction_{A}(\SummaryGraph) 
    + \CostFunction_{B}(\SummaryGraph) - 
    \\ \CostFunction_{AB}(\SummaryGraph)  
    - \CostFunction_{A\cup B}(\MergeAB), \label{eq:CostReduction}
\end{multline}
where $(\CostFunction_{A}(\SummaryGraph) + \CostFunction_{B}(\SummaryGraph) - \CostFunction_{AB}(\SummaryGraph))$ is the sum of costs for supernodes $A$ and $B$ in  $\SummaryGraph$ (after removing duplicates), and $\CostFunction_{A\cup B}(\MergeAB)$ is the cost for the merged supernode $A\cup B$ after merging $A$ and $B$.
Based on $\CostReductionAB$ and $(\CostFunction_{A}(\SummaryGraph) + \CostFunction_{B}(\SummaryGraph) - \CostFunction_{AB}(\SummaryGraph))$, we define the relative cost reduction as
\begin{multline}
    \RelReductionAB := \\
    \frac{\CostReductionAB}{\CostFunction_{A}(\SummaryGraph)  + \CostFunction_{B}(\SummaryGraph)  - \CostFunction_{AB}(\SummaryGraph)}. \label{eq:RelReduction}
\end{multline}
As described in the following subsection, \OurModel uses Eq.~\eqref{eq:RelReduction} when deciding supernodes to be merged. 

For two nodes with dissimilar connectivity patterns, if they are distant from target nodes (that is, if the personalized weights of their incident edges are small), the absolute cost reduction in Eq.~\eqref{eq:CostReduction} can be small, while the relative cost reduction in Eq.~\eqref{eq:RelReduction} should be large.
Thus, when Eq.~\eqref{eq:CostReduction} is used, such nodes can be easily merged (myoypically even when there exist pairs of nodes with similar connectivity patterns), and thus summary graphs with large personalized error can be obtained.
We demonstrate experimentally in the online appendix~\cite{appendixurl} that using Eq.~\eqref{eq:RelReduction} results in better summary graphs where queries can be answered accurately, compared to using Eq.~\eqref{eq:CostReduction}.

\subsection{Candidate Generation}
\label{sec:candidategeneration}

In this subsection, we describe the candidate generation step, which accelerates \OurModel by reducing the search space.
\OurModel groups supernodes with similar connectivity (so that only pairs of nodes within the same group are considered to be merged) based on the following grounds:
\begin{itemize}
    \item It is impractical to consider all pairs of supernodes, whose number is ${S \choose 2}$, whenever deciding a node pair to be merged.
    \item Uniform sampling is likely to result in pairs of supernodes whose merger does not reduce the personalized cost much.
    \item Generally, if supernodes with similar connectivity are merged, encoding their connectivity together using superedges incurs little reconstruction error. For example, in Fig.~\ref{fig:minhashing}, merging supernodes $A$ and $B$ (and $C$ and $D$), which share exactly the same neighbors does not incur any reconstruction error, while merging $A$ and $D$ (and $B$ and $C$) with dissimilar connectivity leads to missing or incorrect edges. 
\end{itemize}

In order to group supernodes with similar connectivity, we refer to the fact that the probability of two nodes having the same shingle \cite{broder2000min} is equal to the jaccard similarity of their neighbor sets.
Specifically, we extend the notion of shingles to supernodes and define the shingle of each supernode $U\in S$ as
\begin{equation}
    \label{eq:minhash}
    F(U) := \min_{u \in U}\left(\min_{v\in N_{u}\cup \{u\}}f(v)\right),
\end{equation}
where $N_{u}$ is the set of neighboring nodes of $u\in V$ in the input graph $\InputGraph$ and $f:V \rightarrow \{1,2,...,|V|\}$ is a uniform random hash function.
Note that two supernodes are more likely to have the same shingle if their members' connectivities are more similar.

\begin{example}[Shingle]
    In Fig.~\ref{fig:minhashing}, suppose that $f(a)=5$, $f(b)=4$, $f(c)=3$, $f(d)=2$, and $f(e)=1$ in the input graph (leftmost). 
    After initialization, $F(A)=\min(f(a),f(c),f(d))=2$, and
    $F(B)=\min(f(b),f(c),f(d))=2$.
    Similarly, $F(C)=F(D)=F(E)=1$.
    After merging $A$ with $B$ and $C$ with $D$, $F(A\cup B)=\min(f(a),f(b),f(c),f(d))=2$ and similarly $F(C\cup D)=1$.
    Note that supernodes with similar connecitivity (e.g., $A$ and $B$; and $C$ and $D$) have the same shingle.
\end{example}
Then, the supernodes with the same shingle are grouped together as a candidate group.
\method further divides each candidate group recursively by repeating the above process at most a constant (spec., $10$) times.
After that \method ensures that the size of each candidate group is at most a constant (spec., $500$) by randomly dividing larger groups.
We use $C=\{C_{1}, C_{2}, \dots, C_{q}\}$ to denote the candidate groups, which are used in a later step (see Sect.~\ref{sec:mergingandencoding}).
In different iterations, \OurModel draws a hash function $f$ using different random seeds, and as a result, obtains different candidate groups to further explore the search space.

\begin{algorithm}[t]
    \small
    \caption{\label{alg:MergeingandEncoding}Merging and Addition Step}
\DontPrintSemicolon
    \KwInput{(1) current summary graph $\SummaryGraph = (S,P)$,\\
    \hspace{9.8mm}(2) candidate group $C_{i}$, (3) adaptive threshold $\threshold$,\\
    \hspace{9.8mm}(4) target node set $\TargetNodeSet$, (5) degree of personalization $\alpha$,\\
    \hspace{9.8mm}(6) input graph $\InputGraph = (V, E)$, (7) current list $L$ 
    }
    \KwOutput{updated summary graph $\SummaryGraph$ and list $L$}
    \SetAlgoLined
        $\#fails \leftarrow 0$\\
        \While{$|C_{i}|> 1$ \text{and} $\#fails \leq \log_{2}|C_{i}|$}{ \label{alg:stopCondition}
            $I \leftarrow$ sample $|C_{i}|$ pairs of supernodes in $C_{i}$ \label{alg:line:sampleI}\\
            $\{A,B\} \leftarrow \argmax\limits_{\{X,Y\}\in I} \label{alg:line:greedy} \RelReductionAB$ \Comment*[r]{Eq.~\eqref{eq:RelReduction}}
            \eIf{$\RelReductionAB \geq \threshold$ \label{alg:line:compare}}{
                remove $A$ and $B$ from $S$ and $C_{i}$ \label{alg:line:merge1}\\ 
                add $A\cup B$ to $S$ and $C_{i}$  \label{alg:line:merge2}\\
                remove superedges incident to $A$ or $B$ from $P$ \label{alg:line:encode1}\\
                add superedges incident to $A\cup B$ to $P$
                selectively to minimize $\CostFunction_{A\cup B}(\SummaryGraph)$  \label{alg:line:encode2} \Comment*[r]{Eq.~\eqref{eq:nodecost}}
                $\#fails \leftarrow 0$ \label{alg:line:encode3}\\
            }{
                add $\RelReductionAB$ to $L$ \label{alg:AdaptiveThresholdArrayAdd} \label{alg:line:notmerge1} \Comment*[r]{Eq.~\eqref{eq:RelReduction}}
                $\#fails \leftarrow \#fails + 1$ \label{alg:line:notmerge2}\\
            } 
        }
        \textbf{return} $\SummaryGraph = (S, P)$  and $L$
\end{algorithm}

\subsection{Merging and Addition (Alg.~\ref{alg:MergeingandEncoding})}
\label{sec:mergingandencoding}

In this subsection, we describe how \OurModel merges supernodes within each candidate group $C_{i}$ (see Sect.~\ref{sec:candidategeneration}) in a greedy manner
and creates superedges accordingly. See Alg.~\ref{alg:MergeingandEncoding} for the pseudocode.
\OurModel first draw $|C_i|$ pairs of supernodes uniformly at random within $C_{i}$ (line~\ref{alg:line:sampleI}). Then, it chooses a pair, which we denote by $\{A,B\}$, that maximizes the relative reduction in the personalized cost (i.e., Eq.~\eqref{eq:RelReduction} in Sect.~\ref{sec:costfunction}) (line~\ref{alg:line:greedy})
If the relative reduction $\RelReductionAB$ is at least the threshold $\theta$ (see Sect.~\ref{sec:thresholding} for how to decide $\theta$) (line~\ref{alg:line:compare}), then the chosen pair is merged (lines~\ref{alg:line:merge1}-\ref{alg:line:merge2}).
When $A$ and $B$ are merged, the superedges incident to $A$ or $B$ are removed from $P$ as they are no longer valid (line~\ref{alg:line:encode1}).
For the merged supernodes $A\cup B$, \OurModel adds superedges incident to it selectively to $P$ so that the personalized cost $\CostFunction_{A\cup B}(\SummaryGraph)$ (see Eq.~\eqref{eq:nodecost}) for $A\cup B$ is minimized, while fixing all non-incident superedges (line~\ref{alg:line:encode2}). It should be noted that, since each supernode is s set of subnodes, merging the two supernodes $A$ and $B$ results in $A\cup B$ (i.e., the union of the two sets).
Specifically, \OurModel adds each potential superedge $\{(A\cup B),X\}$ to $P$ if and only if it decreases the personalized cost $\CostFunction_{(A\cup  B)X}(\SummaryGraph)$ (see Eq.~\eqref{eq:CostAB})  for it. 
It should be noted that $X$ can be $A\cup B$, and thus a self-loop can be added to $A\cup B$.
The detailed procedure with the time complexity are provided in Lemma~\ref{lemma:linearScalability} and its proof.
We denote the updated $\SummaryGraph$ by $\MergeAB$.

\begin{lemma}\label{lemma:linearScalability} For any supernodes $A,B \in S$ in a summary graph $\SummaryGraph$, the time complexity of
updating $\SummaryGraph$ to $\MergeAB$ is 
$$O\left(\sum_{u\in A} |N_{u}|+\sum_{v\in B} |N_{v}|\right),$$
where $N_{x}$ is the set of neighbors of a node $x\in V$ in the input graph $\InputGraph$.
\end{lemma}
\begin{proof}
A proof can be found in the online appendix~\cite{appendixurl}.
\end{proof}

For each candidate group $C_i$, \method repeats the above process until (a) only one supernode is left in the group or (b) it fails to merge supernodes $\log(|C_i|)$ times in a row.
That is, if the relative reduction by a chosen pair is smaller than the threshold $\theta$, $\log(|C_i|)$ times in a row, \method concludes that no promising supernode pairs are left in $C_{i}$. 
The relative reduction smaller than $\theta$ is stored in a list $L$ (line~\ref{alg:line:notmerge1}), which is later used to adjust $\theta$, as described in the following subsection.

\subsection{Adaptive Thresholding}
\label{sec:thresholding}

In this subsection, we present how \method adjusts the threshold $\theta$ for relative reduction, which is used in line~\ref{alg:line:compare} of Alg.~\ref{alg:MergeingandEncoding}.
The threshold $\theta$ balances exploitation and exploration.
Specifically, if $\theta$ becomes smaller, supernode pairs are merged more aggressively within the candidate groups in the current iteration.
However, if $\theta$ becomes larger, \method merge pairs less aggressively, considering the possibility that better pairs can be found within the candidate groups in future iterations. 
Recall that \method obtains different candidate groups in different iterations.

\method intializes $\theta$ to $0.5$ and adjusts it adpatively based on the relative reductions stored in $L$ (line~\ref{alg:line:notmerge1} of Alg.~\ref{alg:MergeingandEncoding}).
Specifically, \method sets $\theta$ for the next iteration to the $\lfloor \beta\times|L|\rfloor$-th largest entry in $L$ and then clears $L$ (lines~\ref{alg:line:changingtheta}-\ref{alg:line:resetList} of Alg.~\ref{alg:overview}).
The larger the parameter $\beta$ is, the faster $\theta$ decreases, with a greater emphasis on exploitation.
The empirically effect of $\beta$ is shown in Sect.~\ref{sec:q:params}.
Recall that relative reductions in $L$ are smaller than $\theta$ for the current iteration (see Sect.~\ref{sec:mergingandencoding}).
Hence, \method gradually decreases $\theta$ over iterations, gradually putting more emphasis on exploitation.

\subsection{Further Sparsification}
\label{sec:furthersparsification}

If the size of the summary graph $\SummaryGraph$ stills exceeds the budget $k$ after $\MaxIterations$ iterations (line~\ref{alg:line:further1} of Alg.~\ref{alg:overview}),
\method greedily drops superedges in $P$ (line~\ref{alg:line:furthersparsification})  until the budget is met. 
Specifically, superedges dropped in increasing order of $\CostFunction_{AB}(\SummaryGraph)$ (Eq.~\eqref{eq:CostAB} in Sect.~\ref{sec:costfunction}), where $\{A,B\}\in P$ denotes a dropped superedge.

\subsection{Comparison with \SSumM \cite{lee2020ssumm}}
\label{sec:comparison}
\OurModel is largely based on \SSumM \cite{lee2020ssumm}, which is a state-of-the-art algorithm for non-personalized graph summarization, with the following major differences:
\begin{itemize}
    \item \textbf{Personalizability}: \OurModel yields a  summary graph personalized to given target nodes, while \SSumM produces a non-personalized one. 
    To this end, \method aims to minimize the personalized error (i.e., Eq.~\eqref{eq:personalizederror}), and it generalizes the reconstruction error, which \SSumM aims to minimize.
    Specifically, if $W^{(T)}_{uv}=1$ for all $u\neq v \in V$, then Eq.~\eqref{eq:personalizederror} becomes equal to the reconstruction error.
    \item \textbf{Applicability to Query Answering}: 
    \OurModel can generate multiple summary graphs with focuses on different regions of a graph, while \SSumM cannot.
    Based on the focused regions, we can choose a summary graph where a given query can be answered accurately  (see Sect.~\ref{sec:application}).
    \item \textbf{Adaptive Thresholding}: In order to balance exploitation and exploration, \OurModel controls the threshold $\theta$ adaptively based on statistics collected at runtime (see Sect.~\ref{sec:thresholding}), while \SSumM relies on a fixed rule. Specifically, \SSumM sets $\theta(t)$ to $(1+t)^{-1}$ if $t<\MaxIterations$ and to $0$ otherwise.
    \item \textbf{Minor Differences}:
   \OurModel uses new computational tricks for rapidly computing personalized error between two supernodes, which are not required in \SSumM, and to this end, maintains additional information (see Eqs. (13-15) in the online appendix \cite{appendixurl}).
    When converting reconstruction error between two supernodes into the number of bits, \SSumM assumes the best of two encoding schemes (entropy coding and error corrections\footnote{The number of bits required to encode the location (i.e., row and column) of each erroneous entry in the reconstructed adjacency matrix.}), while for simplicity, \OurModel assumes error corrections (see Eq.~\eqref{eq:costfunc} and Footnote~\ref{footnote:encode}).
    
\end{itemize}
Notably, the experiments in Sect.~\ref{sec:q:effectsofpersonalization} show that \method outperforms \SSumM in non-persoalized cases (i.e., when $T=V$), as well as in personalized cases, 

\subsection{Complexity Analysis}
\label{sec:complexityAnalysis}

In this section, we analyze the time and space complexity of \OurModel. We assume $|V|=O(|E|)$ for simplicity. % as in \cite{lee2020ssumm}.

\smallsection{Time Complexity:} \OurModel scales linearly with the size of the input graph, as formalized in Theorem~\ref{thm:linearScalability}.

\begin{theorem}[Linear Scalability of \OurModel] \label{thm:linearScalability}
	The time complexity of Alg.~\ref{alg:overview} is $O(\MaxIterations\cdot |E|)$.
\end{theorem}

\begin{proof} 
    We focus on showing that the differences of \method from \SSumM, whose time complexity is $O(\MaxIterations\cdot |E|)$ (see Theorem~3.4 of \cite{lee2020ssumm}), does not increase the time complexity. 
    According to Lemma~\ref{lemma:linearScalability}, computing the cost reduction for each supernode pair (based on which supernode pairs to be merged are chosen) takes $O(\sum_{u\in A} |N_{u}|+\sum_{v\in B} |N_{v}|)$ time, as in \SSumM. 
    Regarding adaptive thresholding, the $\lfloor \frac{|L|}{10}\rfloor$-th largest entry in $L$ (line~\ref{alg:line:changingtheta} of Alg.~\ref{alg:overview}) can be found in $O(|L|)$ time by the ``median of medians'' algorithm \cite{blum1973time}.
    Since the number of failures cannot exceed $\sum_{i=1}^{q}(|C_i|\log|C_i|)=O(|V|\cdot \max_{i=1}^{q}{\log |C_i|})=O(|V|)$, and thus $O(|L|)=O(|V|)$ holds. Recall that, as described in Sect.~\ref{sec:candidategeneration}, the size of each candidate group is at most a constant, and thus $O( \max_{i=1}^{q}{\log |C_i|})=O(1)$.
    Thus, $\MaxIterations$ updates of the threshold $\theta$ take $O(\MaxIterations\cdot |V|)=O(\MaxIterations\cdot |E|)$ time in total. 
\end{proof}

\smallsection{Space Complexity:}
\OurModel retains an input graph $\InputGraph=(V,E)$, a summary graph $\SummaryGraph=(S,P)$, a hash function $f:V\rightarrow \{1,2,...,|V|\}$, a shingle function $F:S\rightarrow \{1,2,...,|V|\}$, and a list $L$.
Additionally, it retains intermediate results of size $O(|V|+|E|)$ (see the online appendix~\cite{appendixurl} for details).
Since $|S|\leq|V|$, $|P|\leq|E|$, and $|L|=O(|V|)$ (see the proof of Theorem~\ref{thm:linearScalability}), the space complexity is $O(|V|+|E|)=O(|E|)$.

\begin{algorithm}[t]
    \DontPrintSemicolon
    \small
    \caption{\label{procedure:application}Application of \OurModel to Distributed Multi-Query Answering}
    \KwInput{(1) input graph $\InputGraph$, (2) machines $\{M_1, M_2, \dots, M_m \}$\\
    \hspace{9.8mm}(3) size budget $\SizeBudget$, (4) degree of personalization $\alpha$, \\
    \hspace{9.8mm}(5) parameter for adaptive thresholding $\beta$\\
    \hspace{9.8mm}(6) max. number of iterations $\MaxIterations$,\\
    
    }
    \noLineNumber\textbf{Preprocessing:} \\
    $\{V_{1},V_{2}, ..., V_{m}\} \leftarrow$ \normalfont{GraphPartitioning(}$\InputGraph$, $m$\normalfont{}{)}\\
    \For{\normalfont{\textbf{each}}\ $V_{i} \in \{V_{1},V_{2}, ..., V_{m}\}$}{
        $\SummaryGraph_{i}\leftarrow \OurModel(\InputGraph, \SizeBudget, V_{i}, \alpha, \beta, \MaxIterations)$  \\
        Load $\SummaryGraph_{i}$ into the main memory of $M_i$
    }
    
    \noLineNumber\textbf{Distributed Multi-Query Answering:} \\
    \For{\normalfont{\textbf{each}} arrived query on a node $q$}{
        Find $i$ where $q\in V_i$ \\
        Assign the query to $M_i$, which answers the query independently using $\SummaryGraph_{i}$ 
    }
    
\end{algorithm}

\section{Application: ``Communication-free'' Distributed Multi-Query Answering}
\label{sec:application}

In this section, we discuss an application of \OurModel to distributed multi-query answering.

\smallsection{Background.}
Real-time processing of various complex graph queries requires fast random access into memory. Thus, if graphs are distributed across multiple machines, answering such queries incurs significant communication overhead, preventing real-time processing.
As a result, much effort has been made to minimize communication overhead when answering queries on graphs distributed over multiple machines \cite{shao2013trinity, sarwat2013horton+, ngo2018worst, khan2018smart}. 

\smallsection{Intuition.}
\OurModel can be utilized for ``communication-free'' distributed multi-query answering.
Specifically, multiple personalized summary graphs with different target node sets are obtained by \OurModel. Then, different summary graphs are loaded on different machines, which act independently without communication.
If multiple queries are given, each query is assigned to a machine on which it can be processed accurately.

\smallsection{Procedure.}
Assume $m$ machines each of which has main memory of size $k$ are available.
We first divide the node set $V$ into $m$ subsets using the Louvain method \cite{blondel2008fast}, while any graph-partitioning method (e.g., \cite{lancichinetti2009community,hager2013exact,hager1999graph,felner2005finding,brunetta1997branch,donath1972algorithms,donath2003lower}) can be used instead.
Let the $m$ subsets of nodes be $V_1$, $V_2$, ..., $V_m$. 
For each node set $V_i$, we create a summary graph $\overline{G}_i$ personalized to $V_i$ within the budget $k$ using \OurModel, and  $\overline{G}_i$ is loaded into the main memory of the $i$-th machine.
Given multiple graph queries, we assign each query on a node $q$ to the $i$-th machine satisfying $q\in V_i$, and the machine answers the query using $\overline{G}_i$ without communicating with other machines.
Since $\overline{G}_i$ is personalized to $q$, $\overline{G}_i$ is expected to maintain much information relevant to $q$, and thus the answer from it is expected to be accurate.
We confirm this expectation experimentally in Section~\ref{sec:q:application}.
We provide the pseudocode in Alg.~\ref{procedure:application}.

\smallsection{Potential Alternatives.}
As a potential alternative for communication-free distributed multi-query processing, $m$ overlapping subgraphs of size $k$ can be distributed over the main memory of $m$ machines for query answering.
In our experiments in Sect.~\ref{sec:q:application}, we first partition the node set $V$ into $m$ subsets using graph-partitioning methods and compose each $i$-th subgraph of size $k$ (see Eq.~\eqref{eq:size:inputgraph}) using the edges closest to each $i$-th subset. Each query on a node $u$ is assigned to the $i$-th machine where $u\in V_i$.

\section{Experiments}
\label{sec:experiments}

\begin{table}[t]
	\begin{center}
		\caption{Summary of six real-world graphs and one synthetic graph. \label{tab:DatasetTable}}
		\vspace{-2mm}
		\begin{tabular}{r|r|r|r}
			\toprule 
			\textbf{Name} & \textbf{\# Nodes} & \textbf{\# Edges}  & \textbf{Summary}\\
			\midrule
			LastFM-Asia (LA) \cite{feather} & 7,624 & 27,806 & Social\\
			Caida (CA) \cite{leskovec2005graphs} & 26,475 & 53,381 & Internet\\
			DBLP (DB) \cite{yang2015defining} & 317,080 & 1,049,866 & Collaboration\\
			Amazon0601 (A6) \cite{leskovec2007dynamics} & 403,364 & 2,443,311 & Co-purchase\\
			Skitter (SK) \cite{leskovec2005graphs} & 1,694,616 & 11,094,209 & Internet\\
			Wikipedia (WK) \cite{kunegis2013konect}& 3,174,745 & 103,310,688 & Hyperlinks\\
			\midrule
			Synthetic (ST) \cite{barabasi1999emergence} & 10,000,000 & 1,000,000,000 &  BA Model\\
			\bottomrule
		\end{tabular}
	\end{center}
\end{table}

In this section,
we review our experiments to answer Q1-Q4.

\begin{itemize}[leftmargin=*]
	\item \textbf{Q1. Effectiveness}: Does \OurModel provide personalized summary graphs?
	\item \textbf{Q2. Scalability}: Does \OurModel scale linearly with the number of edges in the input graph? 
	\item \textbf{Q3. Comparison with the State of the Art}: Does \OurModel provide better summary graphs to target nodes than the best non-personalized graph summarization methods?
    \item \textbf{Q4. Effect of Parameters}: How do $\alpha$ and $\beta$ affect the output summary graphs?
	\item \textbf{Q5. Application}: Is \OurModel useful for communication-free distributed multi-query answering?
\end{itemize}

\begin{figure*}[t]
    \vspace{-3mm}
    \centering
    \label{fig:personErrRatio}
    \includegraphics[width=0.8\linewidth]{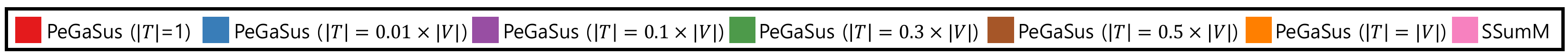} \\
	\subfigure[$\alpha = 1.25$]{
	    \label{fig:personErrRatio125}
		\includegraphics[width=0.31\textwidth]{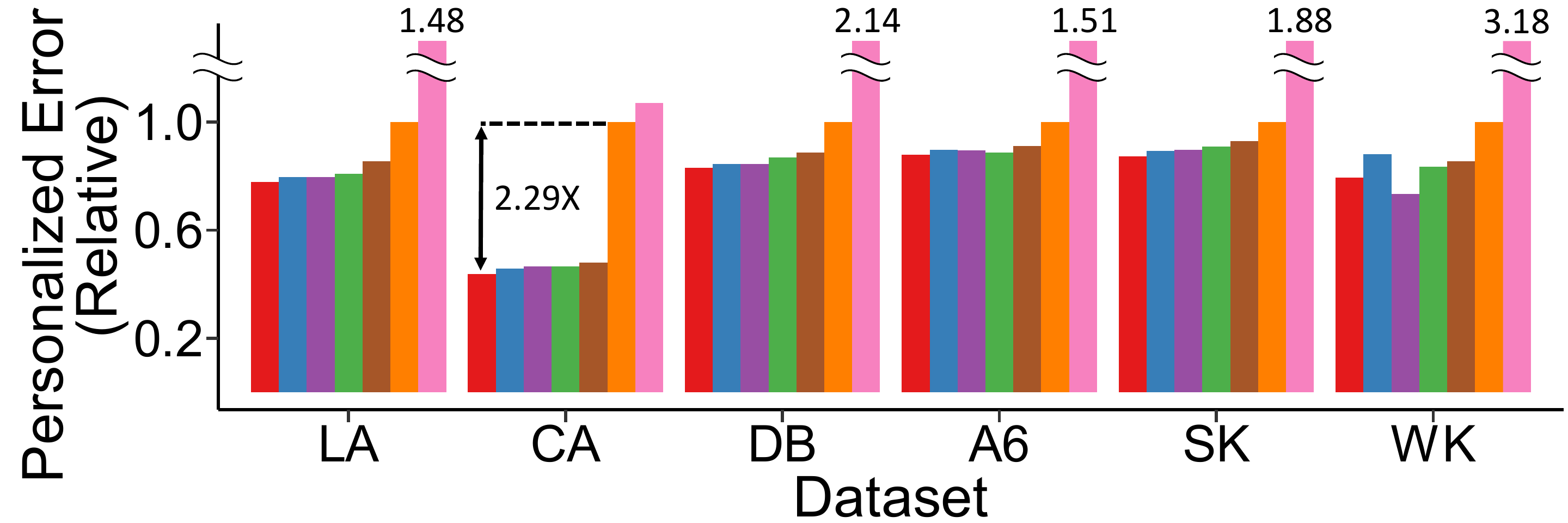}
	}
	\subfigure[$\alpha = 1.5$]{
		\label{fig:personErrRatio15}
		\includegraphics[width=0.31\textwidth]{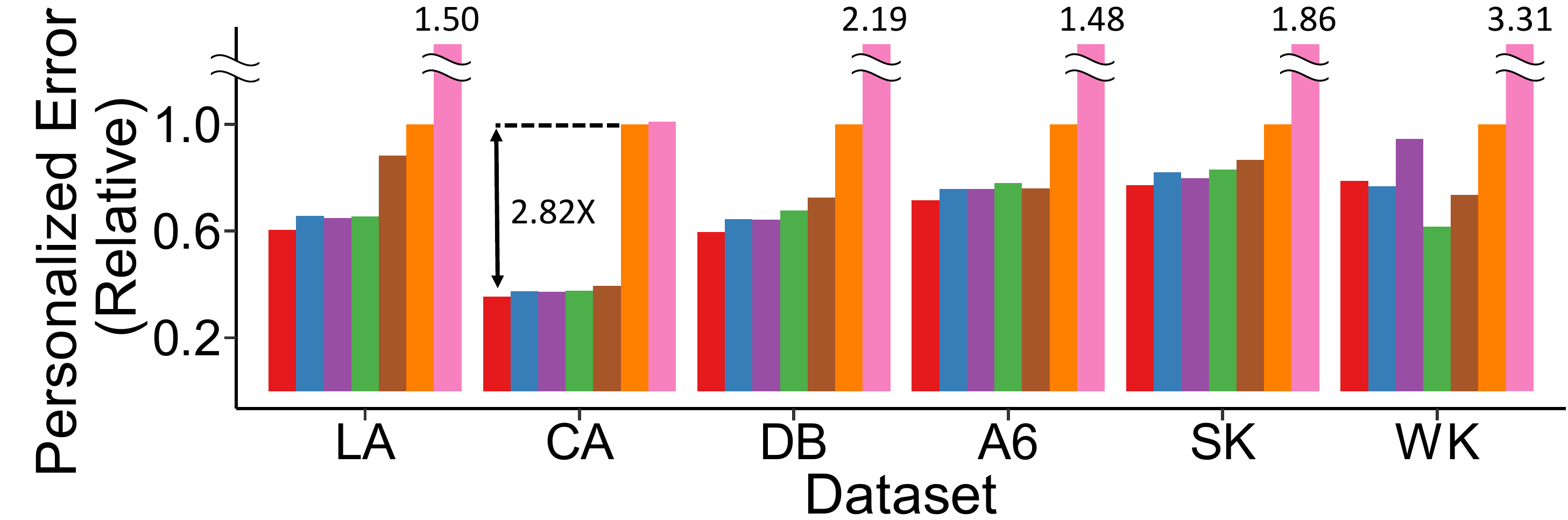}
	}
	\subfigure[$\alpha = 1.75$]{
		\label{fig:personErrRatio175}
		\includegraphics[width=0.31\textwidth]{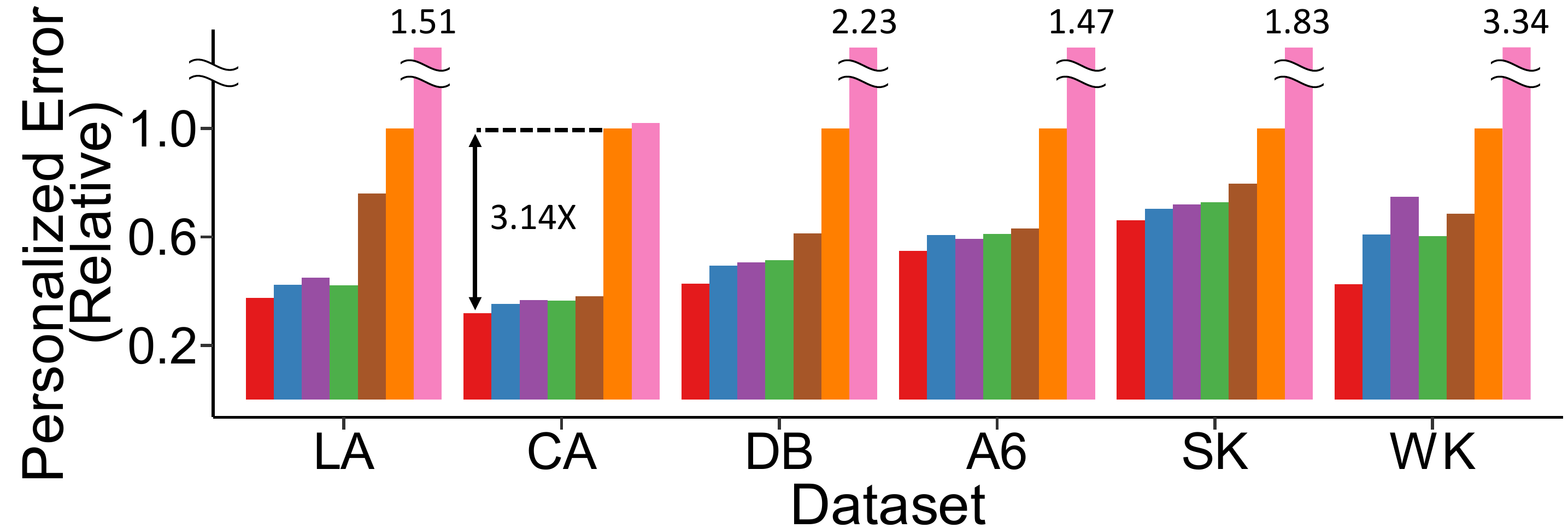}
	}
    \vspace{-2mm}
    \caption{\label{fig:effectiveness}
    \underline{\smash{\textbf{\OurModel provides personalized summary graphs.}}}
    The relative personalized error tends to decrease as we reduce the size of the target node set (i.e., the summary graph is more focused) and grow the degree of personalization $\alpha$.
    }
\end{figure*}

\subsection{Experimental settings}
\label{sec:experimetalsetting}

\smallsection{Machines.} 
We performed our experiments on a desktop with AMD Ryzen 7 3700X CPU and 128GB memory.

\smallsection{Datasets.} 
We used six real-world graphs summarized in Table~\ref{tab:DatasetTable}. 
We removed all self-loops and edge directions in them and used only the largest connected components.

\smallsection{Implementations.} 
We implemented \OurModel, \SSumM \cite{lee2020ssumm}, and \kGrass \cite{lefevre2010grass} in Java. We used the implementations \SAAGs \cite{beg2018scalable} and \SL \cite{riondato2017graph} provided by the authors, and they were implemented in Java and C++, respectively. 
% We used the equation in \cite{lee2020ssumm} for the size of summary graphs obtained from them except \OurModel. 
We set the size budget from $10\%$ to $90\%$ of the size in bits of the input graph at the same interval and set the maximum number of iterations to $20$ for \OurModel and \SSumM; and we set it from $10\%$ to $90\%$ of the number of supernodes at the same interval for \SAAGs, \kGrass, and \SL.
For \method, we set $\alpha$ to $1.25$ and $\beta$ to $0.1$ unless otherwise stated.
For \kGrass, we used the $SamplePairs$ method with $c = 1.0$, as suggested in \cite{lefevre2010grass}.
For \SAAGs, we set the number of sample pairs to $\log n$ and the used the count-min sketch with $w=50$ and $d=2$.
For \SL, we used the type of reconstruction error as $L1$ without dimensionality reduction.
For graph partitioning in Sect.~\ref{sec:q:application}, we implemented the Louvain method \cite{blondel2008fast} in Java and used BLP \cite{ugander2013balanced} and SHP (SHPI, SHPII, and SHPKL) \cite{kabiljo2017social} implemented in NumPy by the authors of \cite{awadelkarim2020prioritized}. 
We set the maximum number of iterations to $10$ and used $8$ shards.

\smallsection{Node-Similarity Queries.}
We considered the following three types of node-similarity queries:
\begin{itemize}
    % \item \textit{Random Walk with Restart} (\RWR) \cite{jung2017bepi,tong2008random}:
    \item \textit{Random Walk with Restart} (\RWR) \cite{tong2008random}:
    The \RWR score of a node w.r.t. a query node $q$ is the stationary probability that a random walker stays at the node when it repeatedly restarts at $q$. We set the restarting probability to $0.05$.
    
    \item 
    \textit{Length of the Shortest Path} (\HOP):
    The \HOP of a node w.r.t. a query noede $q$ is the length of shortest paths from $q$ to the node. If there is no path between them, we used the length of longest path in the given (sub)graph as the \HOP.
    
    \item
    \textit{Penalized Hitting Probability} (\PHP) \cite{zhang2012evaluating,guan2011assessing}:
    The \PHP of a node $u$ w.r.t. a query node $q$ is defined as
    $$
        \PHP_u := \begin{cases}
            1 & \text{ if } u = q, \\
            c\sum_{v\in N_u} \left(\frac{w_{uv}}{w_{u}} \cdot \PHP_v\right) & \text{ if } i\neq q,
        \end{cases}
    $$
    where $N_u$ is the set of neighbors of $u$, $w_{uv}$ is the weight of the edge $\{u,v\}$ (which is $1$ if the graph is unweighted), and $w_u$ is the weighted degree of $u$. We set $c$ to $0.95$.
\end{itemize}
The queries can be answered directly from a summary graph, as described in  Appendix~\ref{sec:appendix:query}. 

\smallsection{Evaluation Measures.}
We measured compression rates in bits. That is, the \textit{compression rate} of a summary graph $\SummaryGraph$ of a graph $\InputGraph$ is  $\frac{Size(\SummaryGraph)}{Size(\InputGraph)}$. 
As in \cite{lee2020ssumm}, for the size in bits of weighted summary graphs with the maximum weight $\omega_{max}$, we used $$|P|(2\log_{2}|S|+\log_{2}\omega_{max}) + |V|\log_{2}|S|.$$

For each query $q$, we measured the accuracy of an approximate answer vector $\hat{x}\in \mathbb{R}^{|V|}$ by comparing it with the ground-truth answer vector $x\in \mathbb{R}^{|V|}$ in two ways:
% \vspace{1mm}
\begin{itemize}
    \item 
    \textit{Symmetric Mean Absolute Percentage Error}  (\SMAPE) \cite{goodwin1999asymmetry} (the lower, the better):
    \begin{equation*}
        \SMAPE(\answerMeasure,\approxMeasure) := \sum_{u\in V} \frac{|\answerMeasure_{u}-\approxMeasure_{u}|}{|\answerMeasure_{u}|+|\approxMeasure_{u}|},
    \end{equation*}
    If $\answerMeasure_{u}=\approxMeasure_{u}=0$, $0$ is used instead of $\frac{|\answerMeasure_{u}-\approxMeasure_{u}|}{|\answerMeasure_{u}|+|\approxMeasure_{u}|}$.
    Note that \SMAPE is well defined even when $\answerMeasure_{u}=0$ or  $\approxMeasure_{u}=0$. 
    \vspace{1mm}
    \item
    % \textit{Spearman Correlation Coefficient} \cite{spearman1961proof} (\Spearman) (the higher the better):
    \textit{Spearman Correlation Coefficients} (SC) \cite{spearman1961proof} (the higher the better):
    It measures the Pearson correlation coefficient between the rankings of the entries of $\answerMeasure$ and the rankings of the entries of $\approxMeasure$. It compares rankings, which are more important than absolute values in many graph applications.
\end{itemize}
We report the average when multiple query nodes were used.

\subsection{Q1. Effectiveness of \OurModel (Fig.~\ref{fig:effectiveness})}
\label{sec:q:effectsofpersonalization}

We demonstrate that \OurModel provides summary graphs personalized to target nodes.
To this end, we sampled $|\TargetNodeSet|$ target nodes uniformly at random.
Then, while varying the degree of personalization $\alpha$ and the target node set $\TargetNodeSet$, we measured the personalized error at $u$ (i.e., Eq.~\eqref{eq:personalizederror} with $T=\{u\}$) relative to that in non-personalized cases (i.e., when $T=V$).
Fig.~\ref{fig:effectiveness} shows the result averaged over three test nodes when the size budget was set so that the compression ratio is $0.5$.
The relative personalized error tended to decrease as we decreased the size of the target node set (i.e., the summary graph is more focused to $i$) and increased the degree of personalization.
\textbf{These results indicate that summary graphs are personalized effectively by \OurModel.}
Notably, even in non-personalized cases (i.e., when $T=V$), \OurModel outperformed \SSumM, as discussed in Sect.~\ref{sec:comparison}.

\begin{figure}[t]
    \centering
    \includegraphics[width=0.7\linewidth]{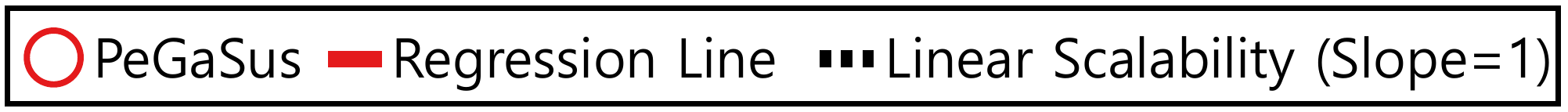}\\
	\subfigure[Skitter ($|\TargetNodeSet|$=$100$)]{
		\includegraphics[width=0.145\textwidth]{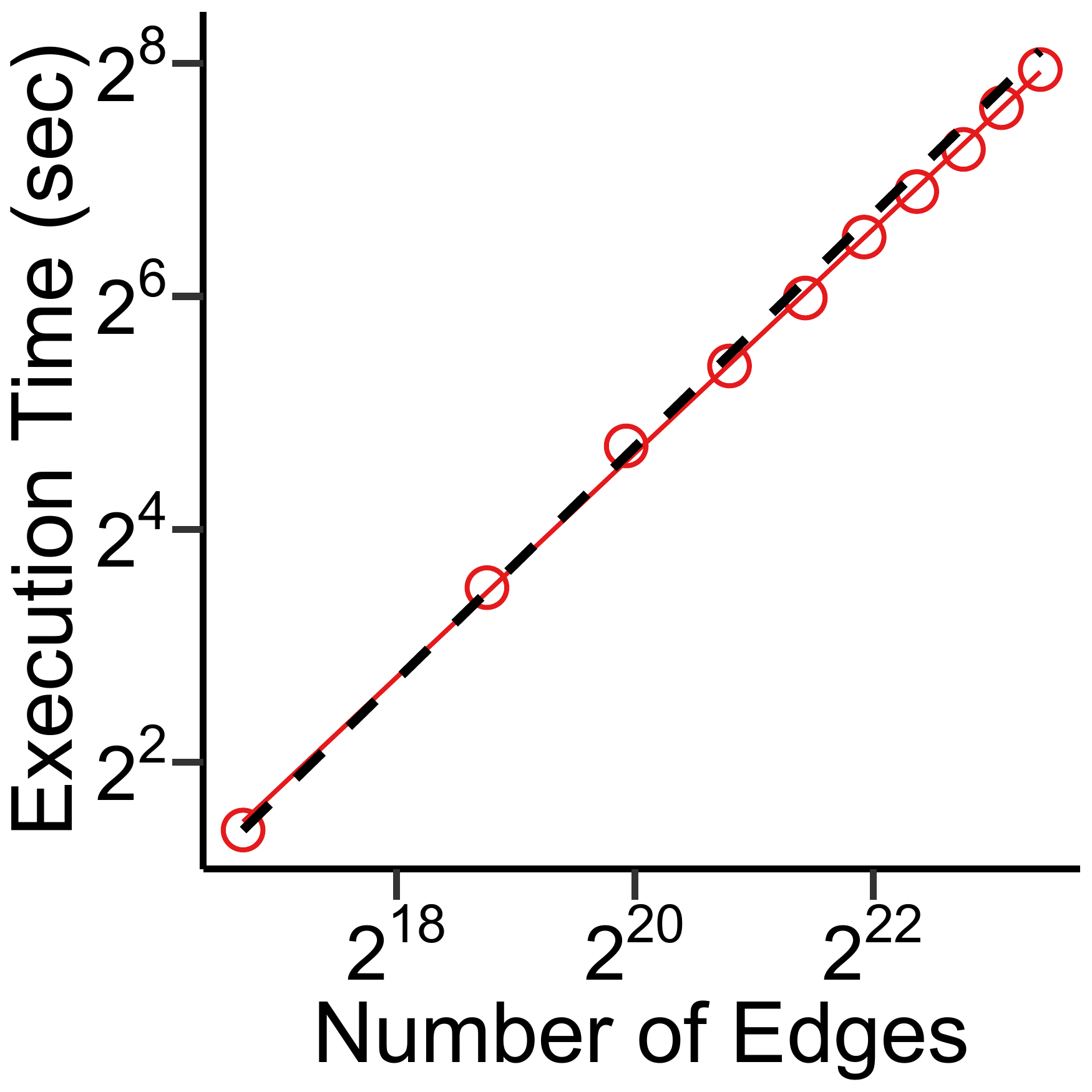}
	}
	\subfigure[Skitter ($|\TargetNodeSet|$=$\frac{|V|}{2}$)]{
		\includegraphics[width=0.145\textwidth]{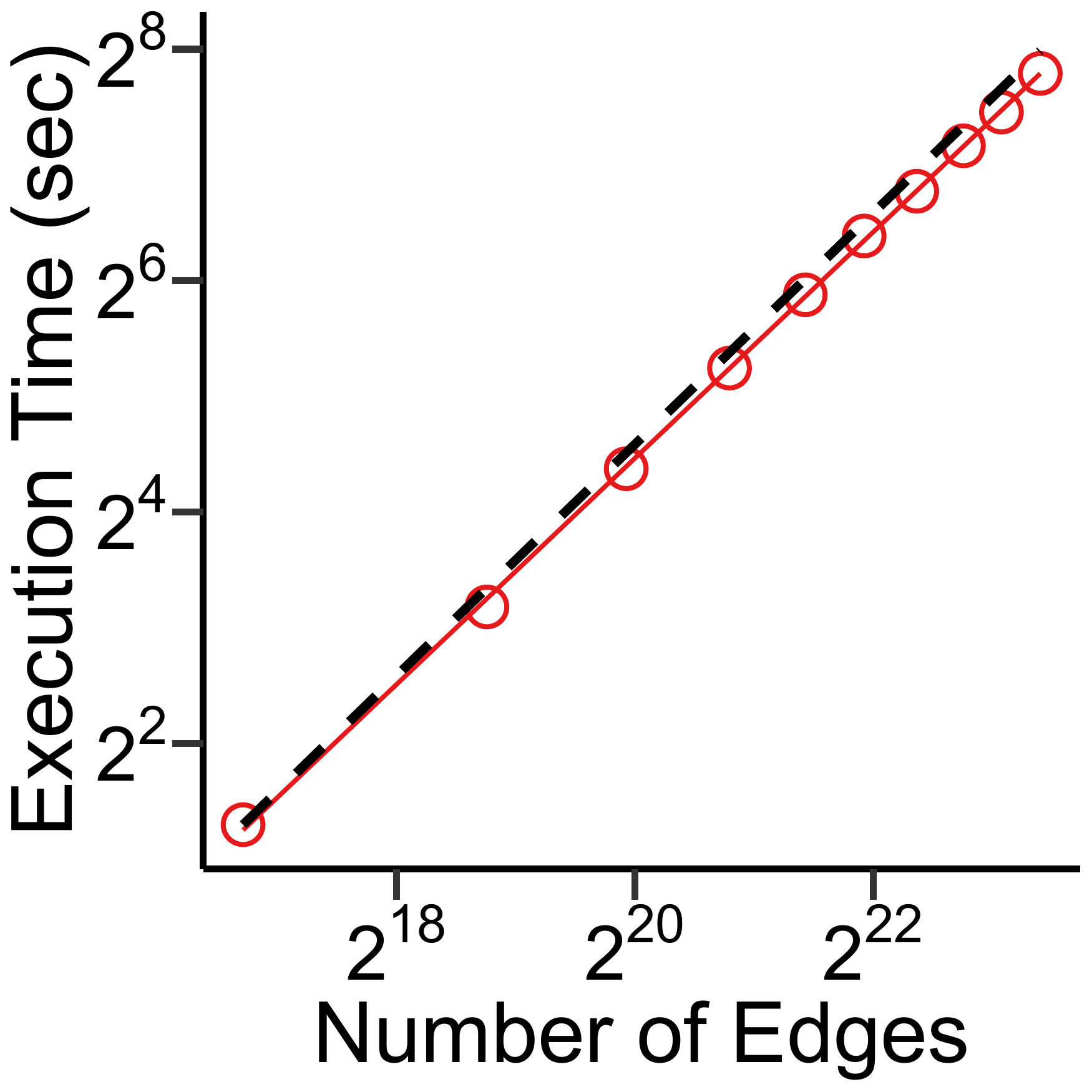}
	}
	\subfigure[Synthetic ($|\TargetNodeSet|$=$100$)]{
		\includegraphics[width=0.145\textwidth]{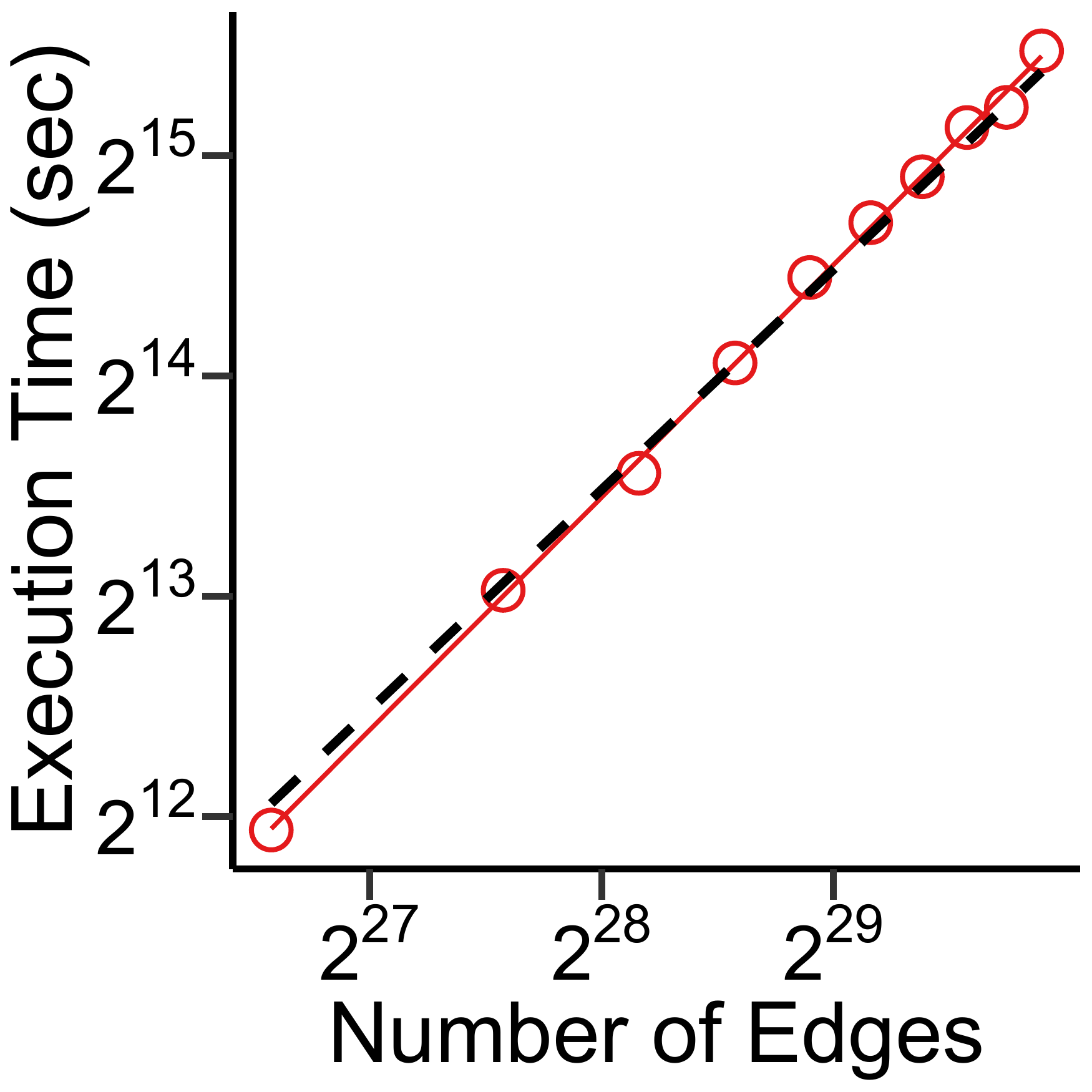}
	} \\
	\vspace{-2mm}
    \caption{\label{fig:random-scalability}
    \underline{\smash{\textbf{\OurModel exhibits linear scalability.}}}
    \OurModel scales linearly with the edge count, \textbf{to one billion edges}, regardless of the target node number.
    Fig.~\ref{fig:crown:scalable} shows the result on the synthetic dataset when $|T|=\frac{|V|}{2}$.
    }
\end{figure}

\begin{figure*}[t!]
    \vspace{-3mm}
	\centering
    \includegraphics[width=0.3\linewidth]{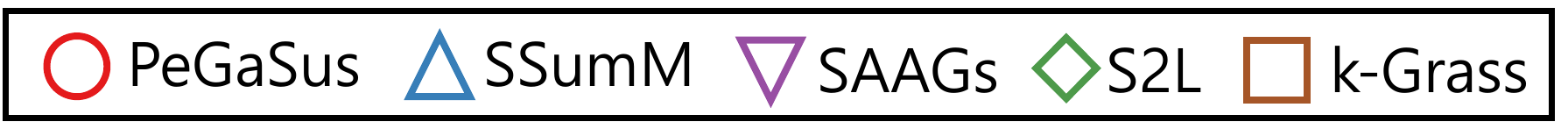}\\
	\textbf{\small Symmetric Mean Absolute Percentage Error (the lower, the better):} \hfill \ \ \\
	\subfigure[LastFM-Asia (\RWR)]{
		\includegraphics[width=0.145\textwidth]{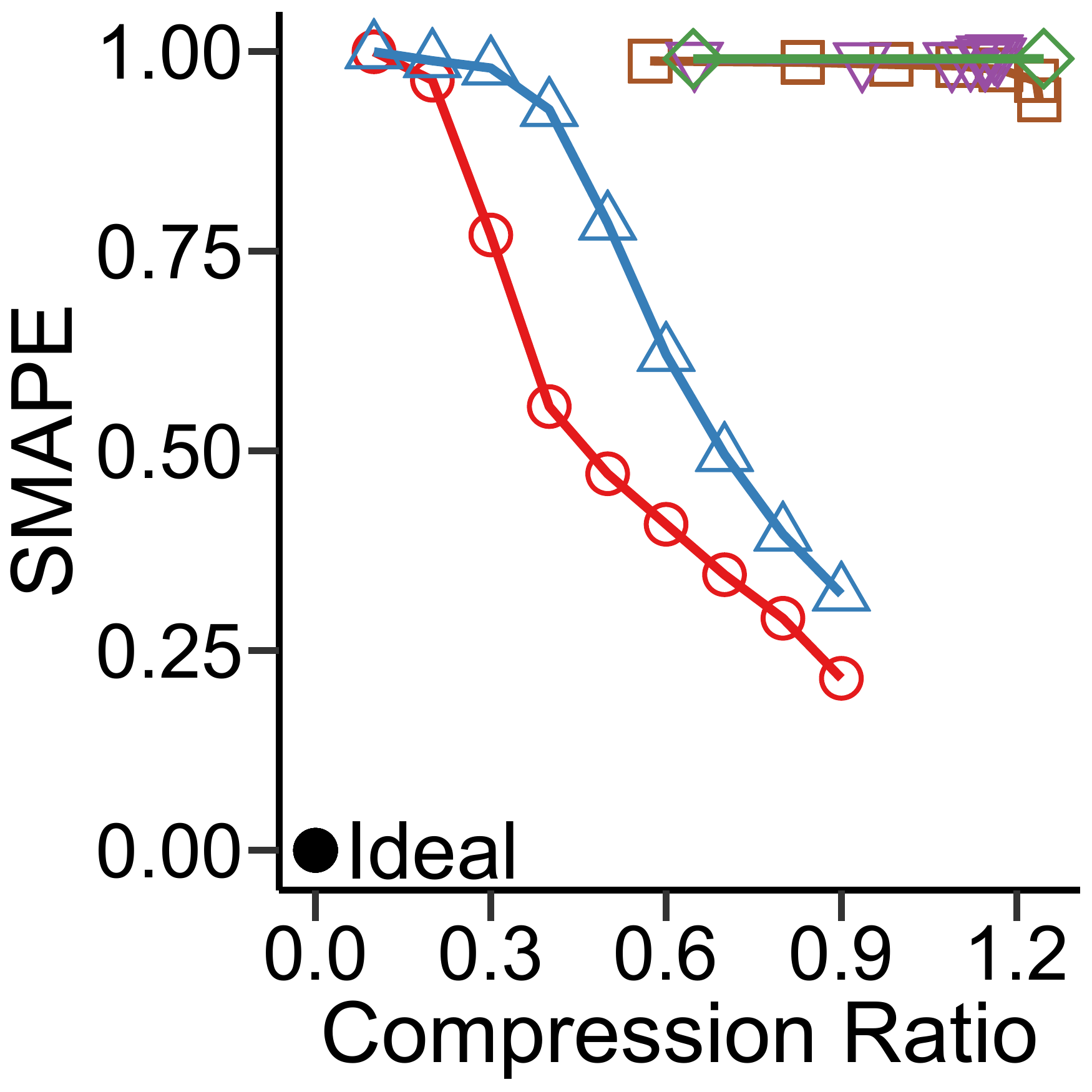}
	}
	\subfigure[Caida (\RWR)]{
		\includegraphics[width=0.145\textwidth]{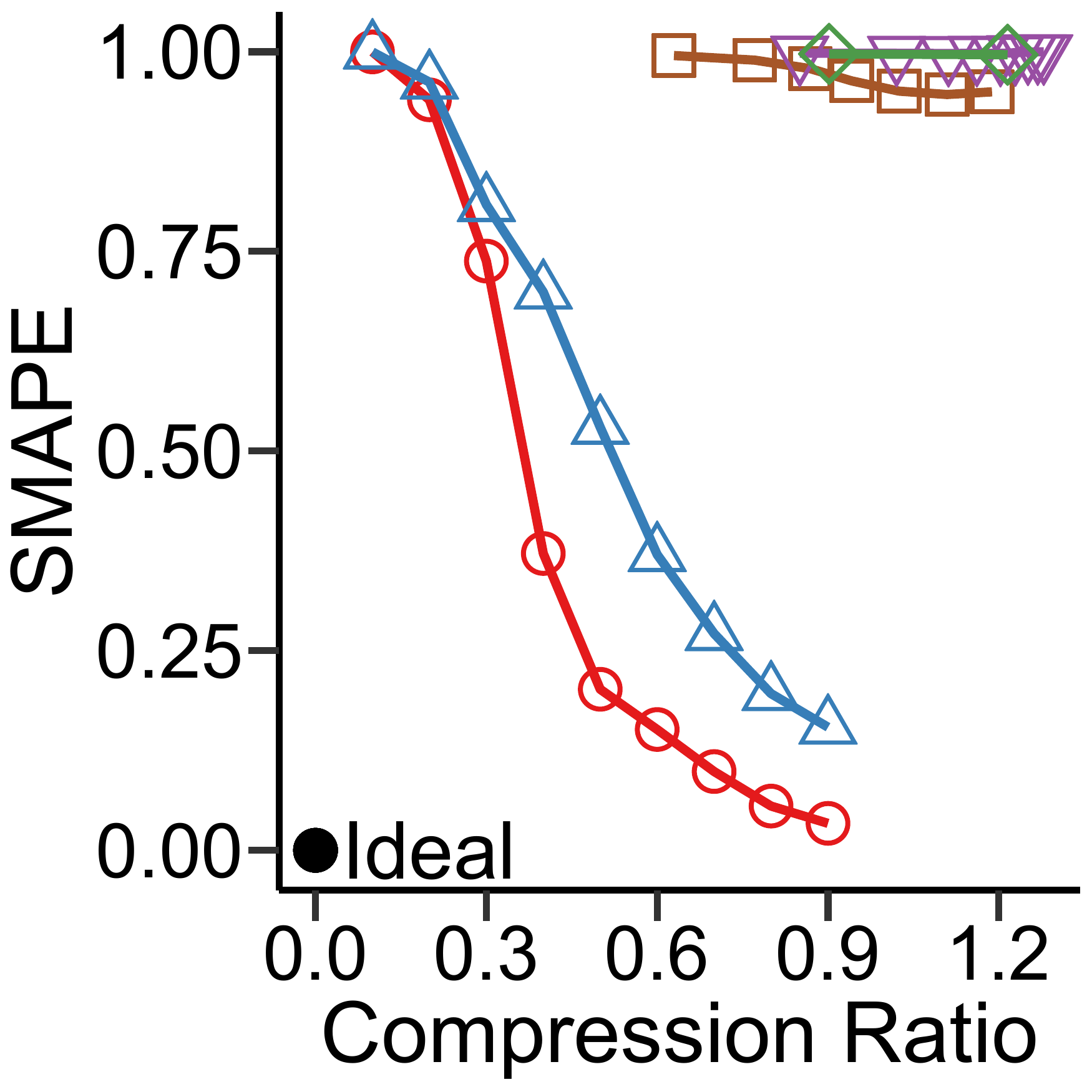}
	}
	\subfigure[DBLP (\RWR)]{
		\includegraphics[width=0.145\textwidth]{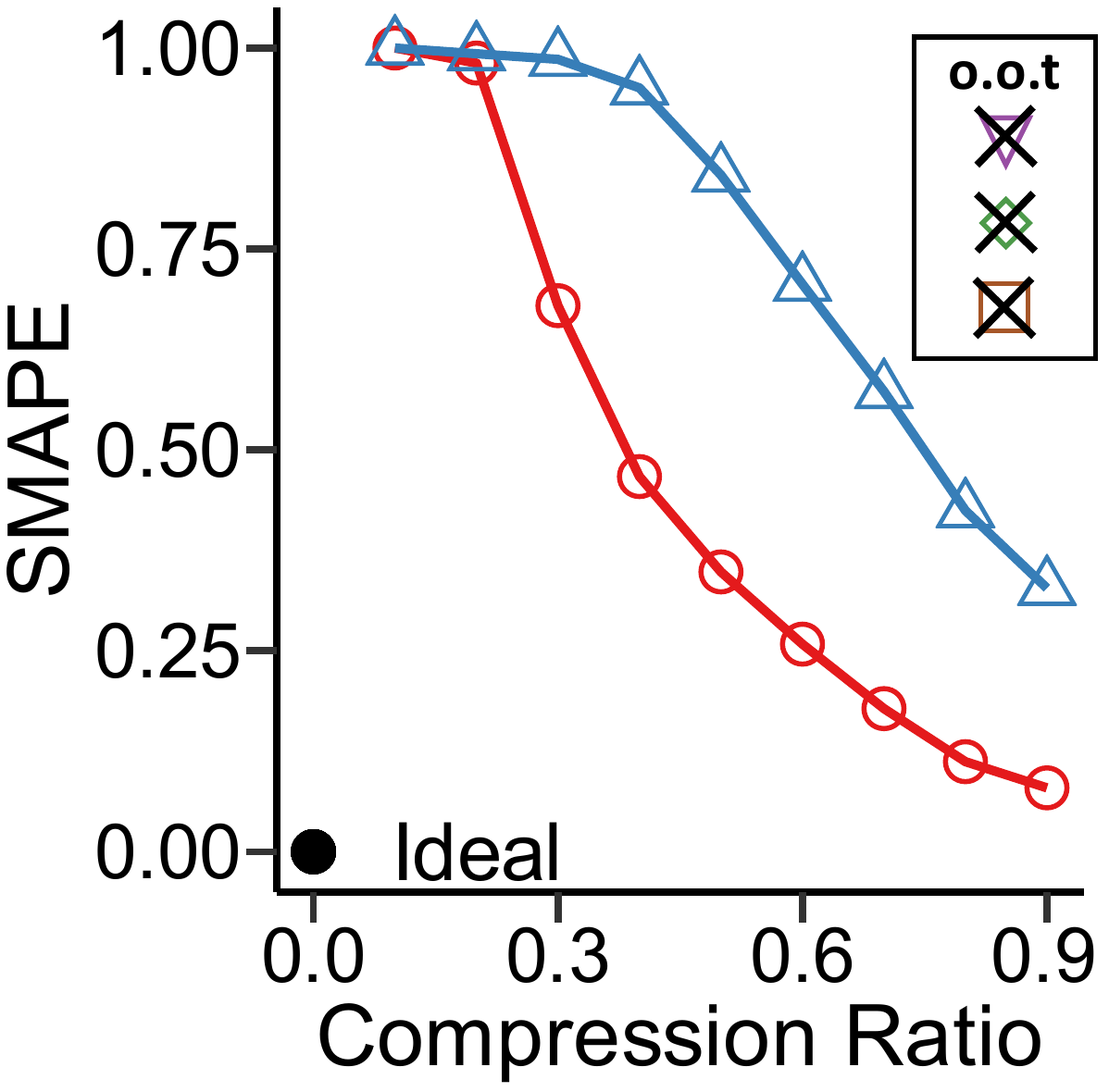}
	}
    \subfigure[Amazon0601 (\RWR)]{
		\includegraphics[width=0.145\textwidth]{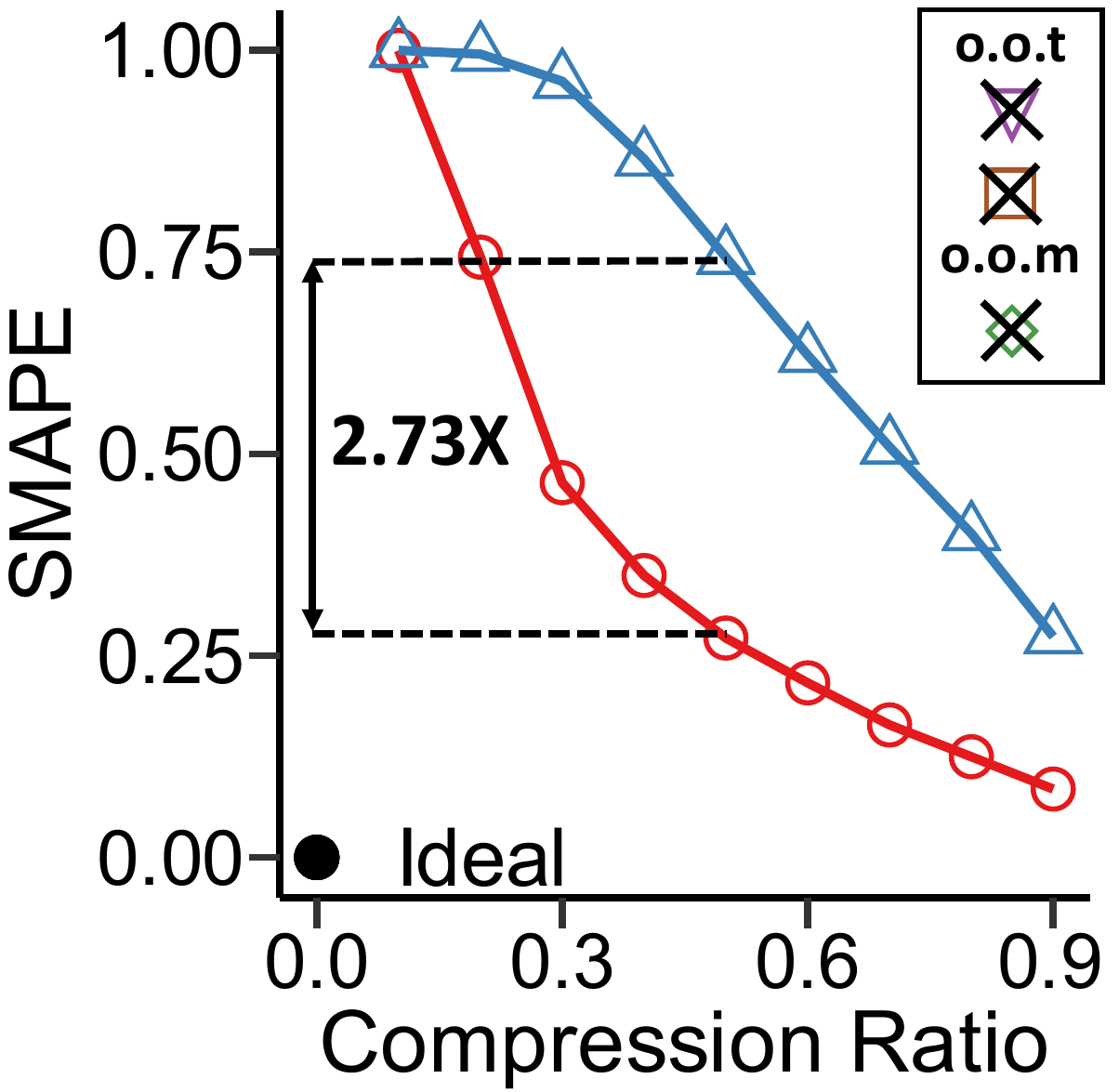}
	}
	\subfigure[Skitter (\RWR)]{
		\includegraphics[width=0.145\textwidth]{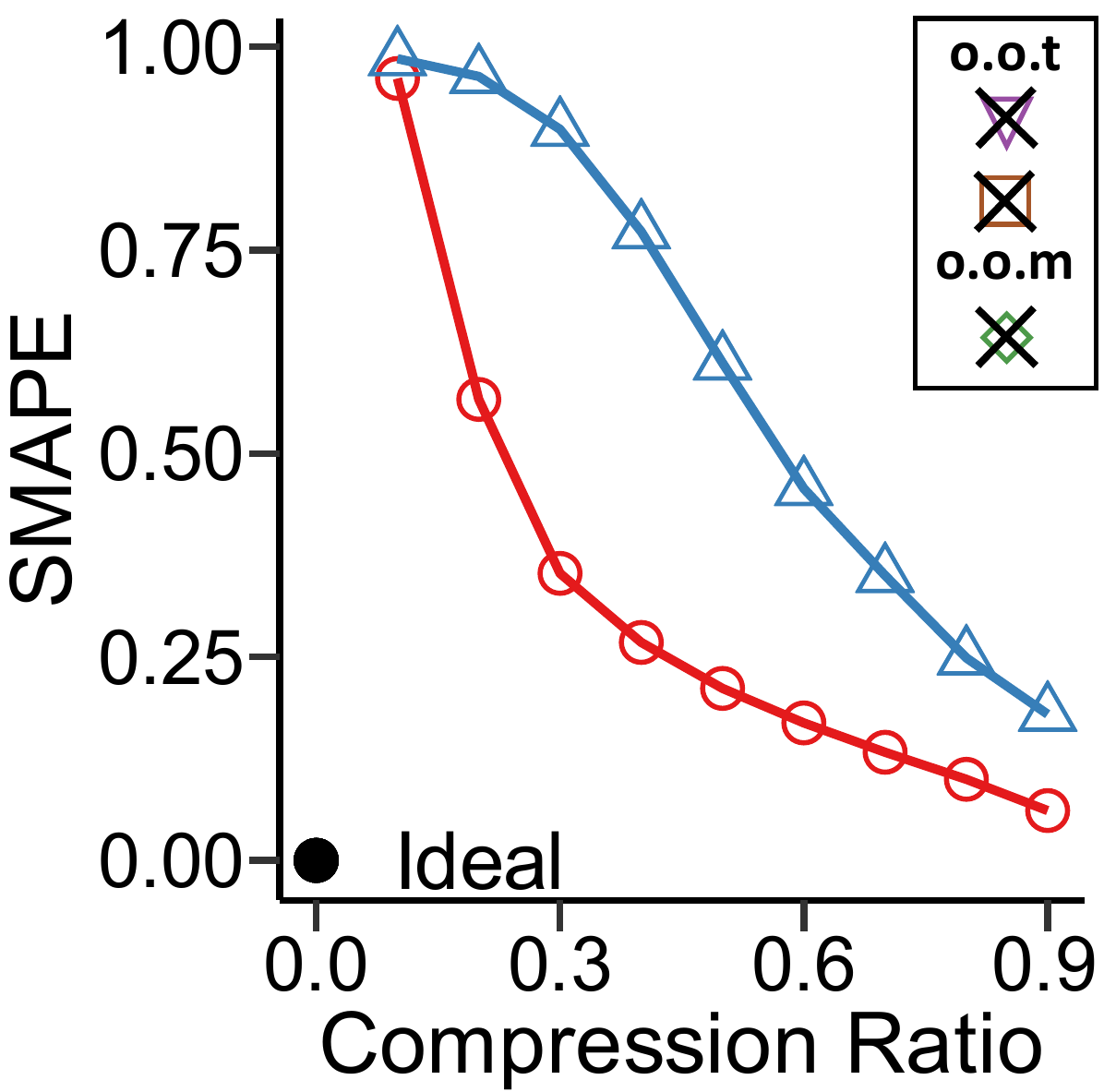}
	}
	\subfigure[Wikipedia (\RWR)]{
		\includegraphics[width=0.145\textwidth]{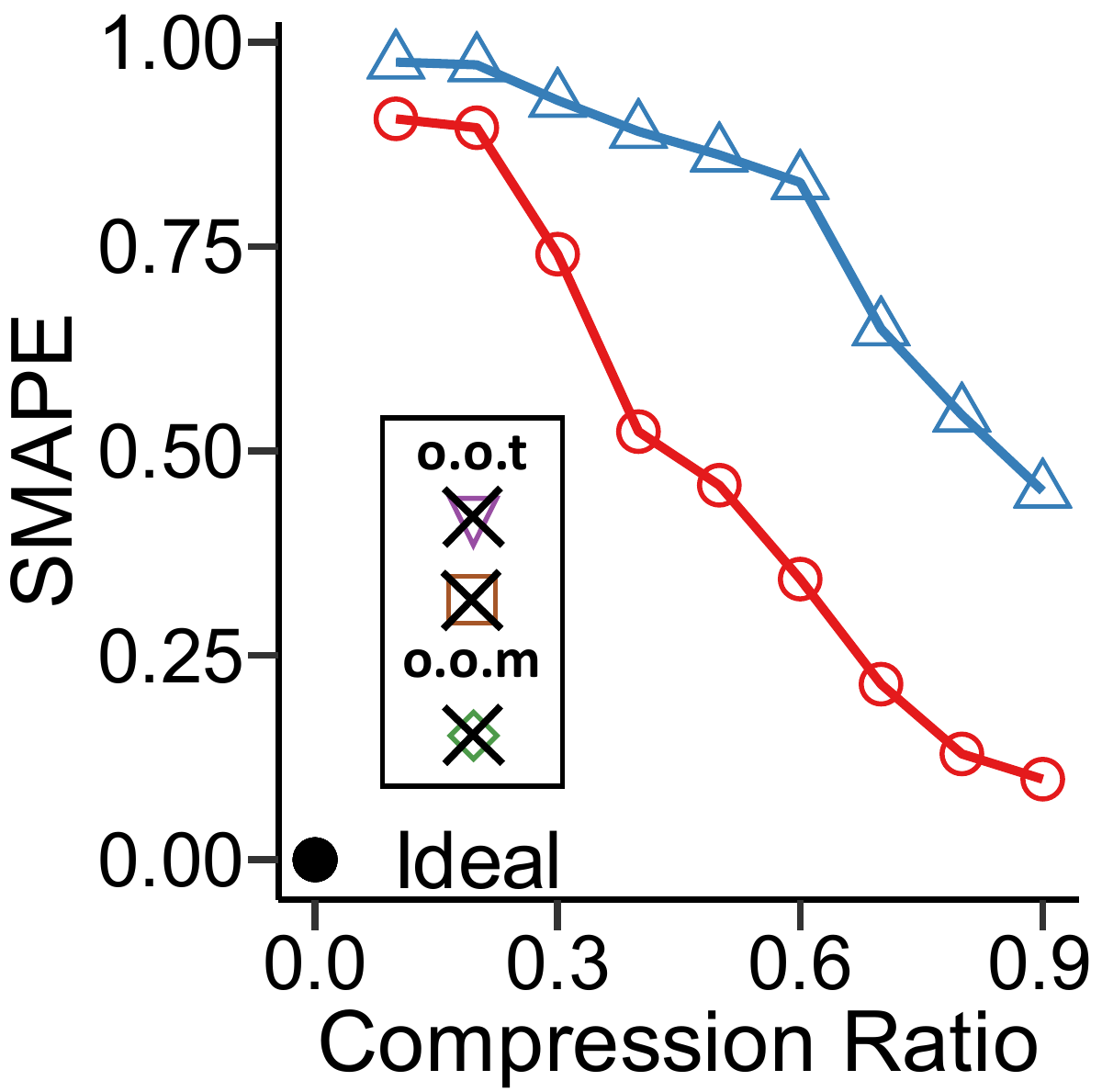}
	}
    \\
    \vspace{-3mm}
    \subfigure[LastFM-Asia (\HOP)]{
		\includegraphics[width=0.145\textwidth]{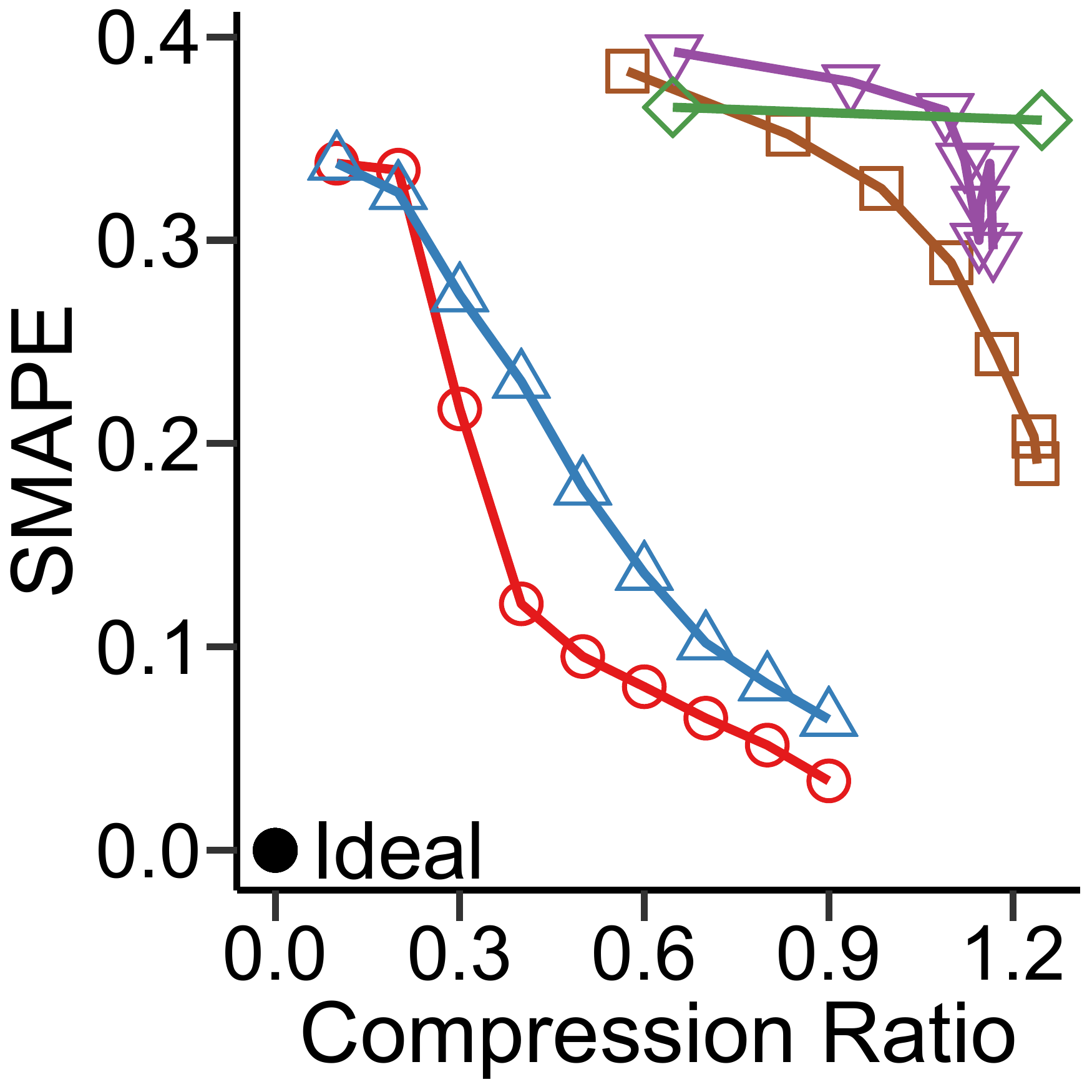}
	}
	\subfigure[Caida (\HOP)]{
		\includegraphics[width=0.145\textwidth]{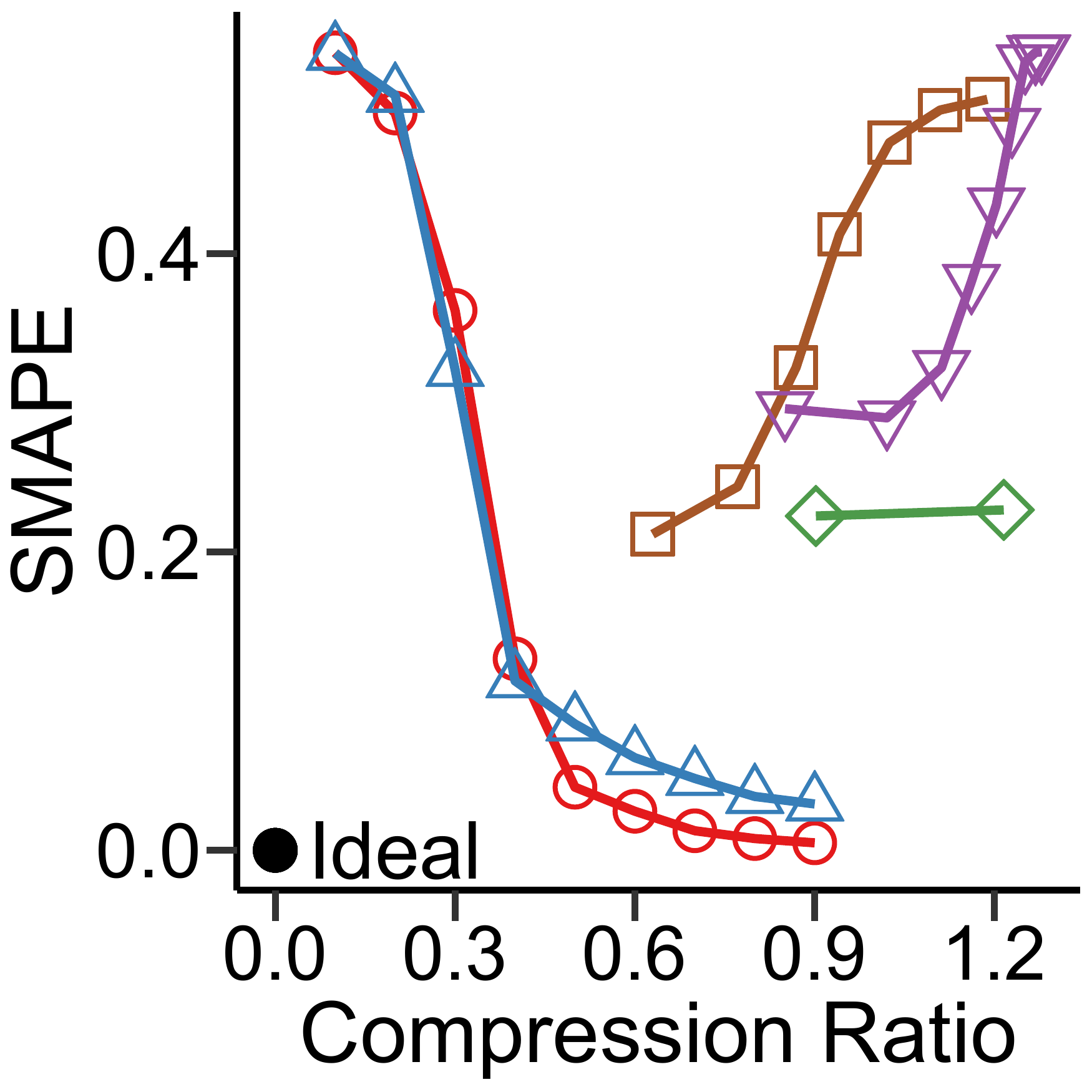}
	}
	\subfigure[DBLP (\HOP)]{
		\includegraphics[width=0.145\textwidth]{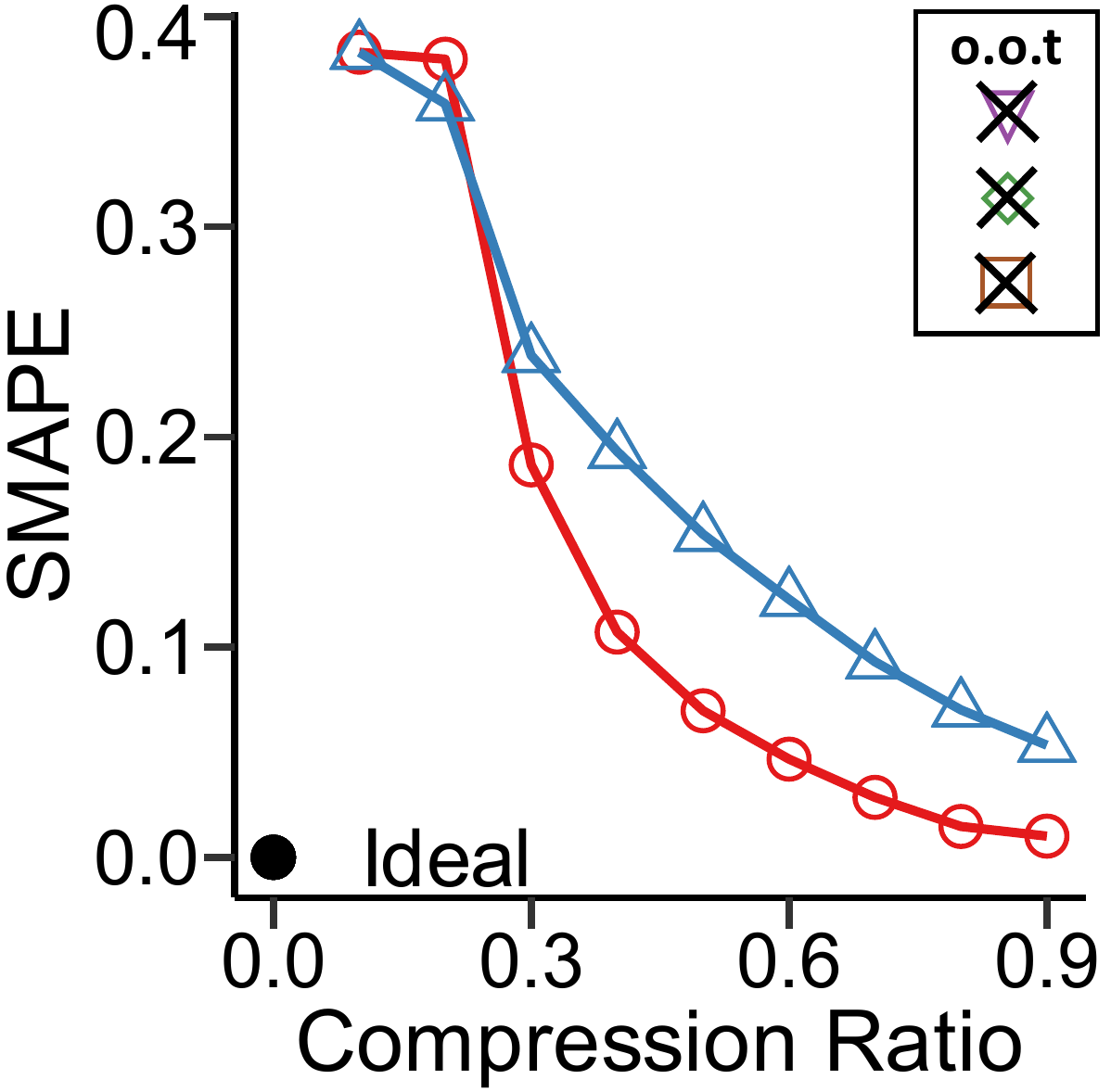}
	}
    \subfigure[Amazon0601 (\HOP)]{
		\includegraphics[width=0.145\textwidth]{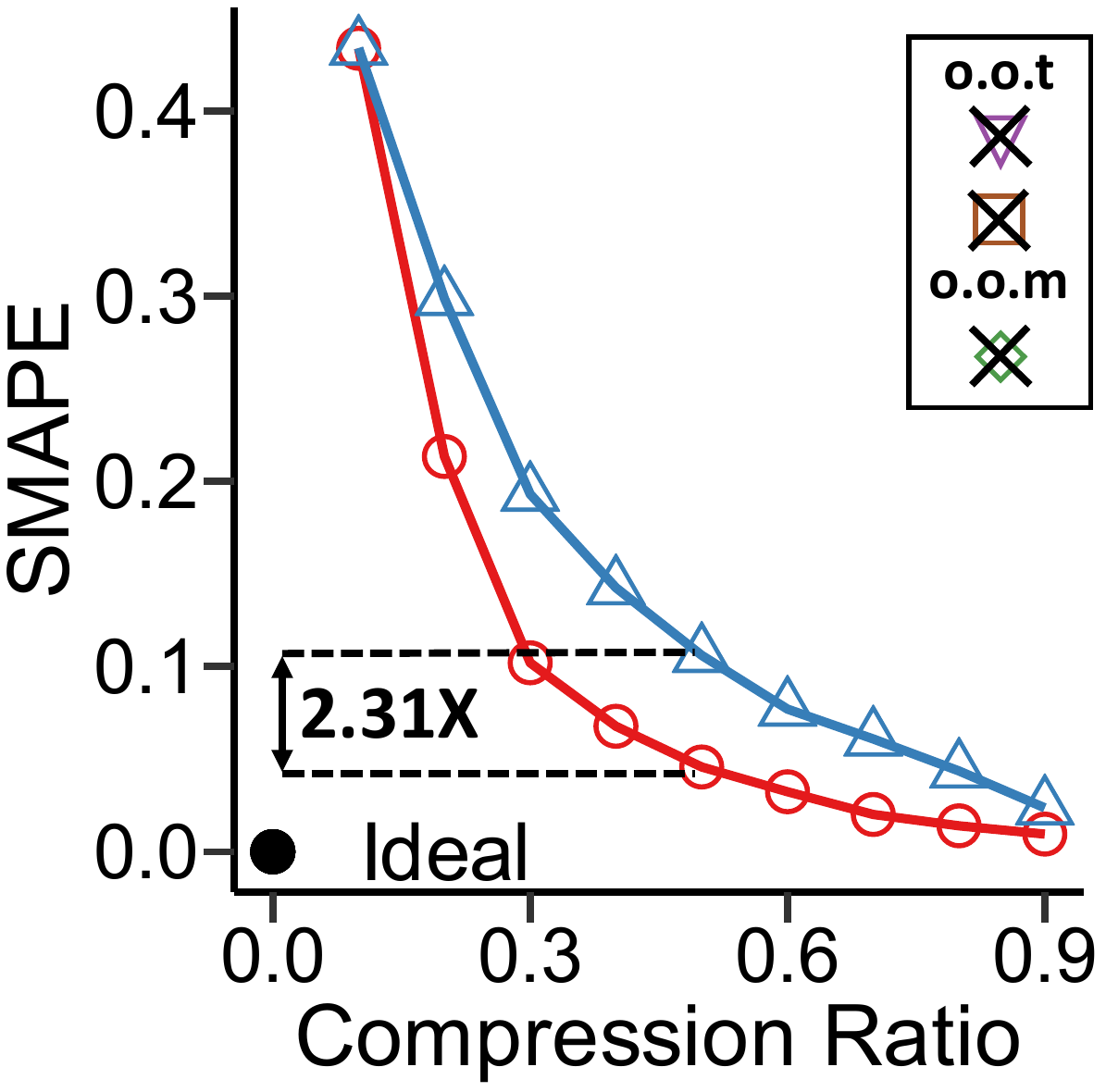}
	}
	\subfigure[Skitter (\HOP)]{
		\includegraphics[width=0.145\textwidth]{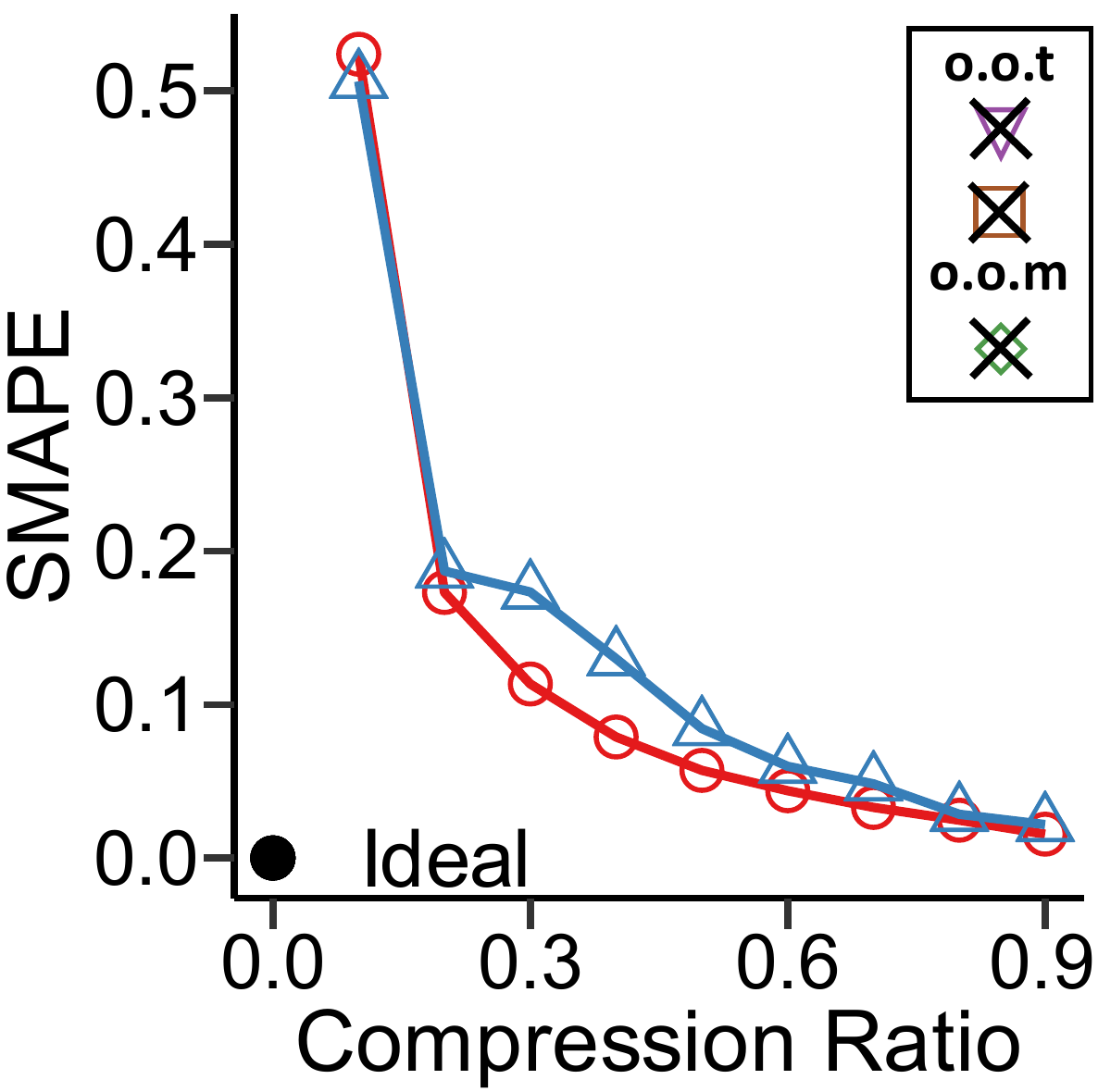}
	} 
	\subfigure[Wikipedia (\HOP)]{
		\includegraphics[width=0.145\textwidth]{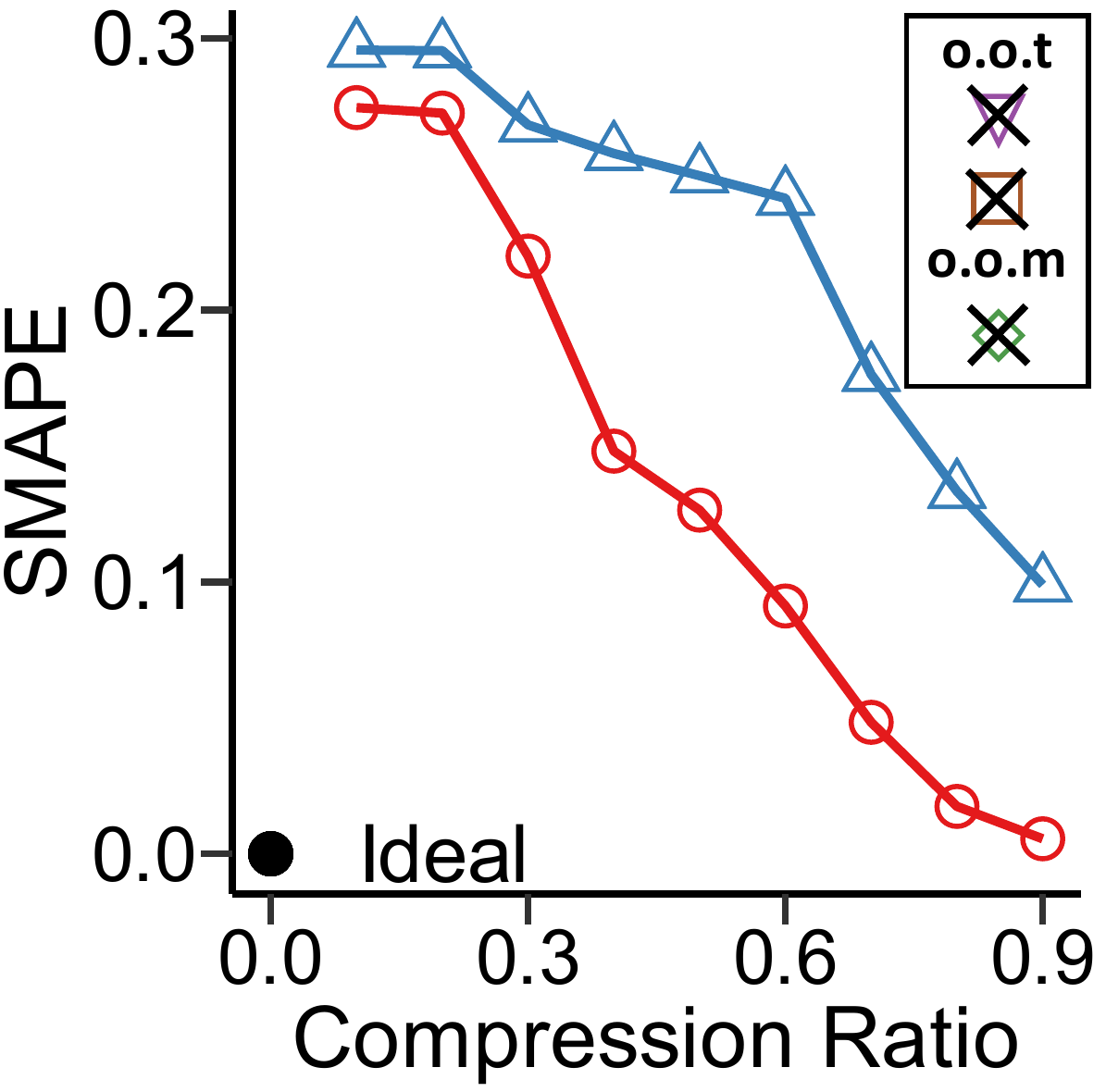}
	}
	\\
	\textbf{\small Spearman Correlation Coefficients (the higher, the better):} \hfill \ \ \\
	\subfigure[LastFM-Asia (\RWR)]{
		\includegraphics[width=0.145\textwidth]{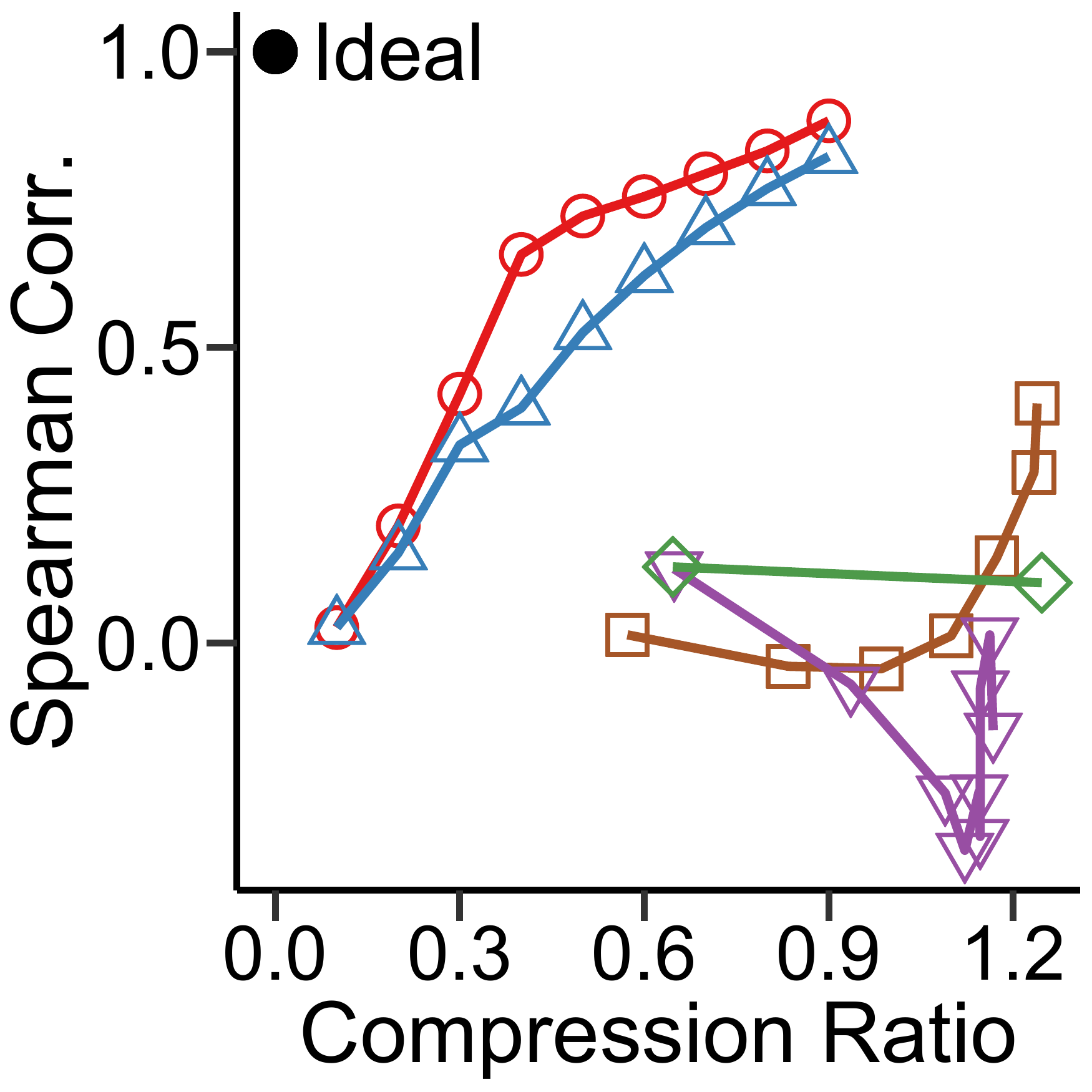}
	}
	\subfigure[Caida (\RWR)]{
		\includegraphics[width=0.145\textwidth]{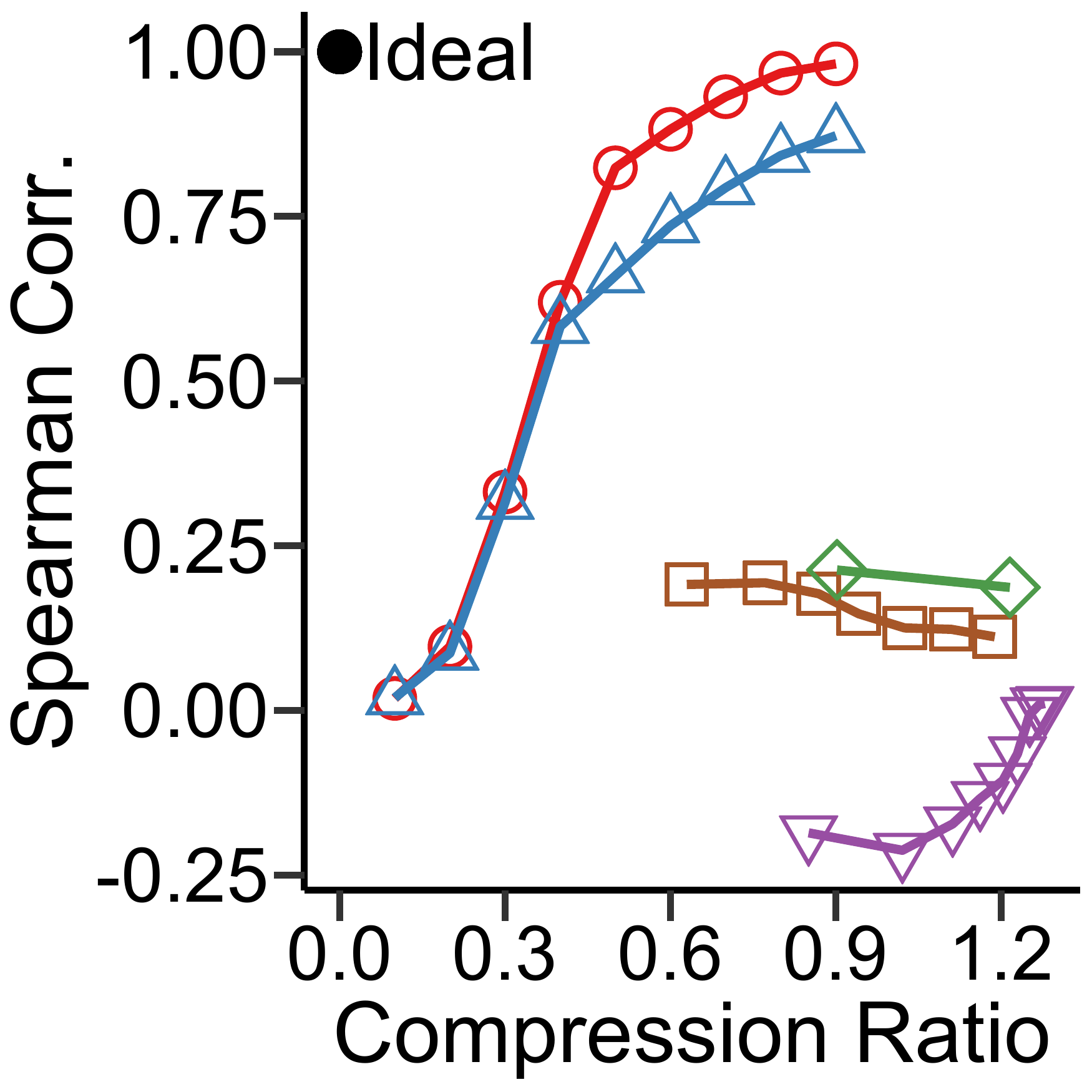}
	}
	\subfigure[DBLP (\RWR)]{
		\includegraphics[width=0.145\textwidth]{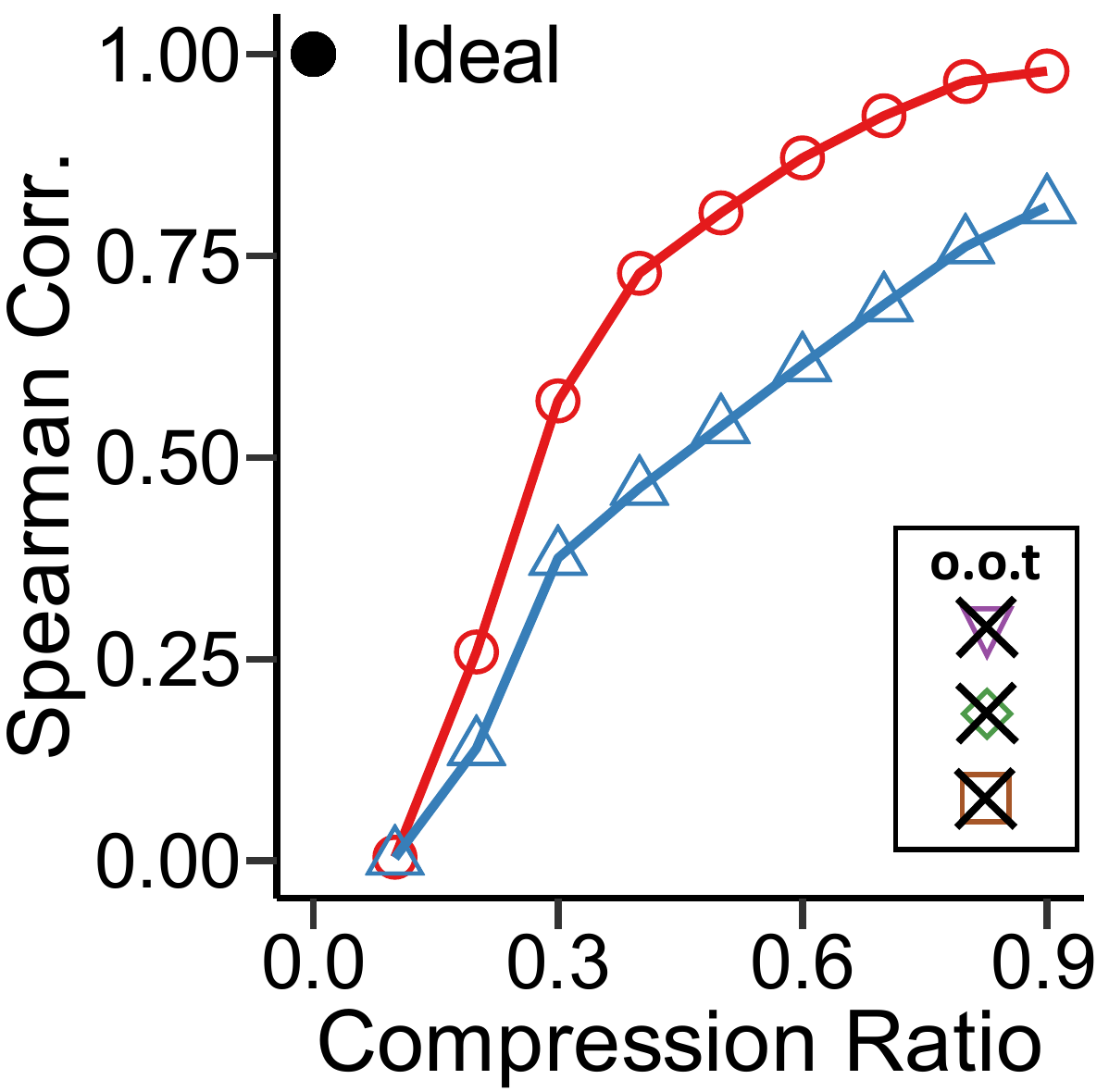}
	}
    \subfigure[Amazon0601 (\RWR)]{
		\includegraphics[width=0.145\textwidth]{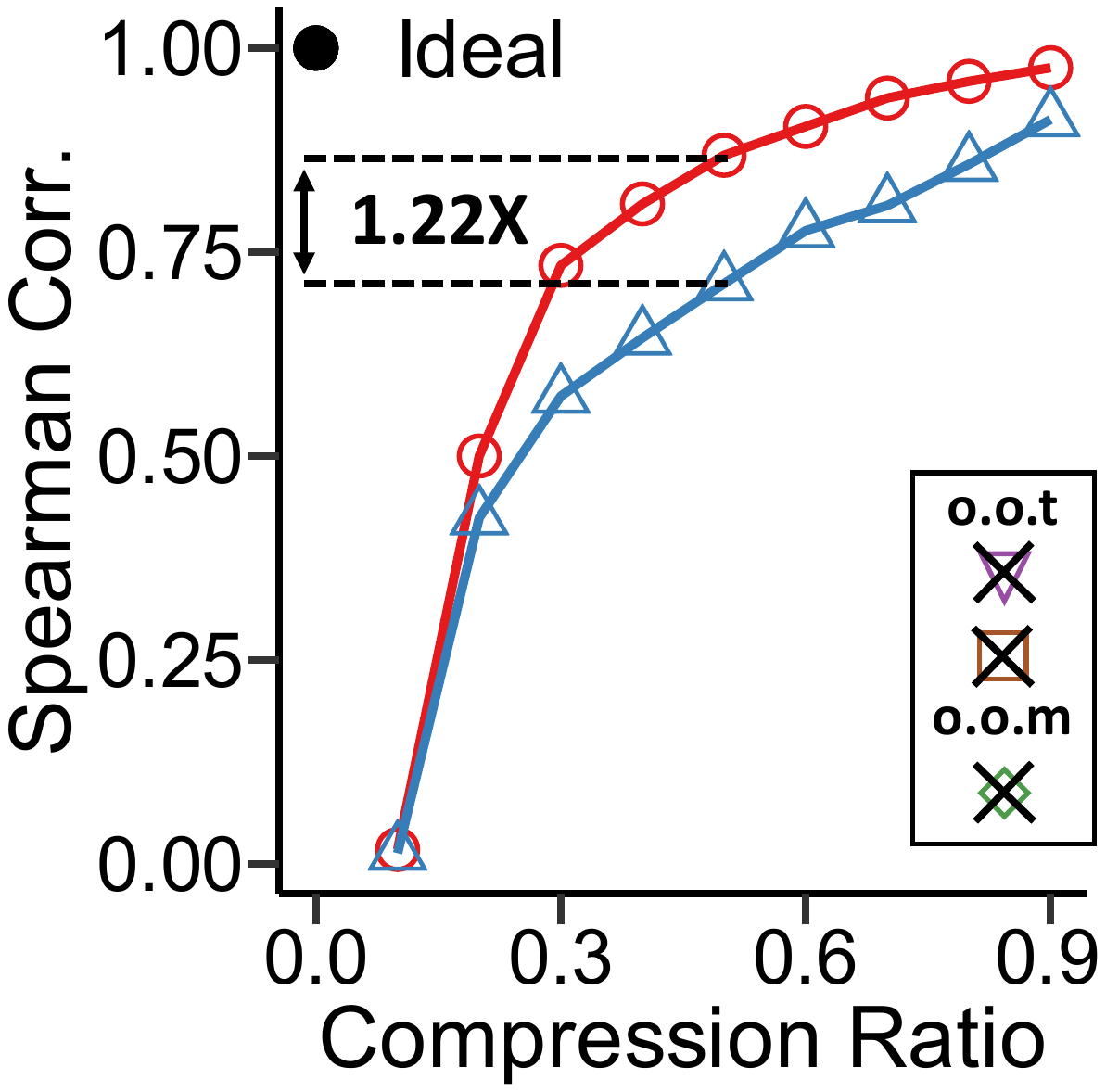}
	}
	\subfigure[Skitter (\RWR)]{
		\includegraphics[width=0.145\textwidth]{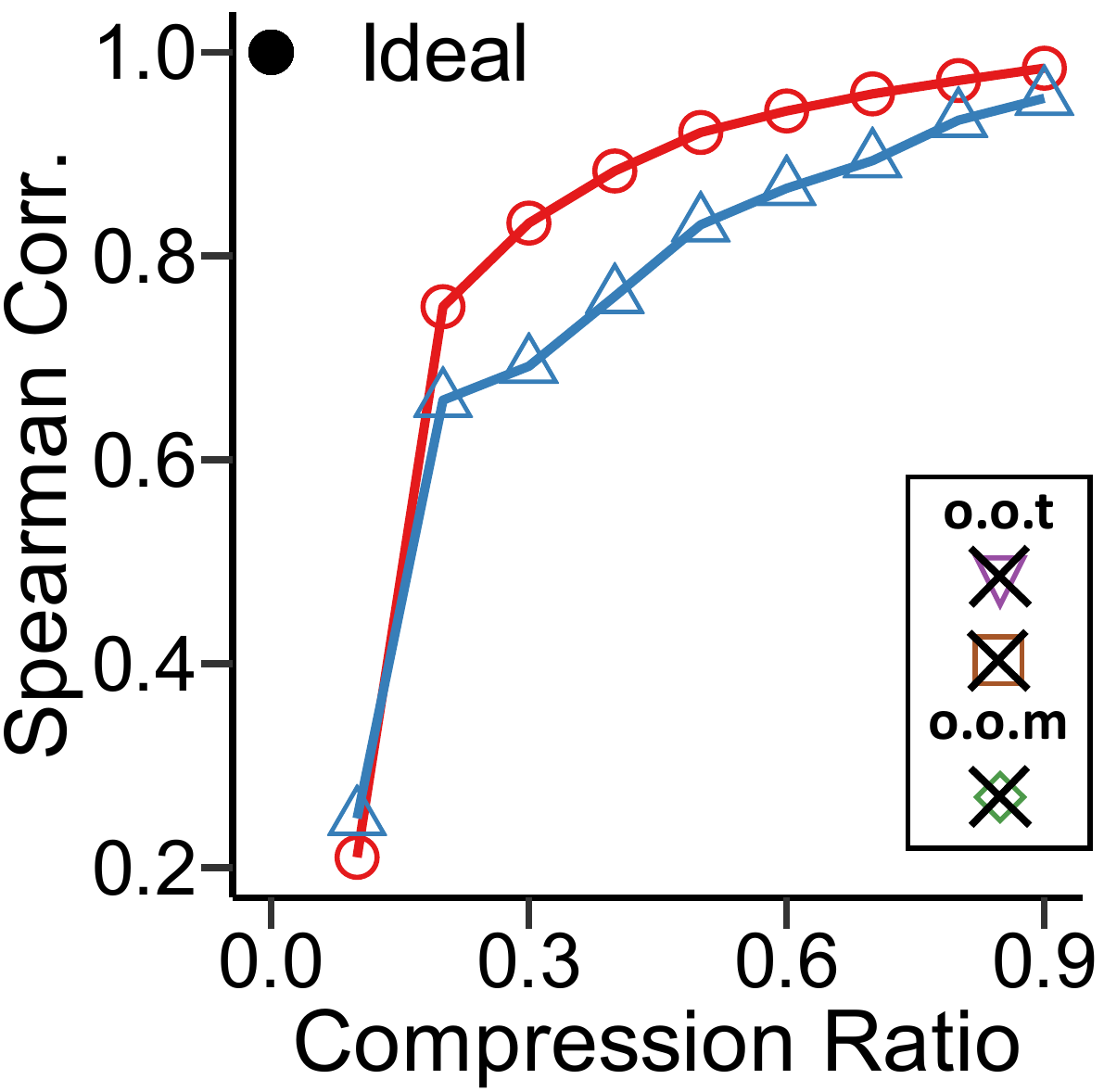}
	}
	\subfigure[Wikipedia (\RWR)]{
		\includegraphics[width=0.145\textwidth]{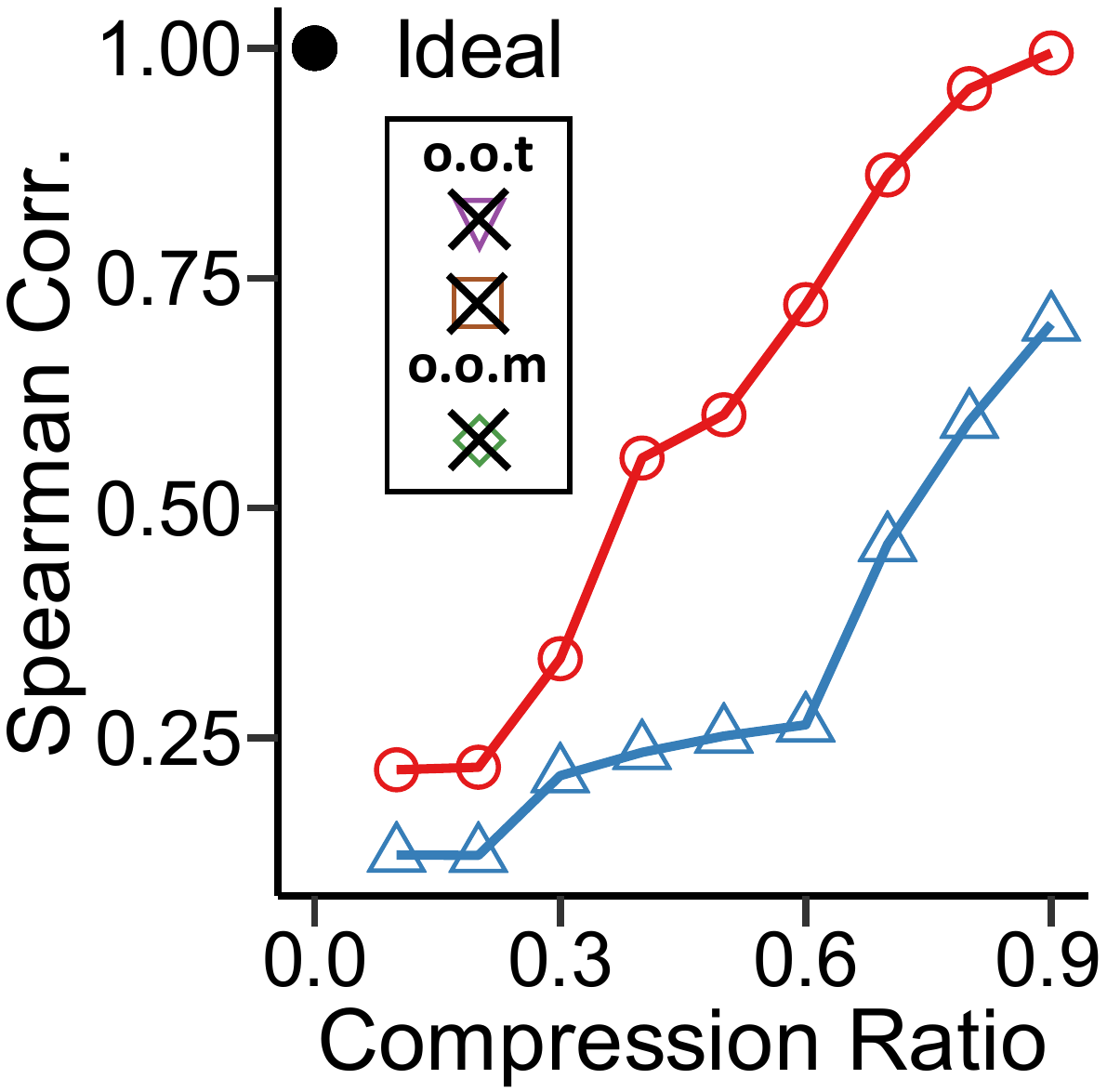}
	}
	\vspace{-3mm}
	\\
	\subfigure[LastFM-Asia (\HOP)]{
		\includegraphics[width=0.145\textwidth]{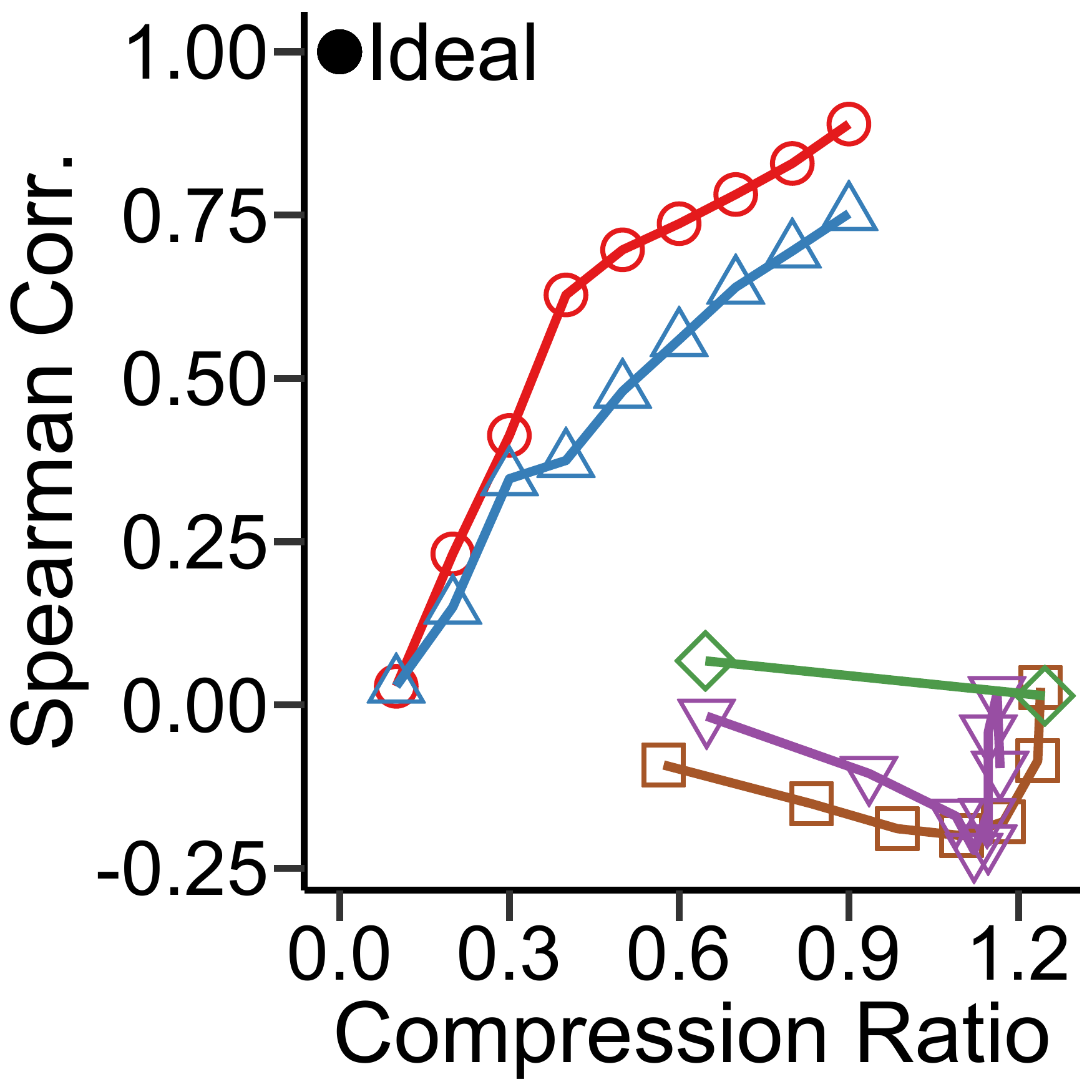}
	}
	\subfigure[Caida (\HOP)]{
		\includegraphics[width=0.145\textwidth]{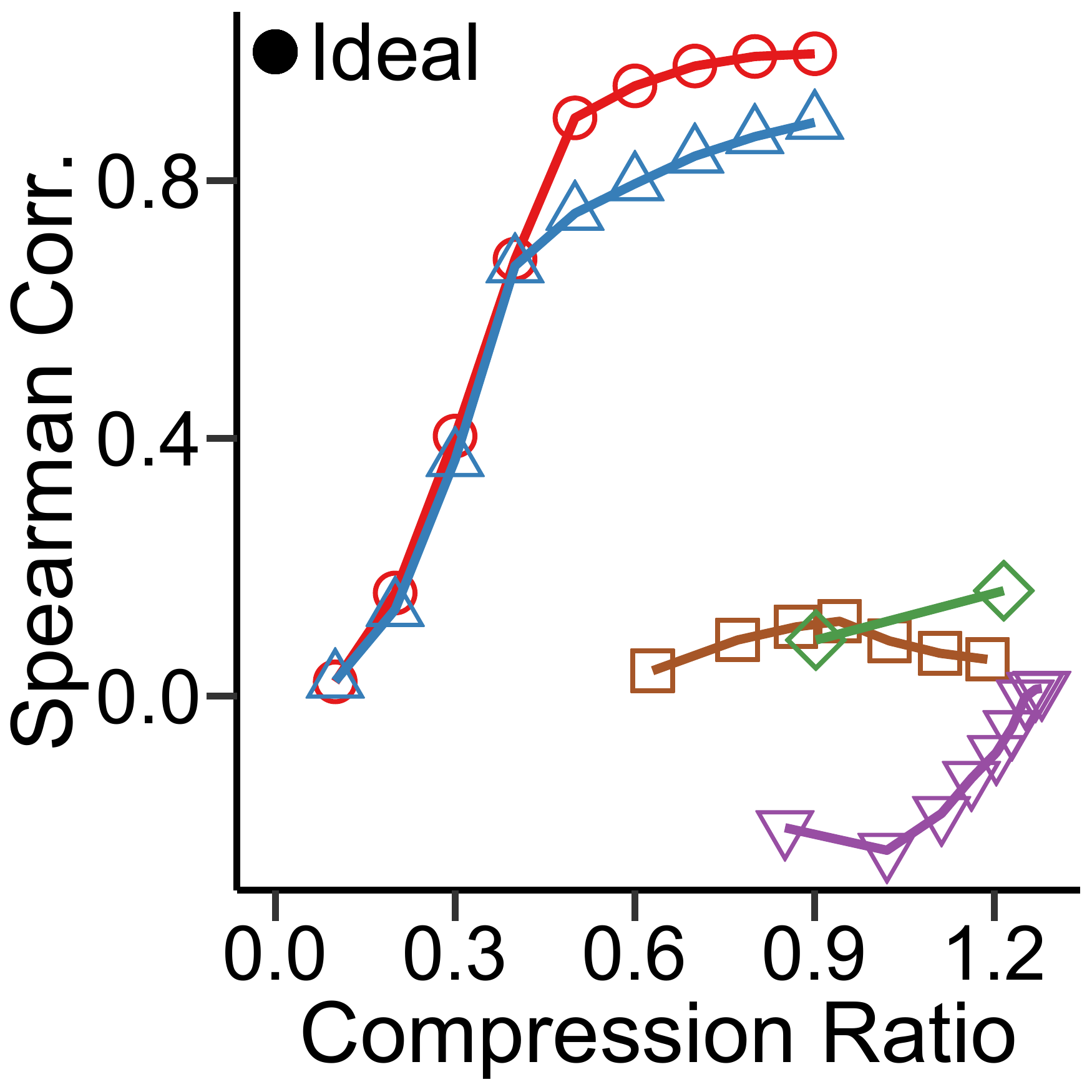}
	}
	\subfigure[DBLP (\HOP)]{
		\includegraphics[width=0.145\textwidth]{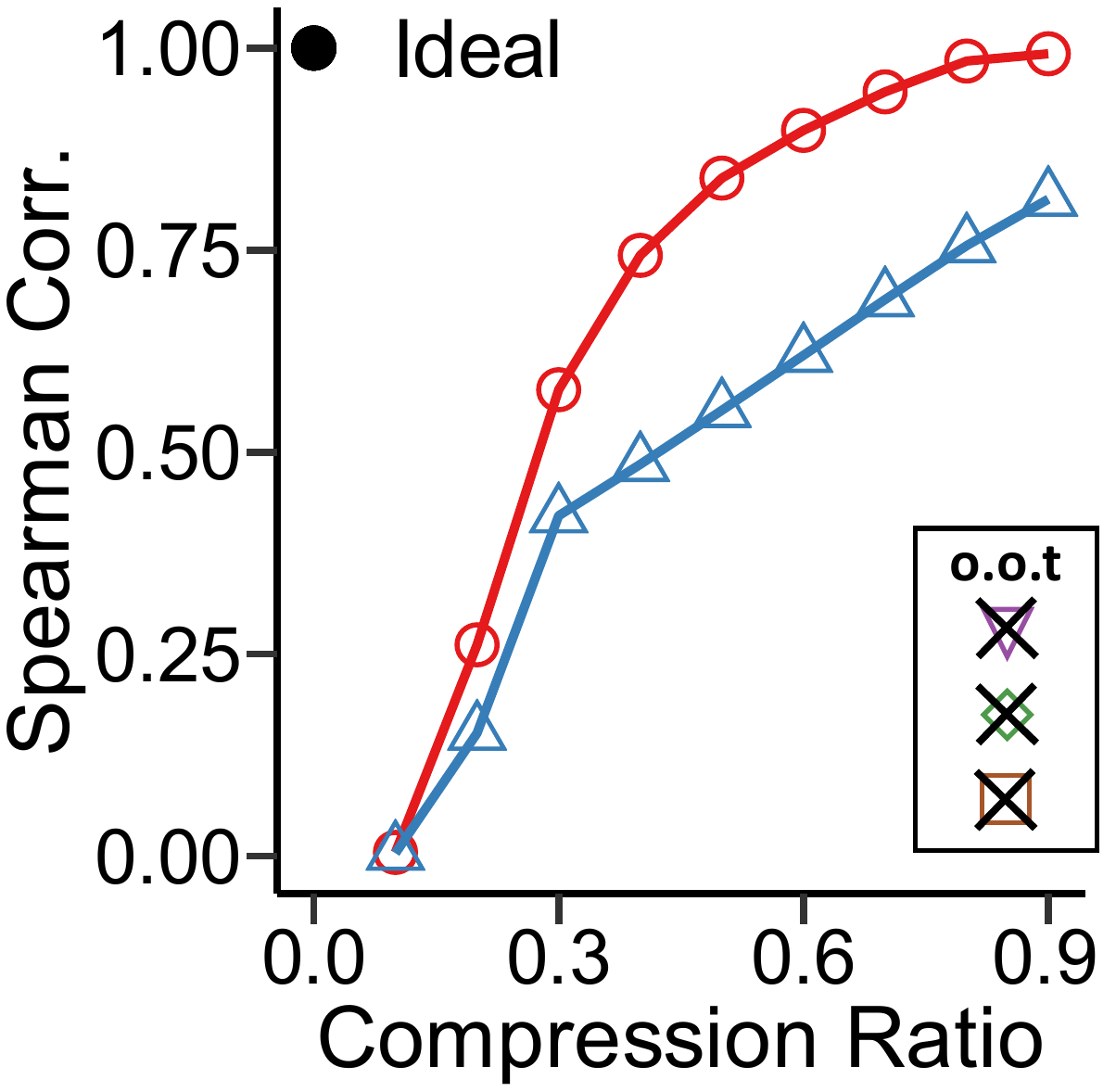}
	}
    \subfigure[Amazon0601 (\HOP)]{
		\includegraphics[width=0.145\textwidth]{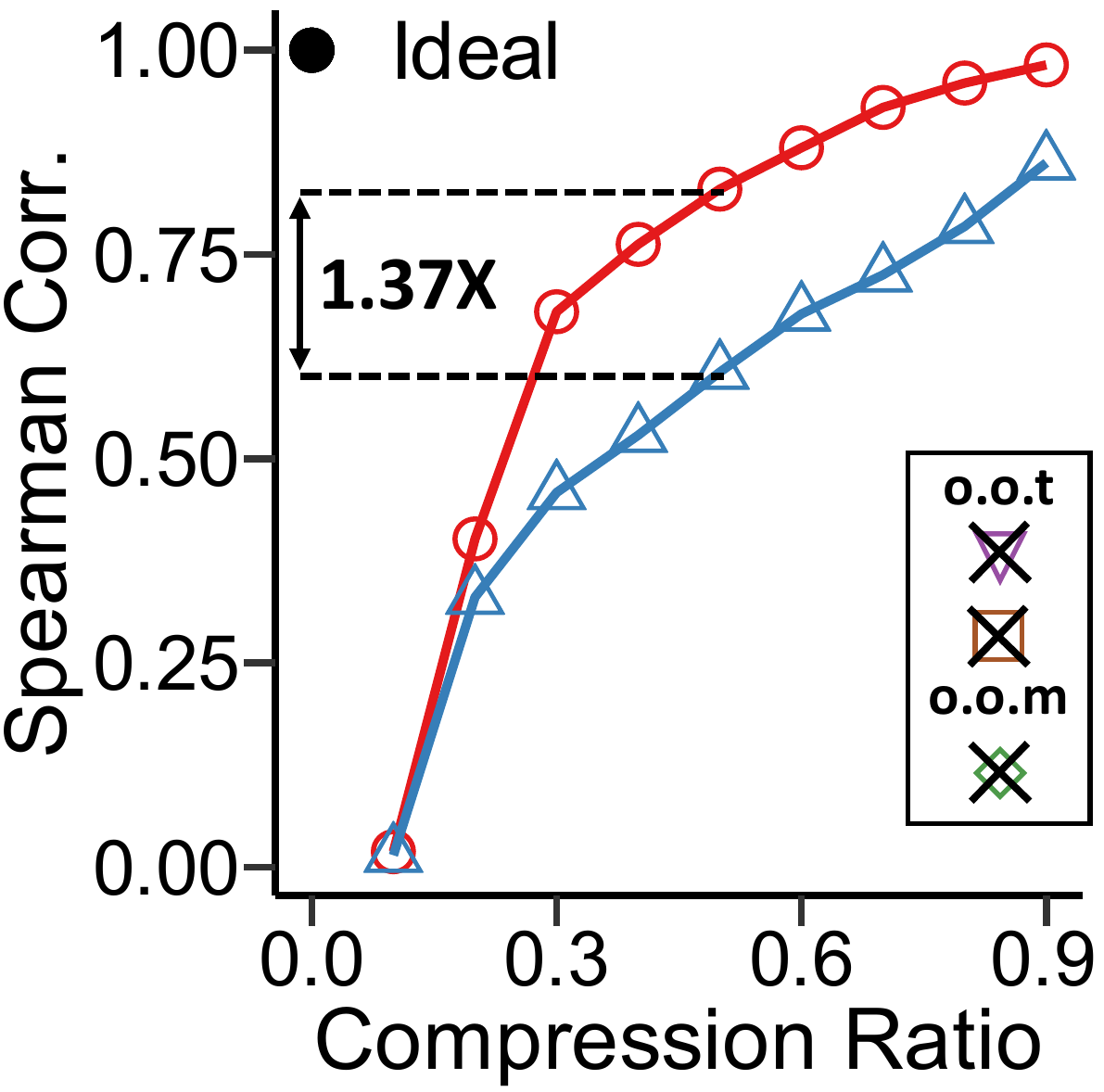}
	}
	\subfigure[Skitter (\HOP)]{
		\includegraphics[width=0.145\textwidth]{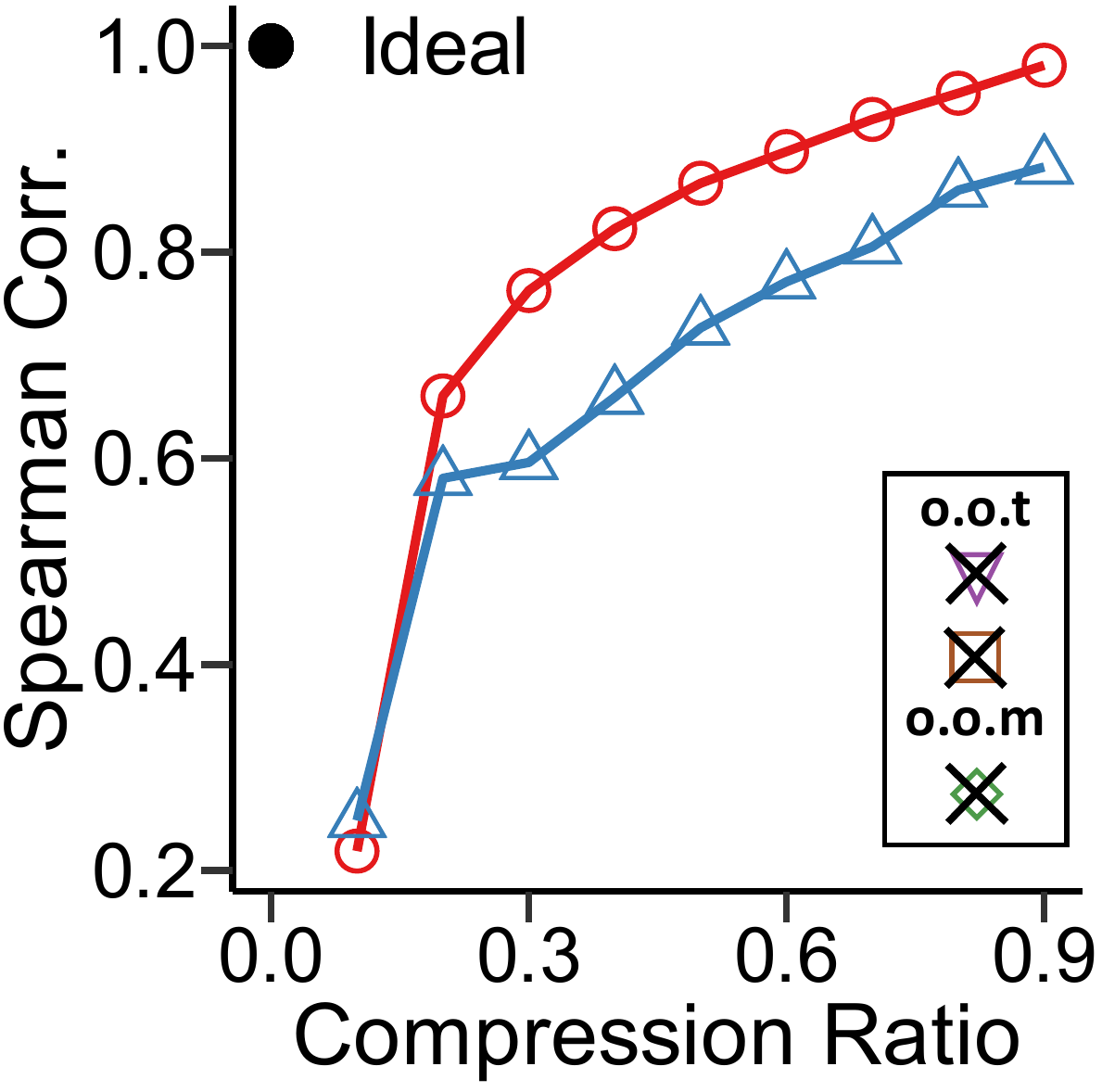}
	} 
	\subfigure[Wikipedia (\HOP)]{
		\includegraphics[width=0.145\textwidth]{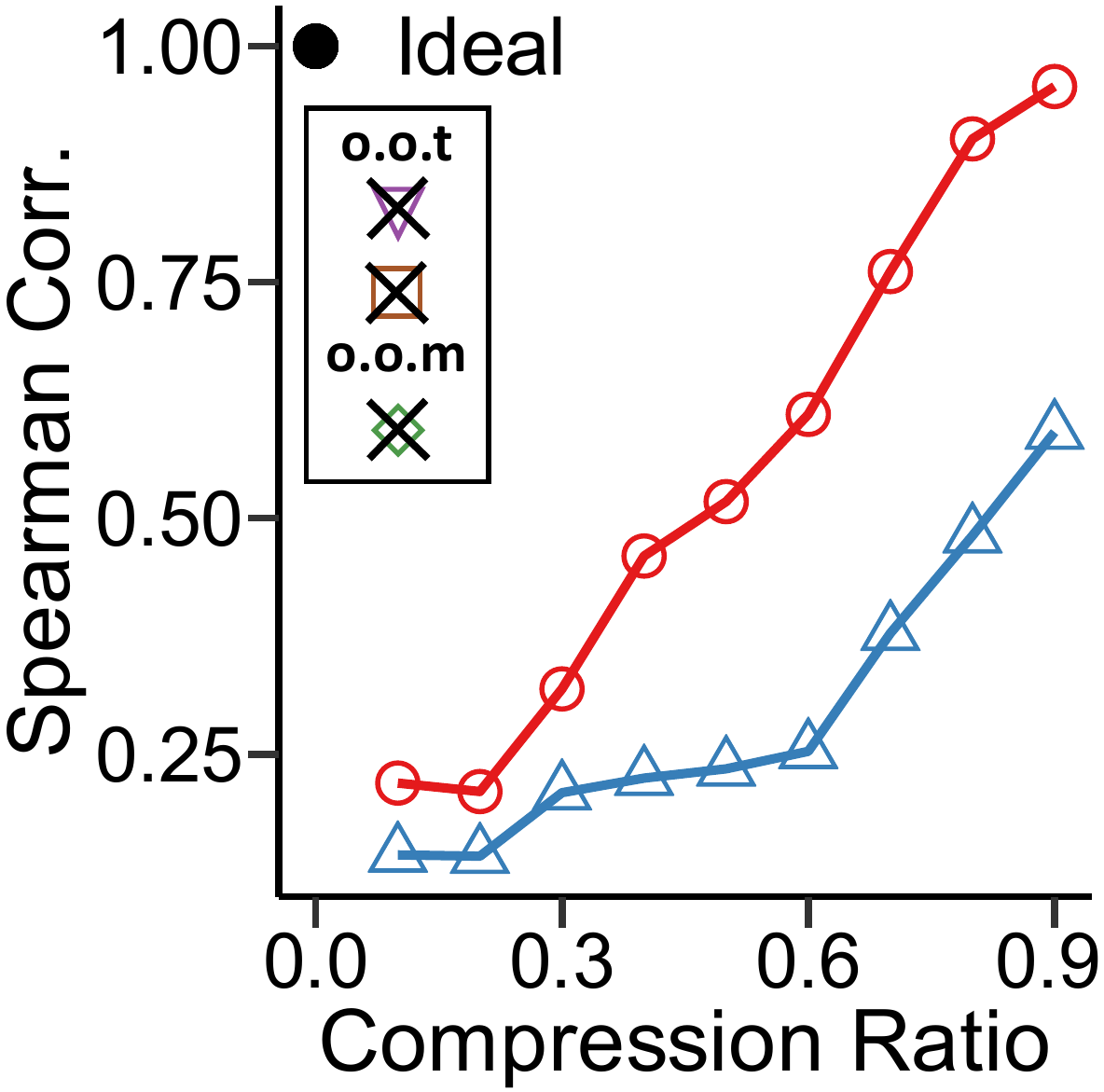}
	}
	\\
    \caption{\label{fig:PSSingleRandomTNS}
    \underline{\smash{\textbf{\OurModel gives summary graphs where queries on target nodes are answered most accurately.}}}
    o.o.t: out of time ($>$ 48hours). o.o.m: out of memory ($>$ 128GB). The degree of personalization $\alpha$ is fixed to $1.25$, and the size of the target node set is fixed to $100$. 
	}
\end{figure*}

\subsection{Q2. Scalability of \method (Fig.~\ref{fig:random-scalability})} 
\label{sec:q:scalability}

We show the linear scalability of \OurModel.
To this end, we measured how the execution time depend on the number of edges in the input graph, while fixing $|\TargetNodeSet|$ to $100$ or $\frac{|V|}{2}$.
We obtained the induced subgraphs of different sizes by randomly sampling different numbers of nodes ranging from $10\%$ to $100\%$ at the same interval from the the Skitter dataset and a synthetic graph ($|V|=10^{7}, |E|=10^{9}$).
The synthetic graph was generated by the the Barabási-Albert model \cite{barabasi1999emergence}, which reflects several structural properties of real-world graphs.
As seen in Figs.~\ref{fig:crown:scalable} and \ref{fig:random-scalability}, \textbf{\OurModel scaled linearly with the number of edges}, regardless of the target node number.

\begin{figure*}[!t]
    \vspace{-3mm}
	\centering
	\includegraphics[width=0.5\linewidth]{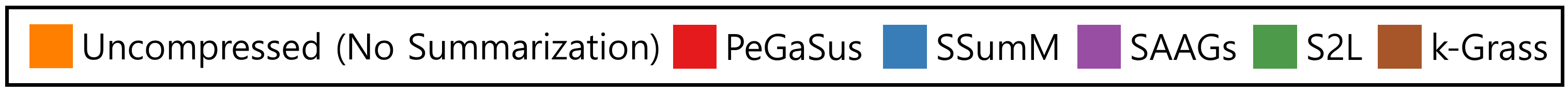} \\
	\vspace{-2mm}
	\subfigure[Summarization Time]{
		\includegraphics[width=0.31\textwidth]{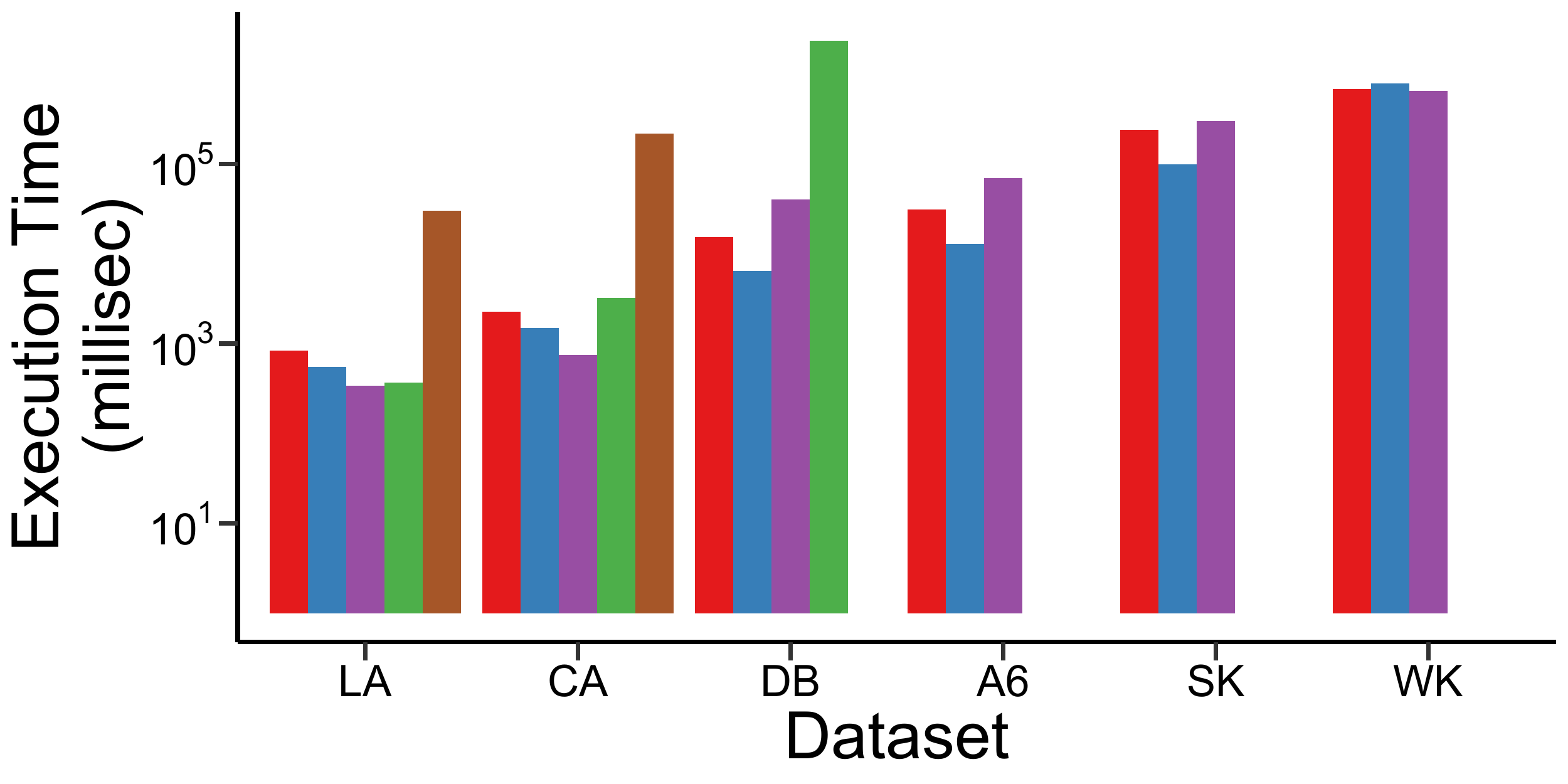}
	}
		\subfigure[Query Time: Breadth First Search]{
		\includegraphics[width=0.31\textwidth]{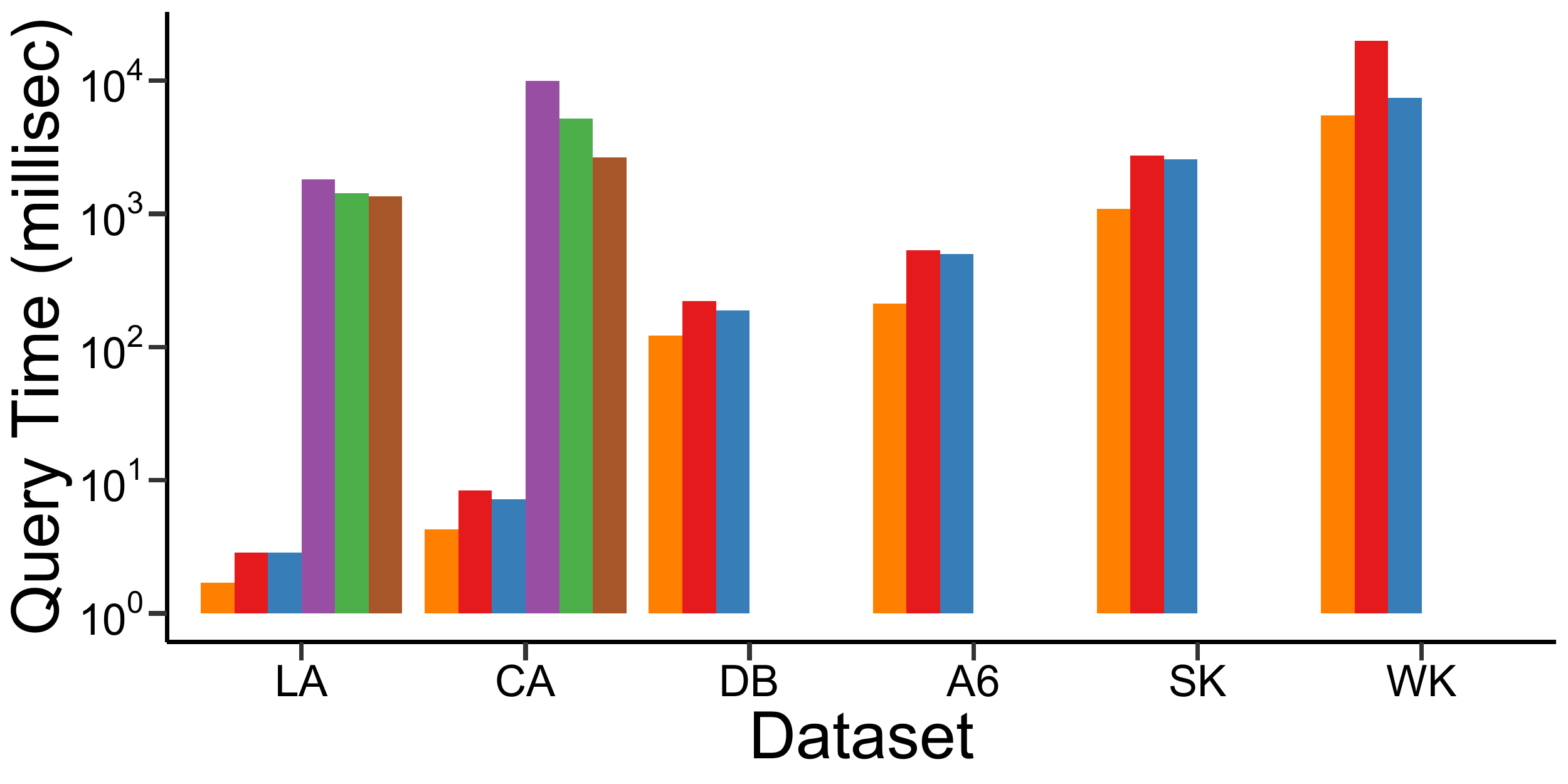}
	}
	\subfigure[Query Time: Random Walk with Restart]{
		\includegraphics[width=0.31\textwidth]{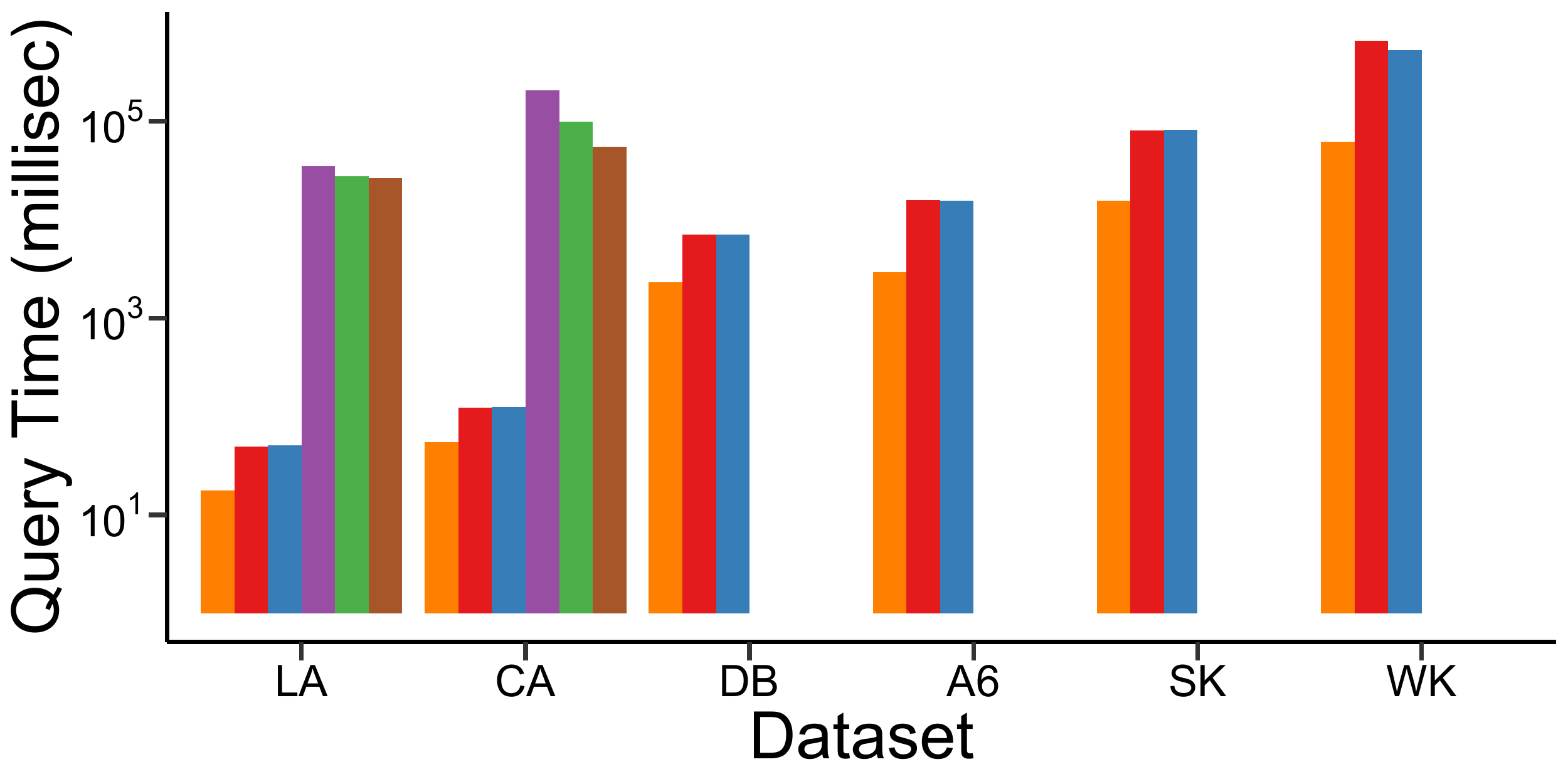}
	}
    \vspace{-1mm}
    \caption{
    \underline{\smash{\textbf{\OurModel is scalable, and it provides sparse summary graphs on which queries are processed efficiently.}}}
    \SL and \kGrass run out of time ($>$ 48 hours) or out of memory ($>$ 128 GB)  for large datasets; and 
    \SAAGs produces dense summary graphs where queries run out of time ($>$ 48 hours). 
    The compression ratio is $0.5$, and it takes almost the same time to process \RWR queries and \PHP queries.
	}
	\label{fig:executionandquery}
\end{figure*}

\subsection{Q3. Comparison with the State of the Art (Figs.~\ref{fig:PSSingleRandomTNS}-\ref{fig:executionandquery})}
\label{sec:q:comparewithSumm}

We compare \OurModel with state-of-the-art non-personalized graph summarization methods in terms of (a) the accuracy of query answers, (b) the conciseness of summary graphs, and (c) speed.
We sampled $100$ query nodes uniformly at random and used them as the target node set $\TargetNodeSet$.

As shown in Fig.~\ref{fig:PSSingleRandomTNS}, 
\textbf{RWR and HOP queries were answered significantly more accurately from personalized summary graphs obtained by \OurModel} than from non-personalized ones obtained by all other algorithms.
We obtained similar results for PHP queries, as reported in the online appendix\cite{appendixurl}.
For example, in the Amazon0601 dataset, queries were answered up to $2.74\times$ and $1.37\times$ more accurate in terms of \SMAPE and \Spearman (see Sect.~\ref{sec:experimetalsetting}), respectively, when the compression rate was $0.5$.

\begin{figure}[t]
    \vspace{-1mm}
    \centering
    \includegraphics[width=0.8\linewidth]{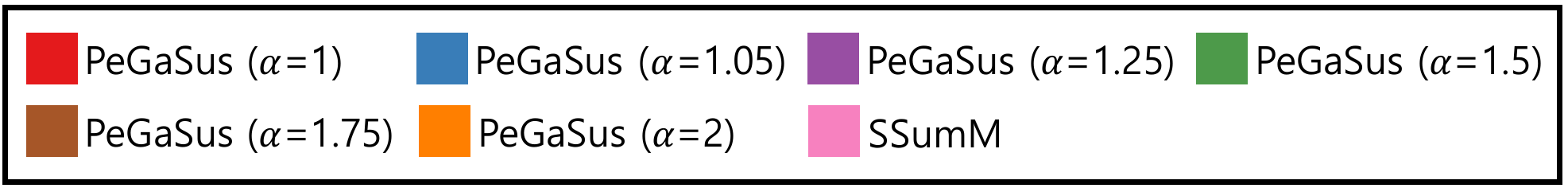} \\
    \vspace{-1mm}
	\subfigure[\hspace{6mm}\SMAPE\hspace{8mm}(Compression Ratio = 0.3)]{
		\includegraphics[width=0.22\textwidth]{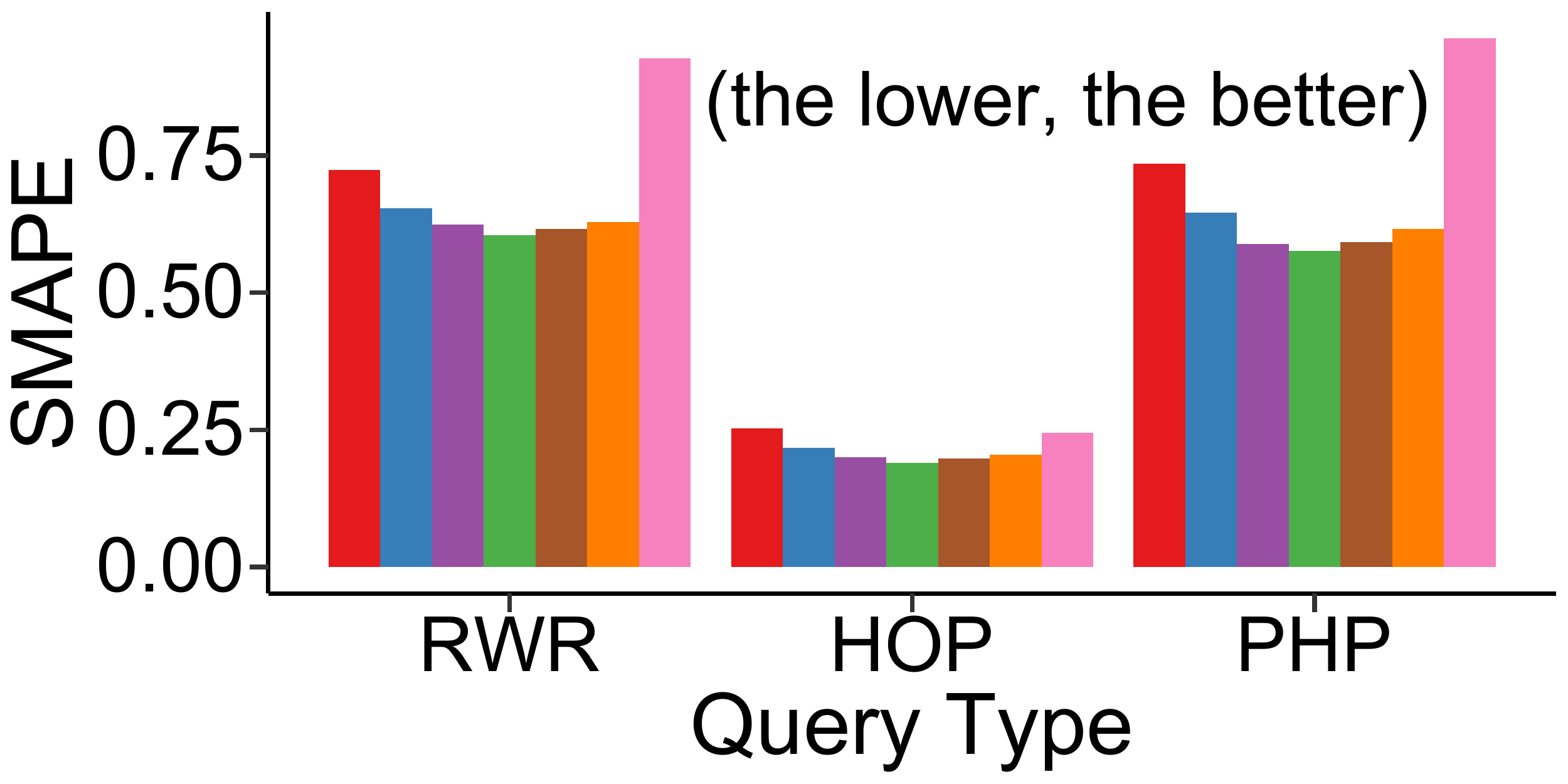}
	}
	\subfigure[\hspace{2mm}Spearman Correlation (Compression Ratio = 0.3)]{
		\includegraphics[width=0.22\textwidth]{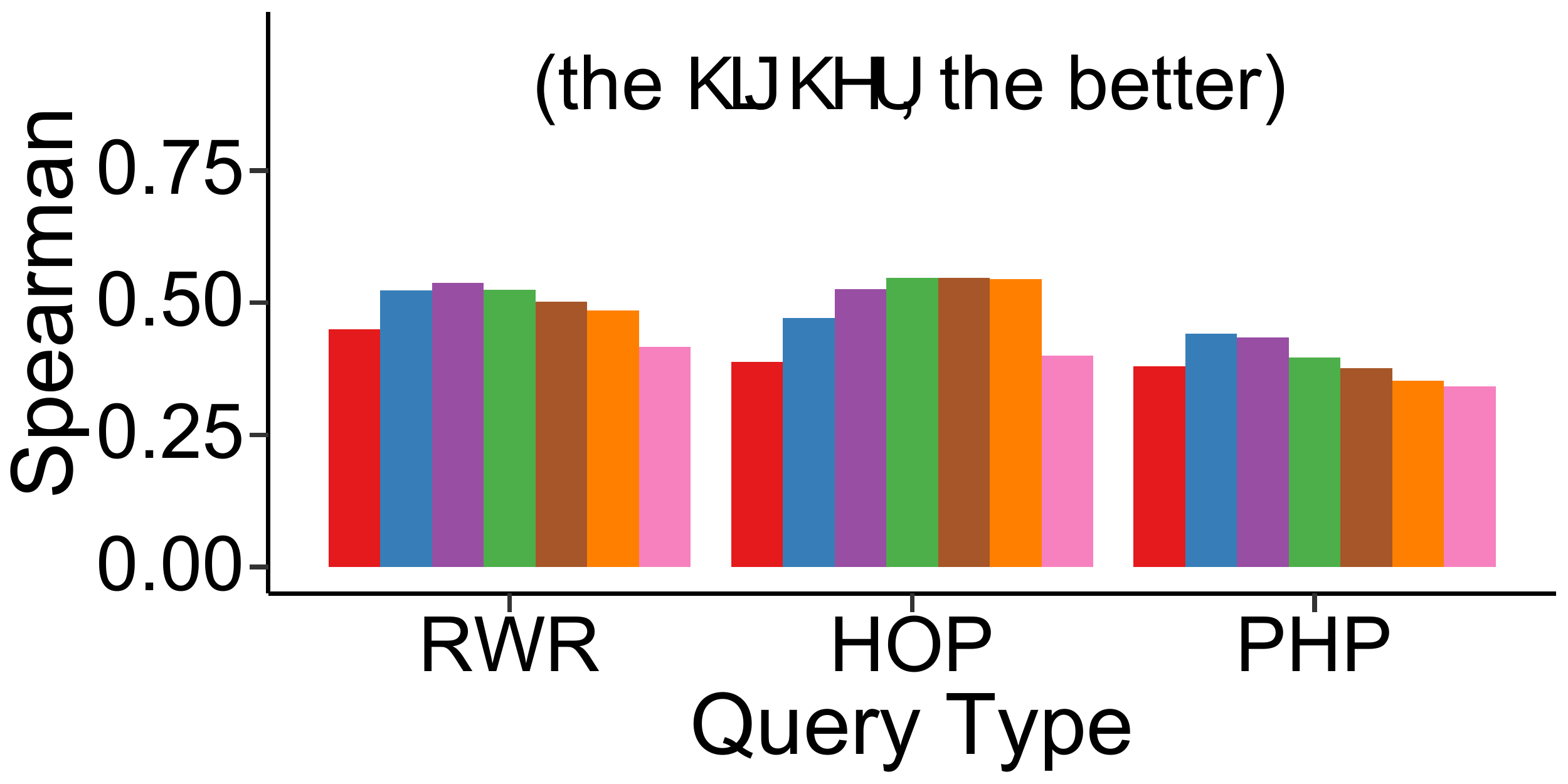}
	}\\
	\vspace{-2mm}
	\subfigure[\hspace{6mm}\SMAPE\hspace{8mm}(Compression Ratio = 0.5)]{
		\includegraphics[width=0.22\textwidth]{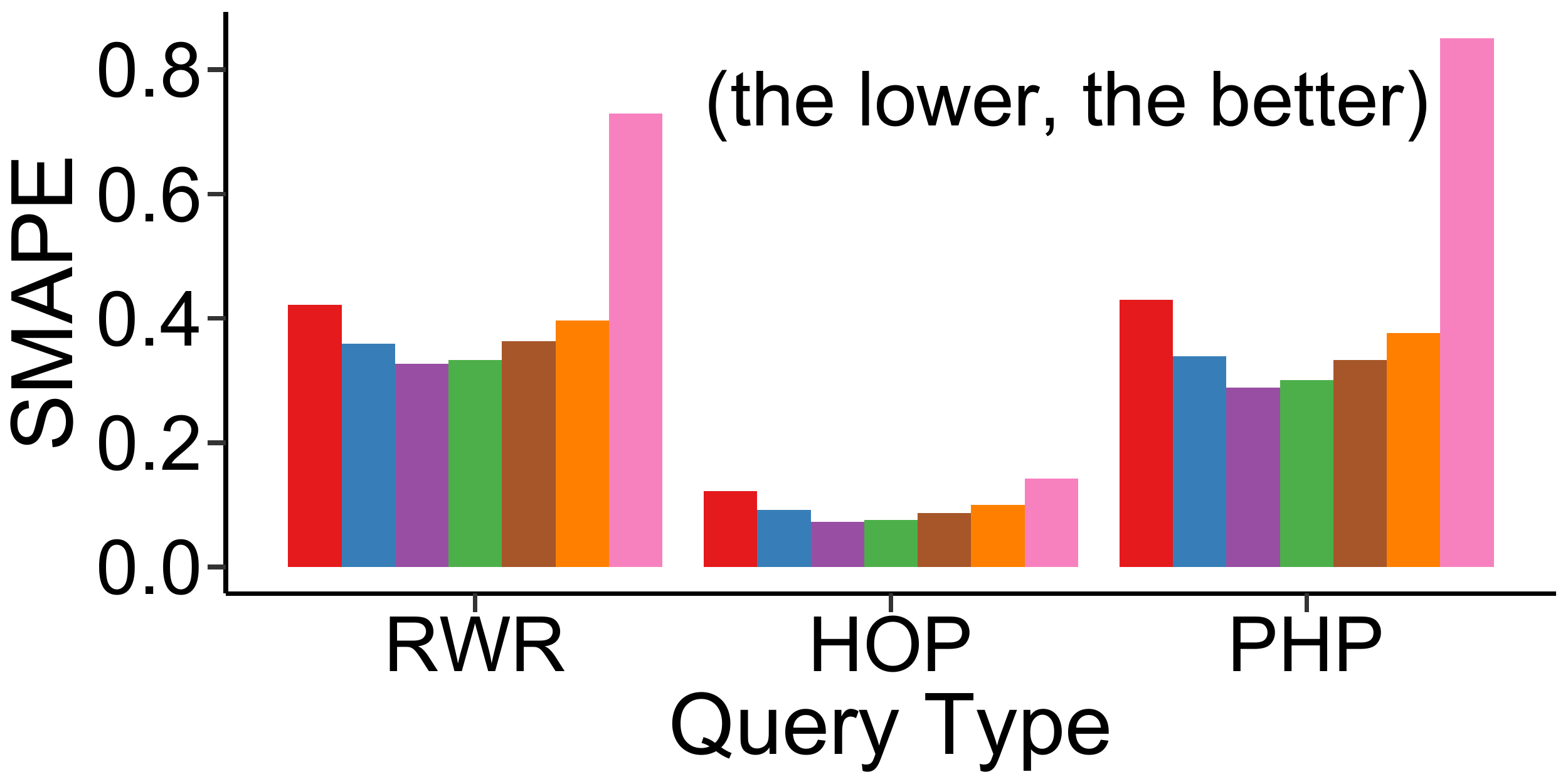}
	}
	\subfigure[\hspace{2mm}Spearman Correlation (Compression Ratio = 0.5)]{
		\includegraphics[width=0.22\textwidth]{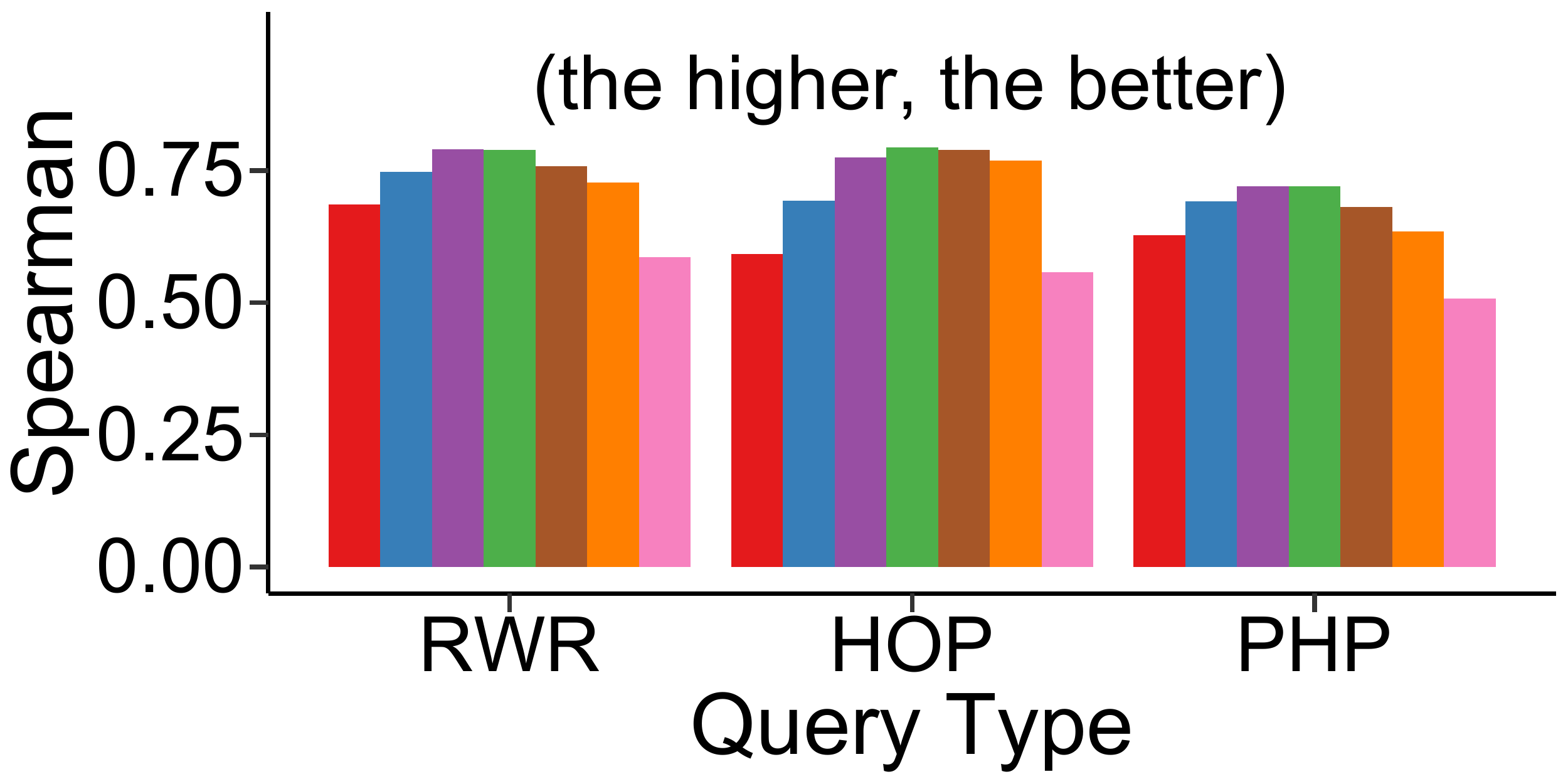}
	}
    \vspace{-2mm}
    \caption{\label{fig:varMetricperAlphaRandomTNS}
    \underline{\smash{\textbf{Queries on target nodes can be answered more accurately from}}}\\ 
    \underline{\smash{\textbf{personalized summary graphs}}} ($\alpha>1$) than from non-personalized ones ($\alpha=1$).
    The results are averaged over all datasets. 
    Note that answers are most accurate when the degree of personalization $\alpha$ is moderate.
    }
\end{figure}

\begin{figure}[t]
    \vspace{-5mm}
	\centering
	\includegraphics[width=0.6\linewidth]{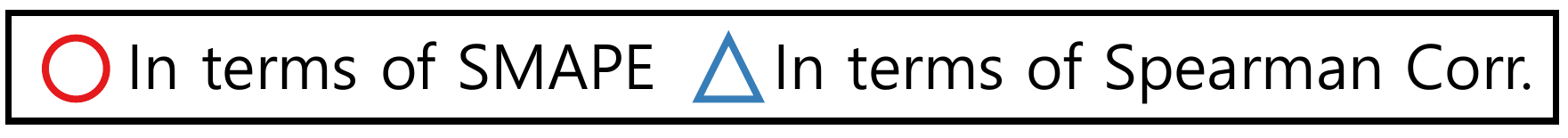} \\
	 \vspace{-1mm}
	\subfigure[\RWR]{
    		\includegraphics[width=0.145\textwidth]{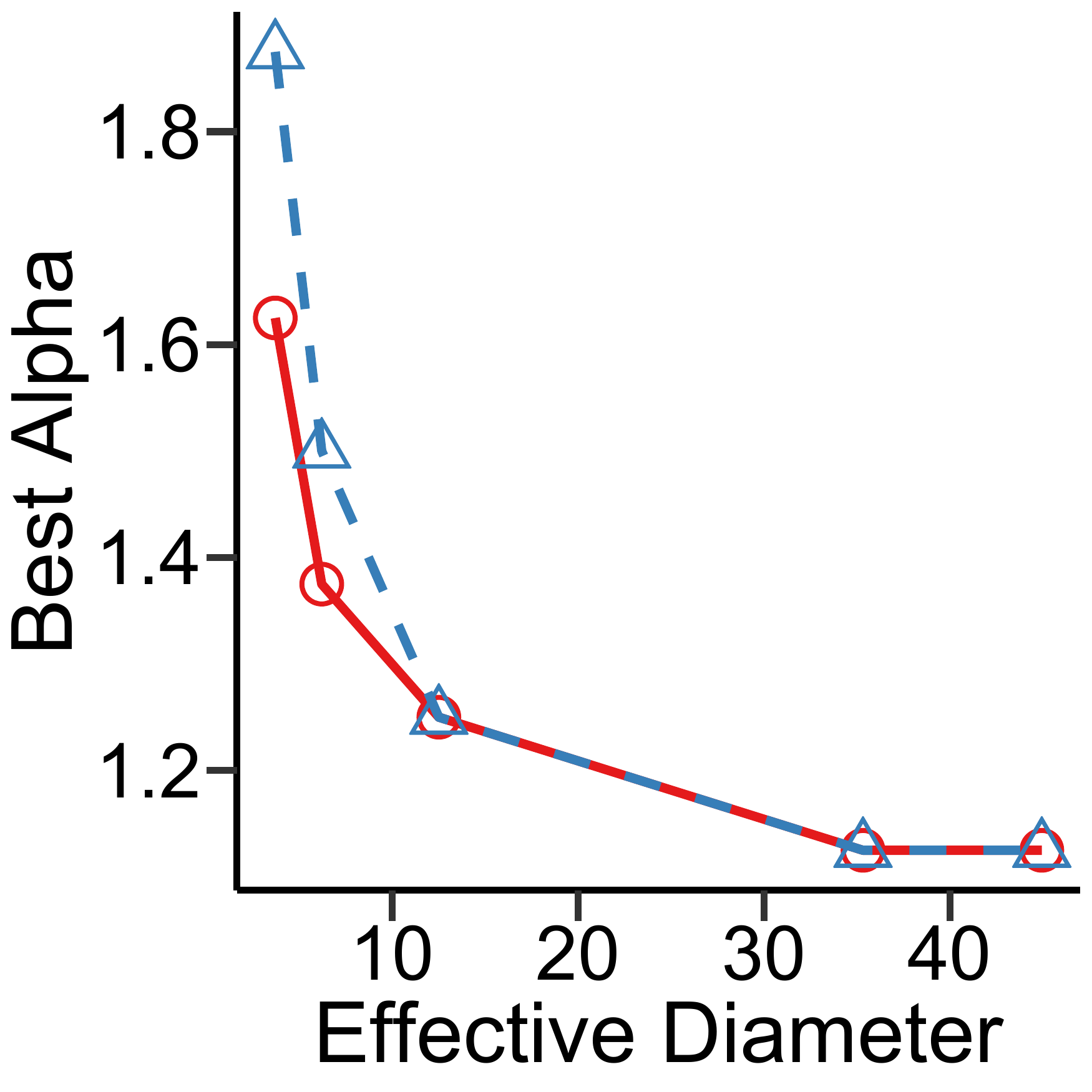}
	}
	\subfigure[\HOP]{
		\includegraphics[width=0.145\textwidth]{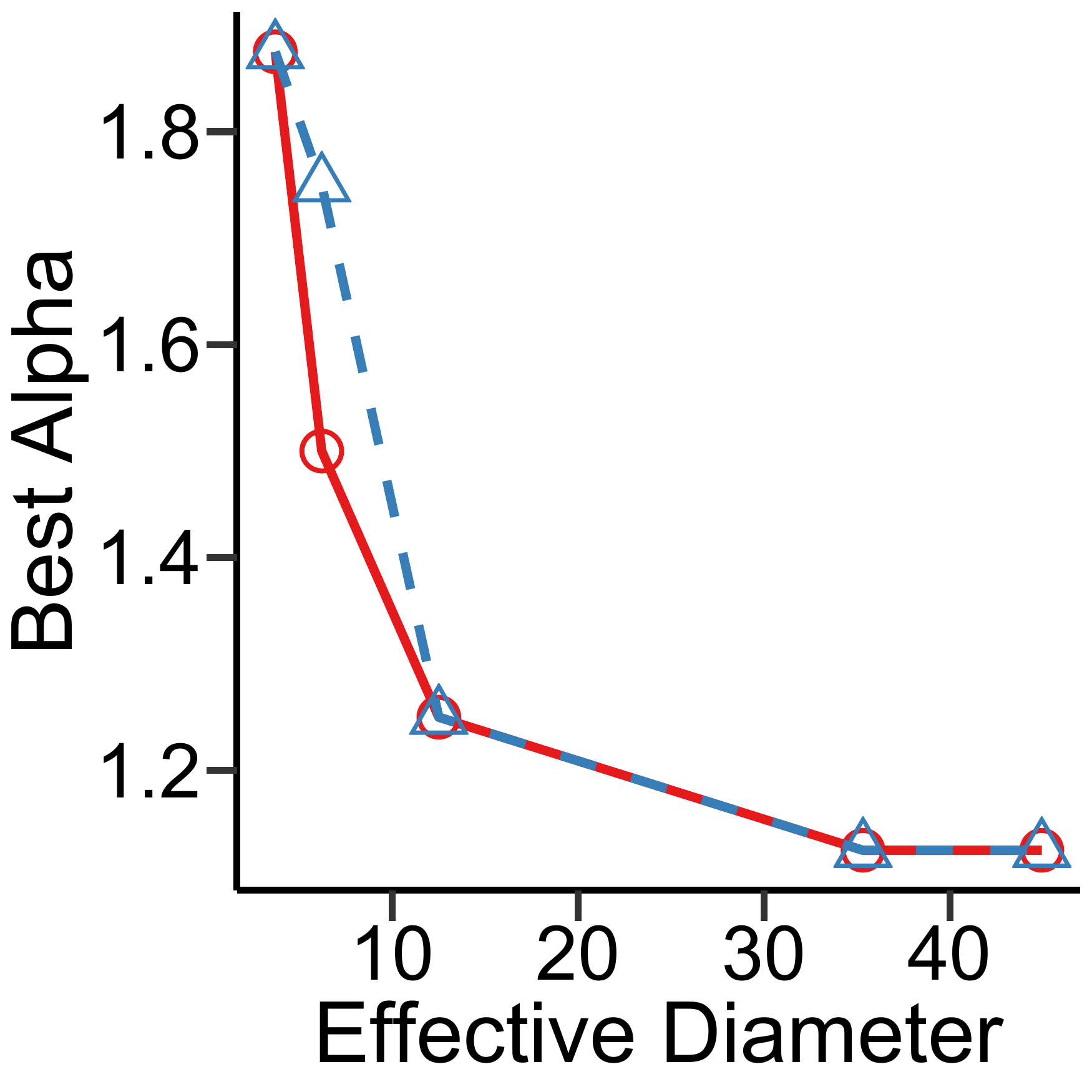}
	}
	\subfigure[\PHP]{
		\includegraphics[width=0.145\textwidth]{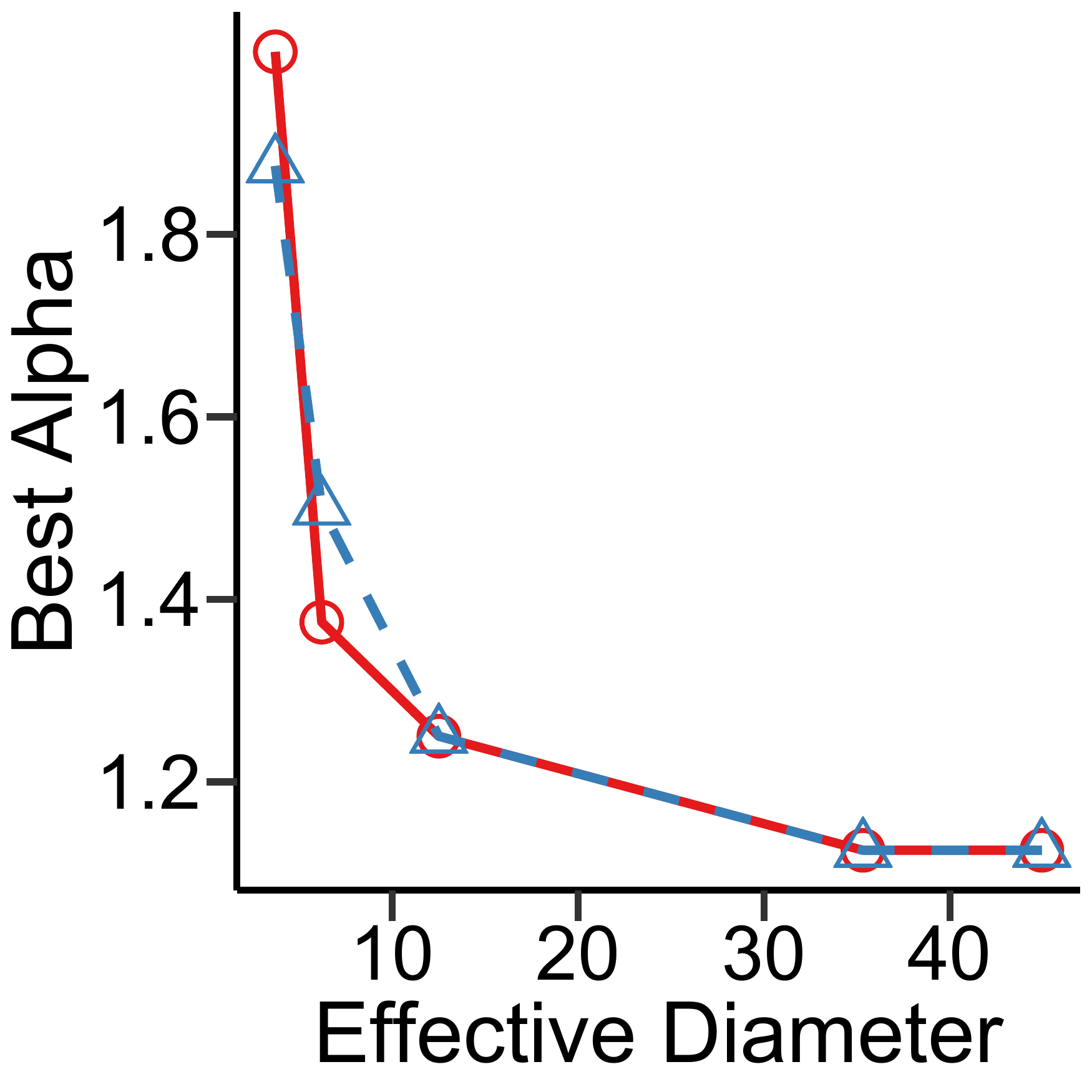}
	}
	\\
	\vspace{-1.5mm}
    \caption{\label{fig:random-diameter-alpha}
    \underline{\smash{\textbf{The best-performing $\alpha$ decreases as the effective diameter of the}}} \underline{\smash{\textbf{input graph increases.}}}  
    The best-performing alphas are chosen based on each evaluation measure (i.e., \SMAPE and \Spearman) when the compression ratio is $0.3$.
	}
\end{figure}

\begin{figure}[t]
    \vspace{-1mm}
    \centering
    \includegraphics[width=0.8\linewidth]{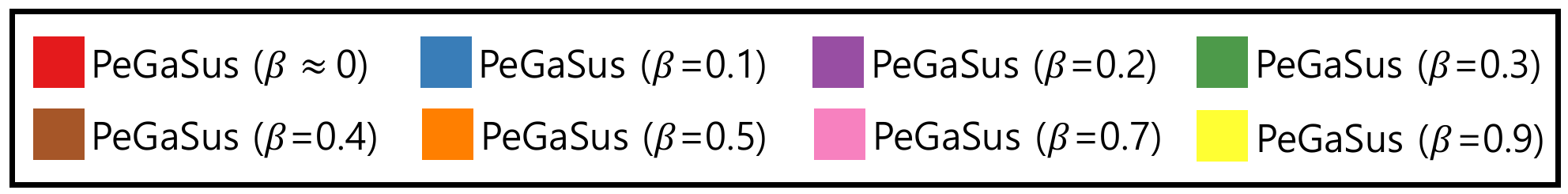} \\
    \vspace{-1mm}
	\subfigure[\hspace{6mm}\SMAPE\hspace{8mm}(Compression Ratio = 0.3)]{
		\includegraphics[width=0.22\textwidth]{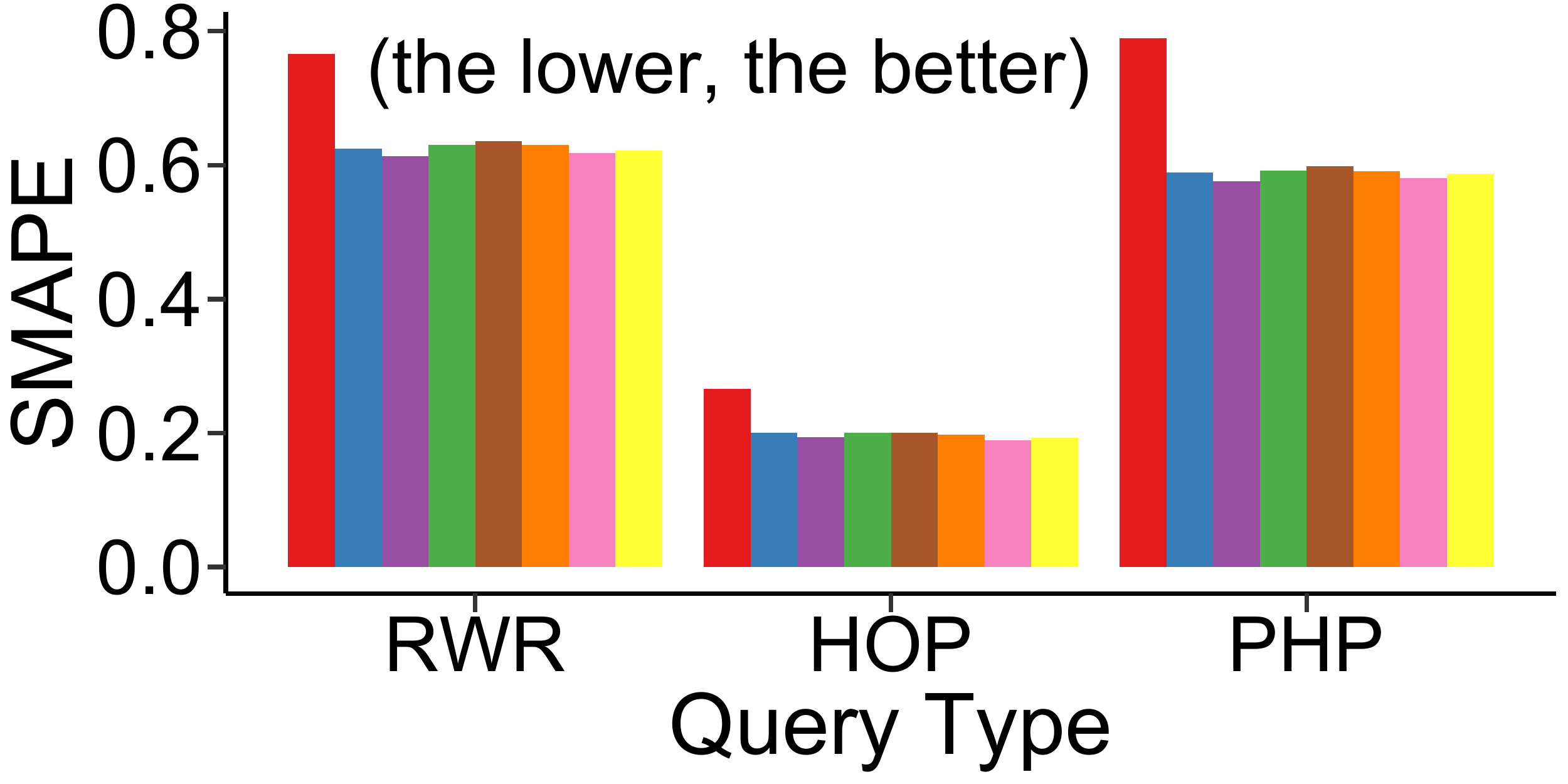}
	}
	\subfigure[\hspace{2mm}Spearman Correlation (Compression Ratio = 0.3)]{
		\includegraphics[width=0.22\textwidth]{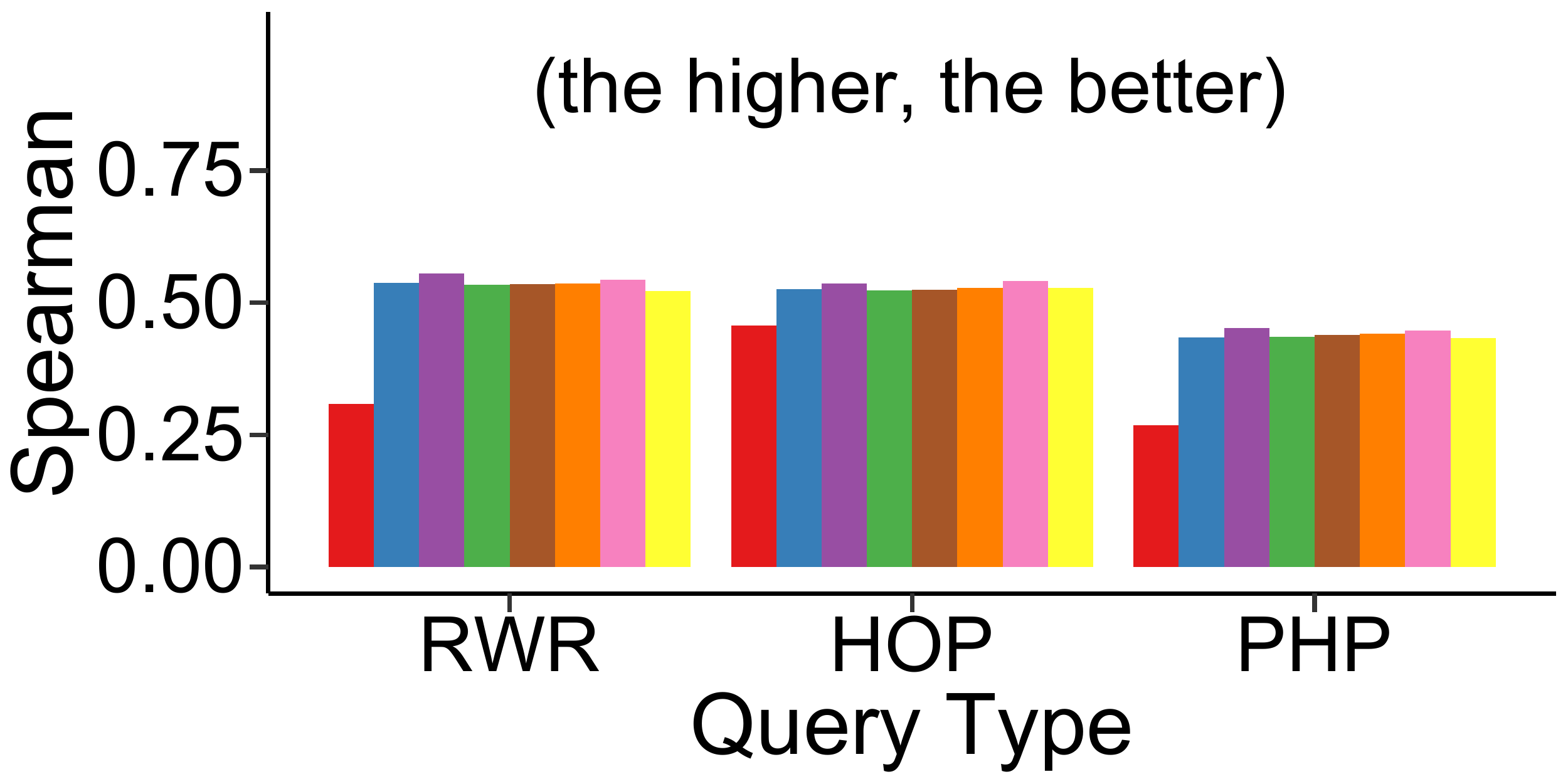}
	}\\
	\vspace{-2mm}
	\subfigure[\hspace{6mm}\SMAPE\hspace{8mm}(Compression Ratio = 0.5)]{
		\includegraphics[width=0.22\textwidth]{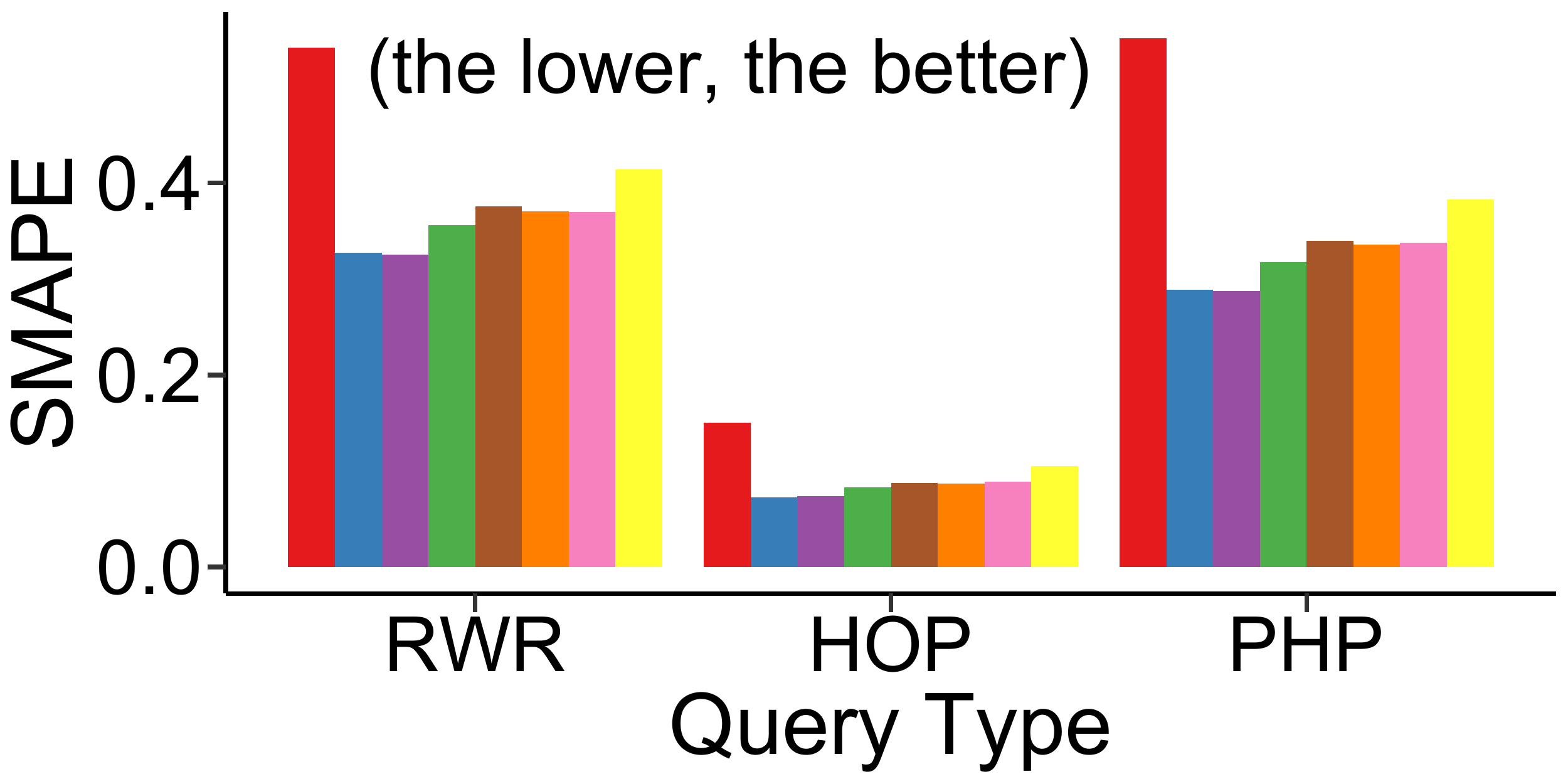}
	}
	\subfigure[\hspace{2mm}Spearman Correlation (Compression Ratio = 0.5)]{
		\includegraphics[width=0.22\textwidth]{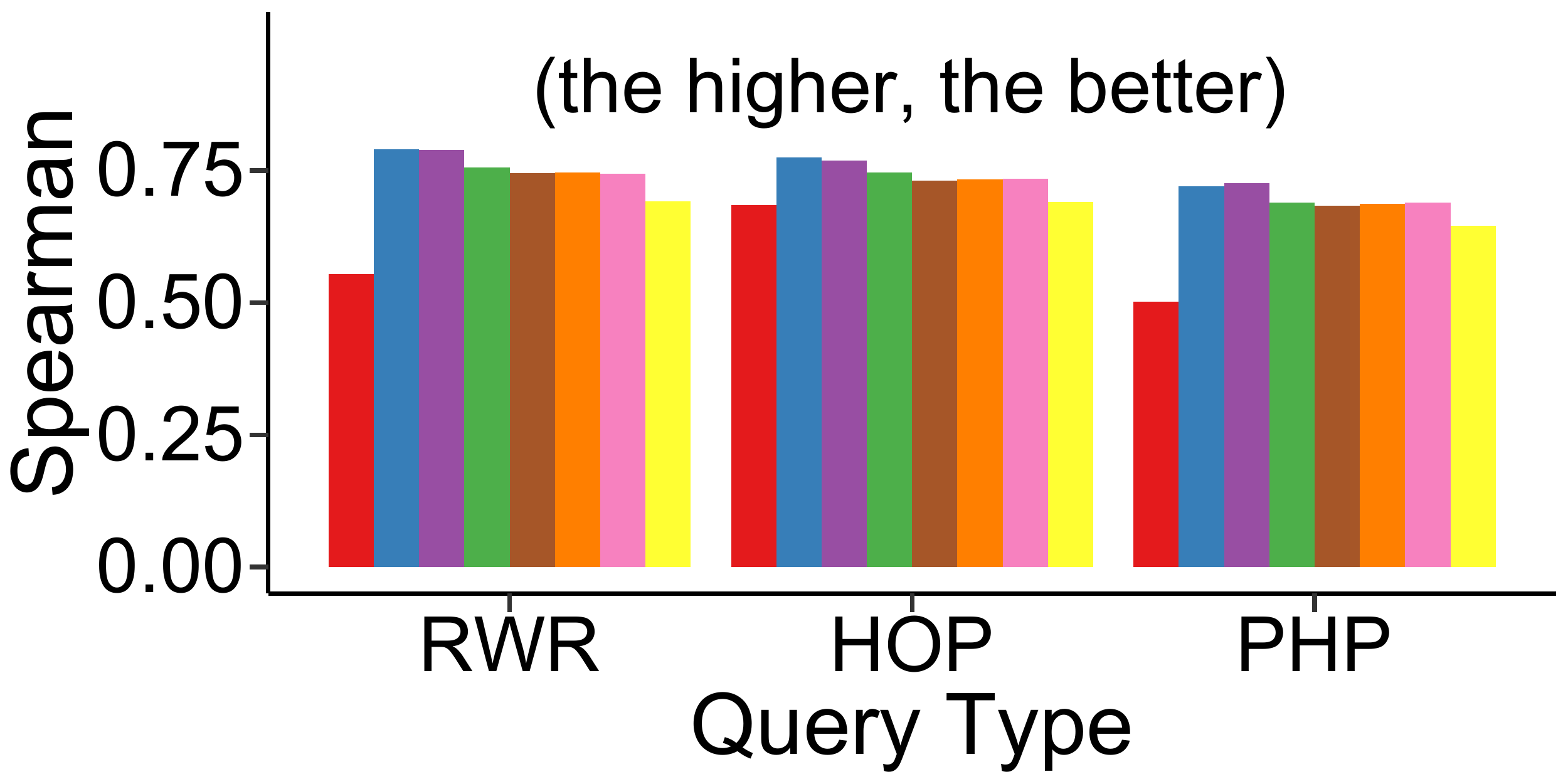}
	}
    \vspace{-2mm}
    \caption{\label{fig:randomTNSvarMetricPerBeta}
    \underline{\smash{\textbf{Queries are answered most accurately when $\mathbf{\beta}$ is moderate}}}, in the majority of the cases. The largest entry in $L$ (see Sect.~\ref{sec:thresholding}) is chosen when $\beta\approx 0$. The reported results are averaged over all datasets. 
    }
\end{figure}

As shown in Fig.~\ref{fig:executionandquery}, \textbf{\method was one of the most scalable algorithms, and queries were processed rapidly on its output} since it selectively adds superedges (see Sect.~\ref{sec:mergingandencoding}).
Query processing took much longer on dense summary graphs obtained by $\kGrass$, $\SL$, and $\SAAGs$, which add superedges without selection.

\subsection{Q4. Effects of Parameters (Figs.~\ref{fig:varMetricperAlphaRandomTNS}-\ref{fig:randomTNSvarMetricPerBeta})}
\label{sec:q:params}
We analyze how the accuracy of the answers of node-similarity queries (see Sect.~\ref{sec:experimetalsetting}) obtained from personalized summary graphs depends on parameters $\alpha$ and $\beta$.
We sampled $100$ query nodes uniformly at random and used them as the target node set $\TargetNodeSet$, unless otherwise stated.

\smallsection{Effect of $\alpha$.}
As seen in Fig.~\ref{fig:varMetricperAlphaRandomTNS}, the answers were most accurate when the degree of personalization $\alpha$ was moderate (spec., when it was $1.25$ or $1.5$);
and they were less accurate when $\alpha$ was higher and more global information was lost. 
Specifically, $\alpha = 1.5$ led to the best accuracy in the Wikipedia dataset, whose effective diameter\footnote{We used the 90-percentile effective diameter \cite{leskovec2005graphs}, i.e., the minimum number of hops such that 90\% of nodes pairs are within the hops from each other.} is remarkably small despite its huge size. In the all other datsets, $\alpha = 1.25$ led to the best accuracy.

In order to systematically analyze the relation between $\alpha$ and the effective diameter, we generated $5$ synthetic graphs of the same size and edges but with different effective diameters using the Watts-Strogatz model \cite{watts1998collective}. 
Specifically, we changed the rewiring probability from $0$ to $0.1$ (spec., we tried $0$, $0.0001$, $0.001$, $0.01$, and $0.1$), while fixing the number of nodes to $1,000$ and the number of edges to $10,000$, and as a result the effective diameter varied from $3.71$ to $44.95$. In these synthetic graphs, due to their large effective diameter, summary graphs cannot be personalized effectively to distant nodes. Thus, we sampled $100$ adjacent nodes by BFS from a random node and used them as query nodes and the target node set $T$.
\textbf{As shown in Fig.~\ref{fig:random-diameter-alpha}, the best-performing degree of personalization $\alpha$ decreased as the effective diameter increased.}
This is because, when the effective diameter is large, a large fraction of edges are distant from a target node set, and large $\alpha$ understates their weight, which depends on the distance, too much.

\begin{figure*}[!t]
    \vspace{-3mm}
	\centering
	\includegraphics[width=0.4\linewidth]{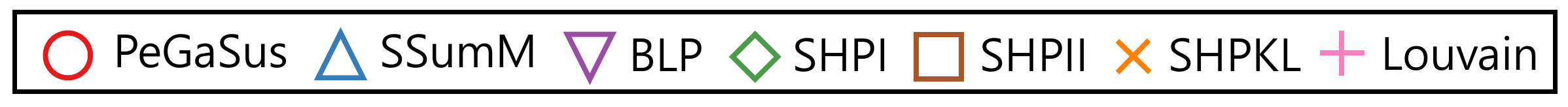} \\
	\textbf{\small Symmetric Mean Absolute Percentage Error (the lower, the better):} \hfill \ \ \\
	\subfigure[LastFM-Asia (\RWR)]{
		\includegraphics[width=0.145\textwidth]{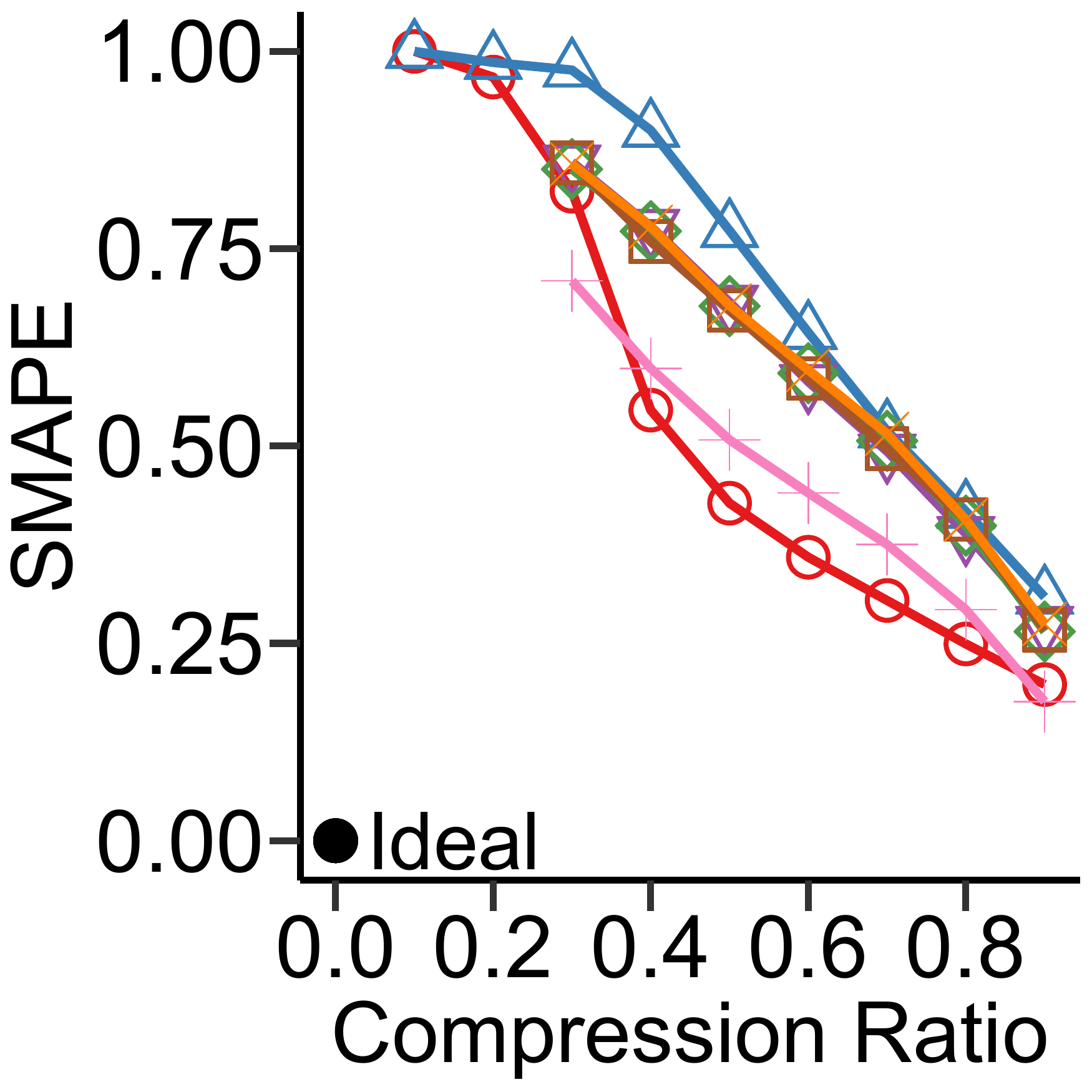}
	}
	\subfigure[Caida (\RWR)]{
		\includegraphics[width=0.145\textwidth]{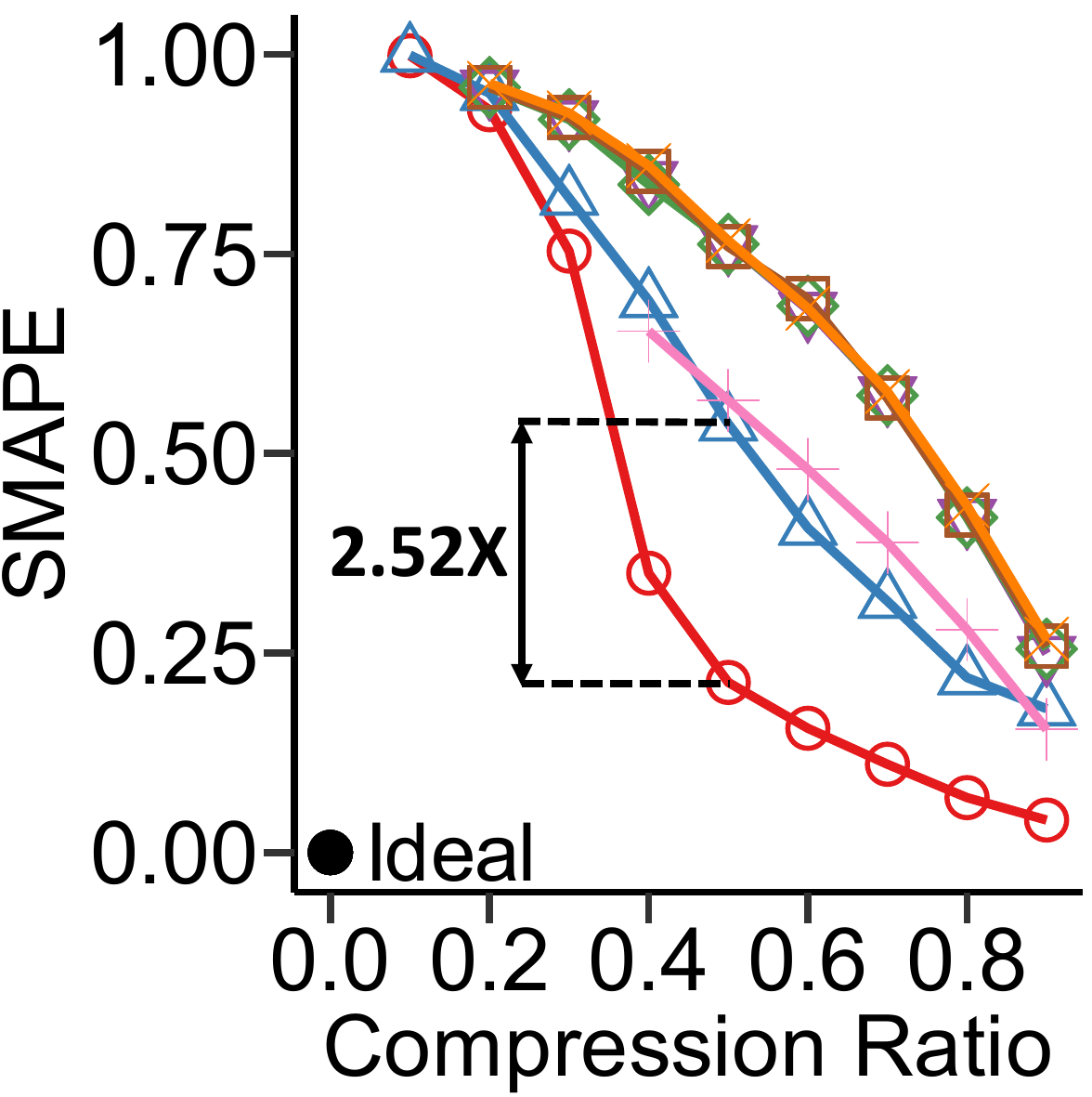}
	}
	\subfigure[DBLP (\RWR)]{
		\includegraphics[width=0.145\textwidth]{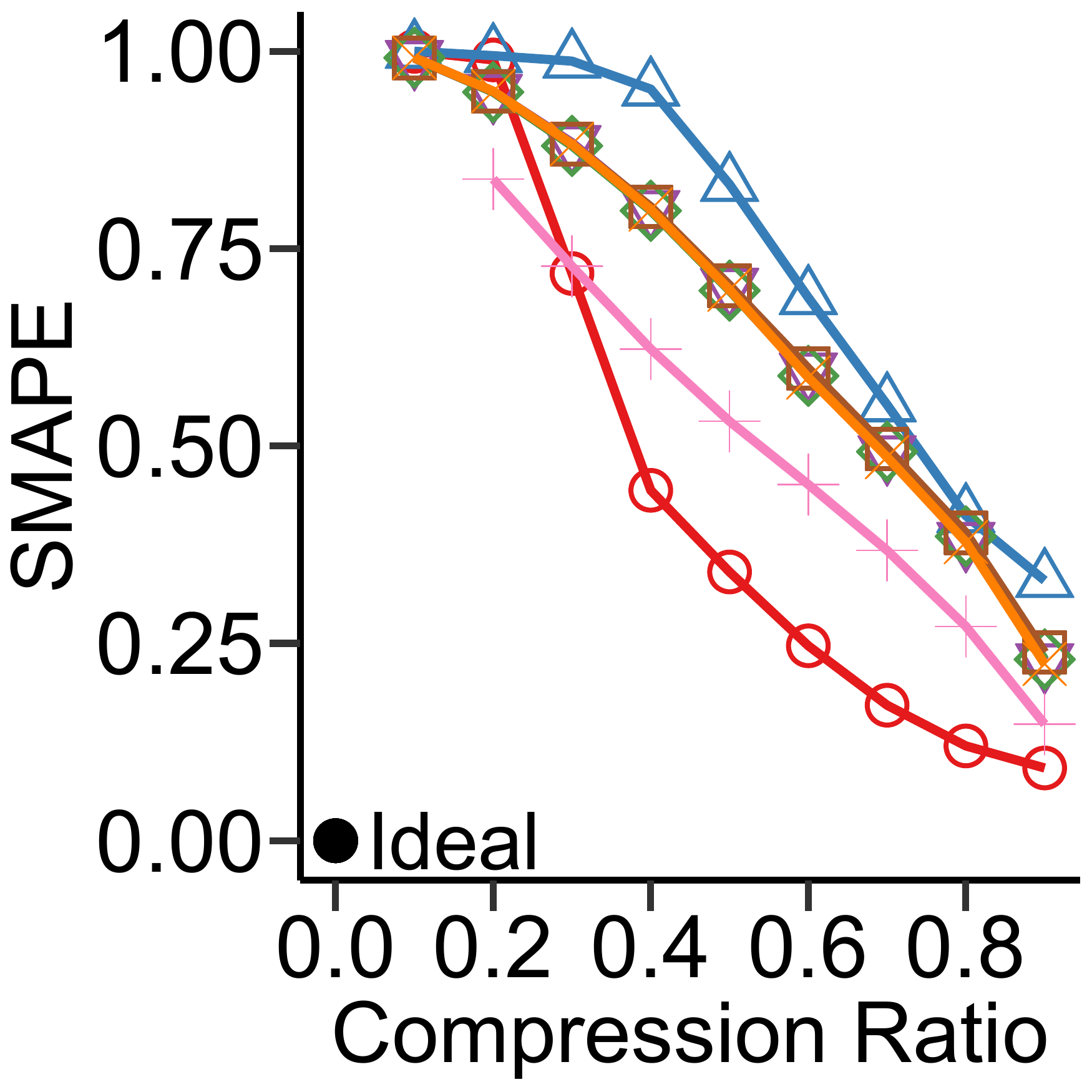}
	} 
	\subfigure[Amazon0601 (\RWR)]{
		\includegraphics[width=0.145\textwidth]{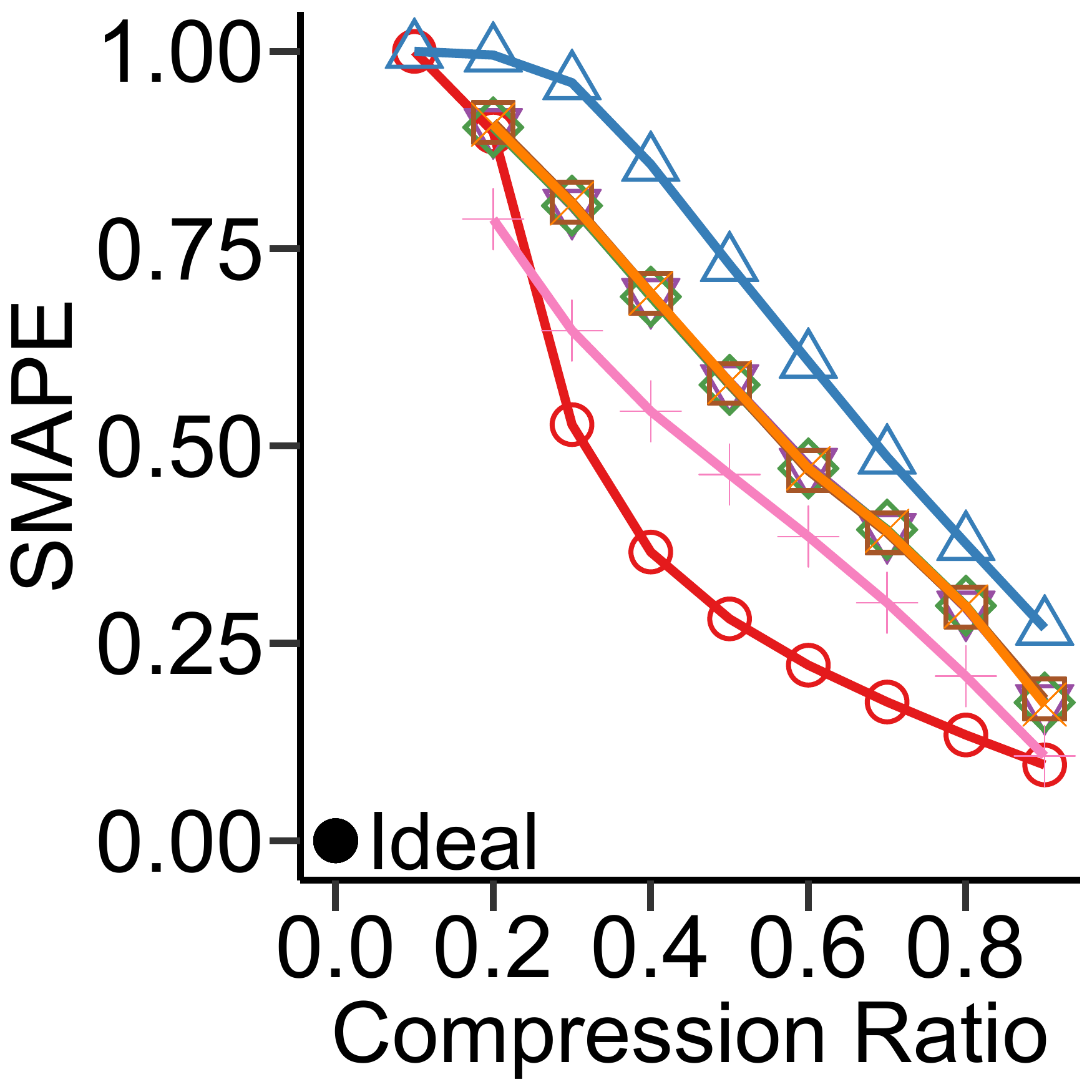}
	} 
	\subfigure[Skitter (\RWR)]{
		\includegraphics[width=0.145\textwidth]{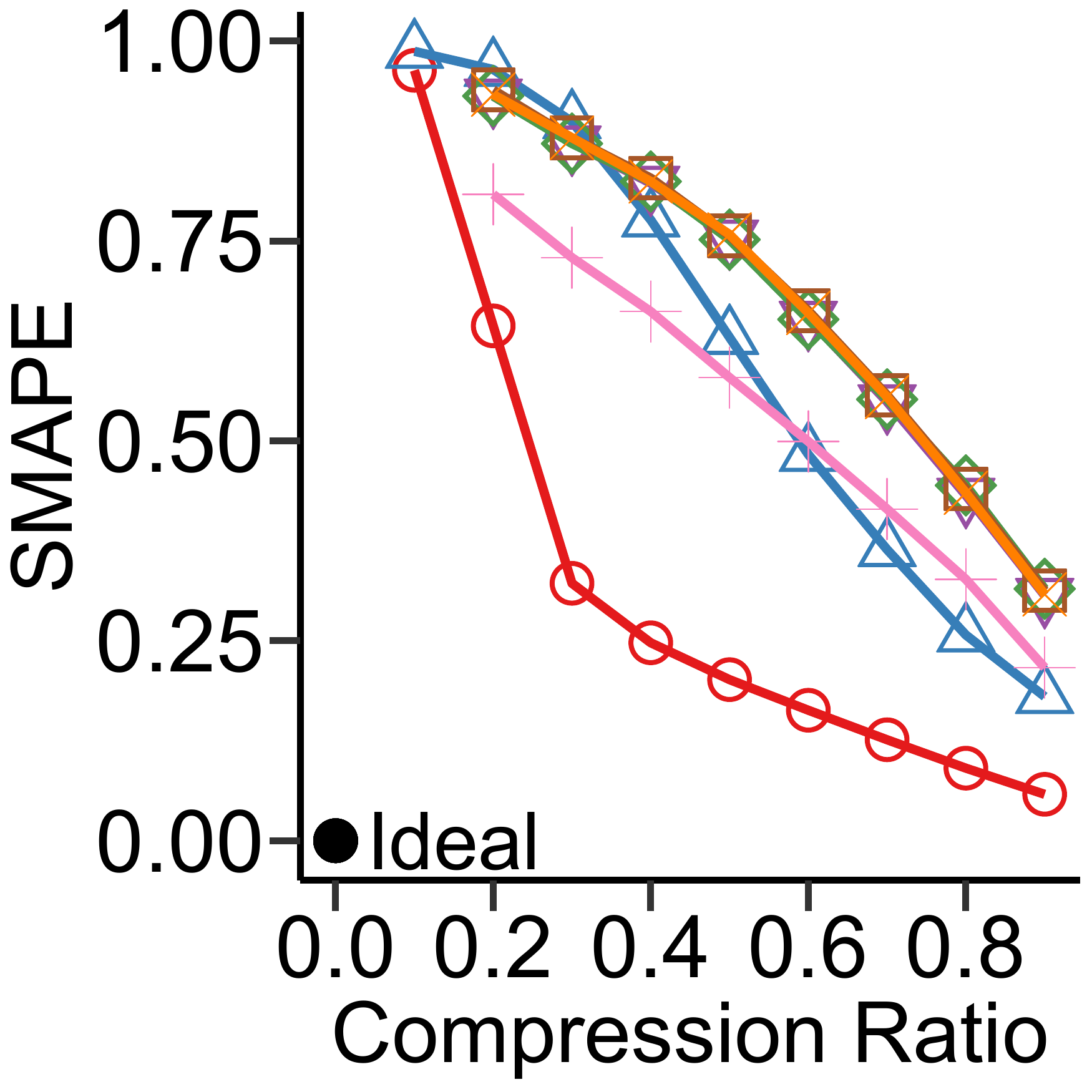}
	} 
	\subfigure[Wikipedia (\RWR)]{
		\includegraphics[width=0.145\textwidth]{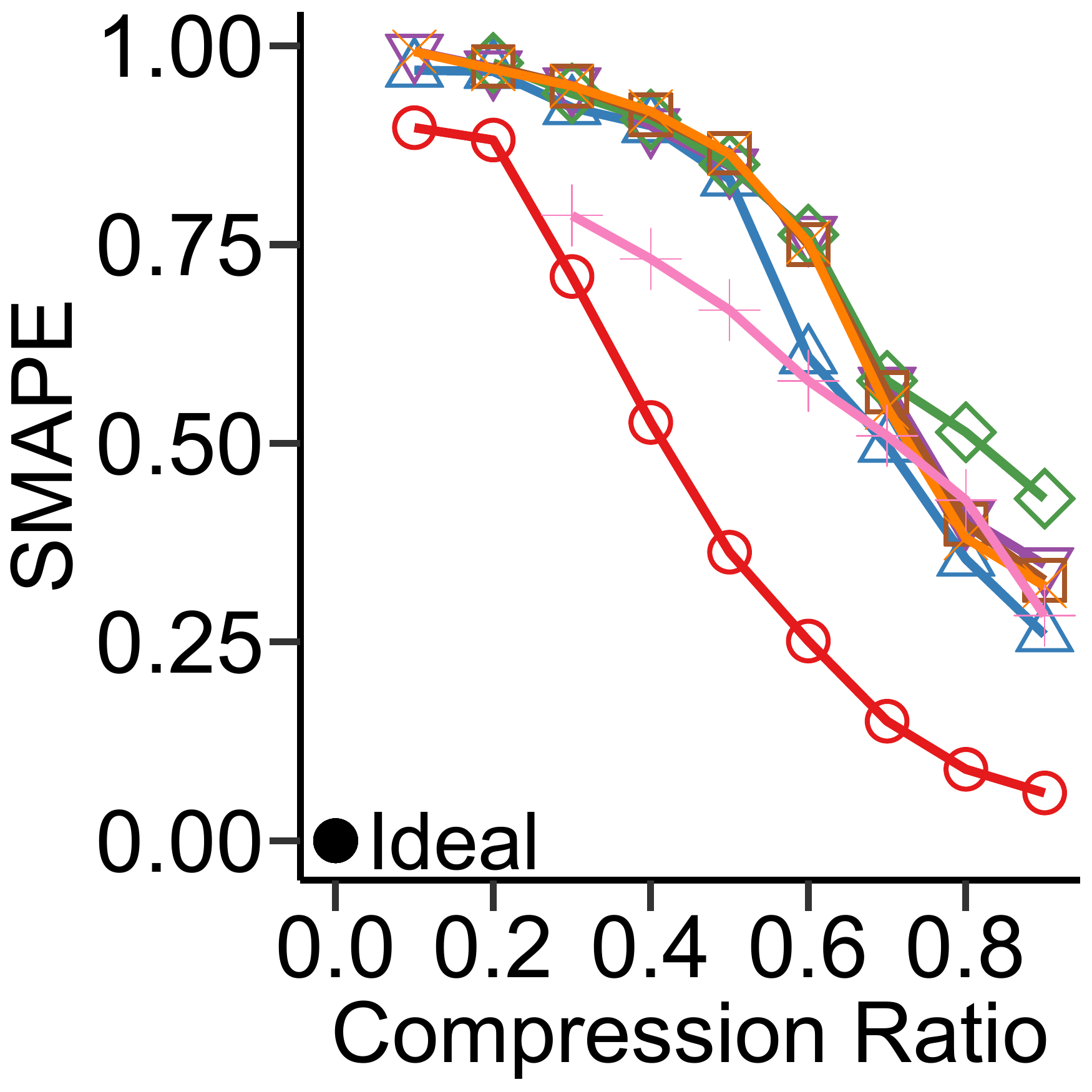}
	}
	\vspace{-3mm}
	\\
	\subfigure[LastFM-Asia (\HOP)]{
		\includegraphics[width=0.145\textwidth]{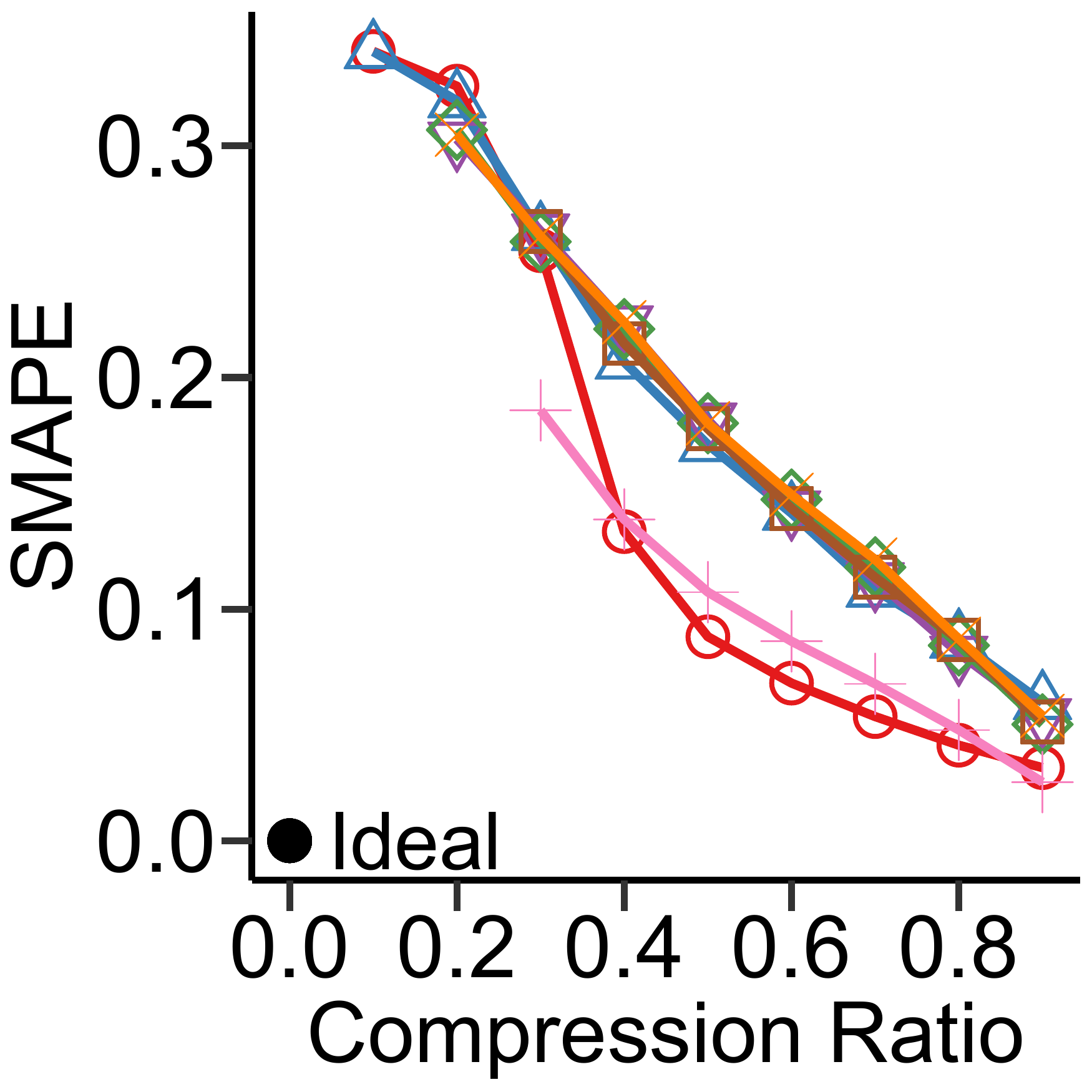}
	}
	\subfigure[Caida (\HOP)]{
		\includegraphics[width=0.145\textwidth]{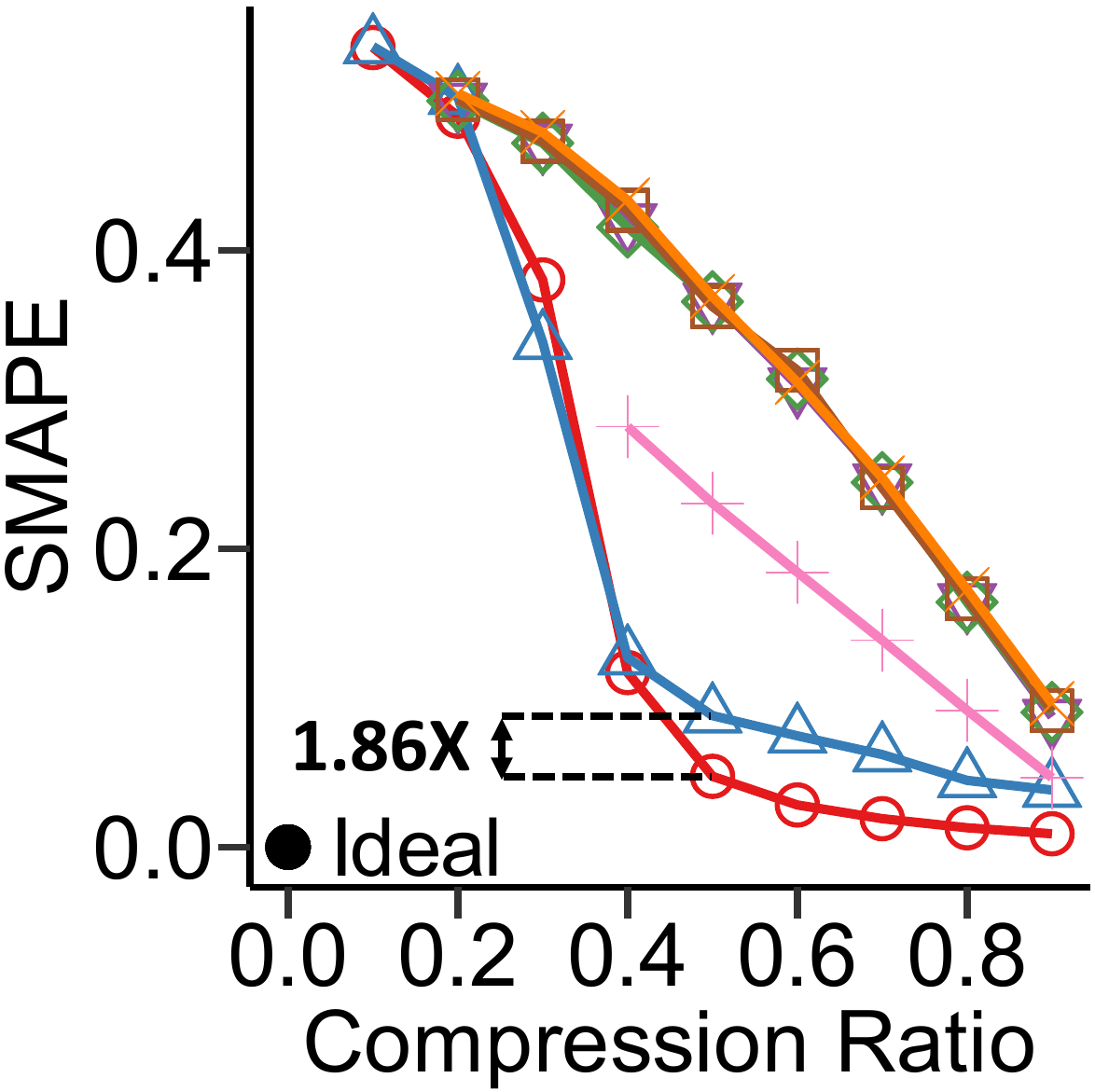}
	}
	\subfigure[DBLP (\HOP)]{
		\includegraphics[width=0.145\textwidth]{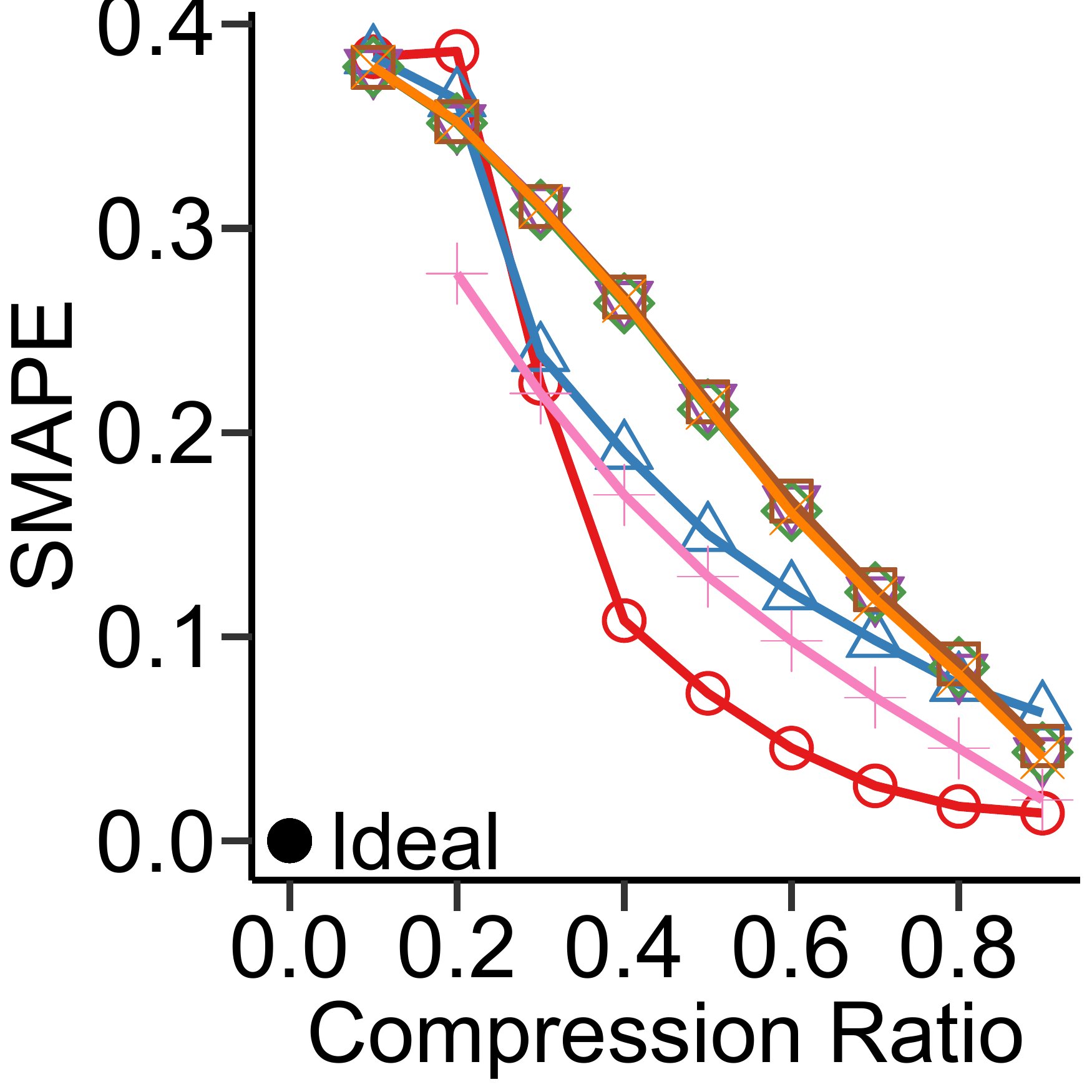}
	} 
	\subfigure[Amazon0601 (\HOP)]{
		\includegraphics[width=0.145\textwidth]{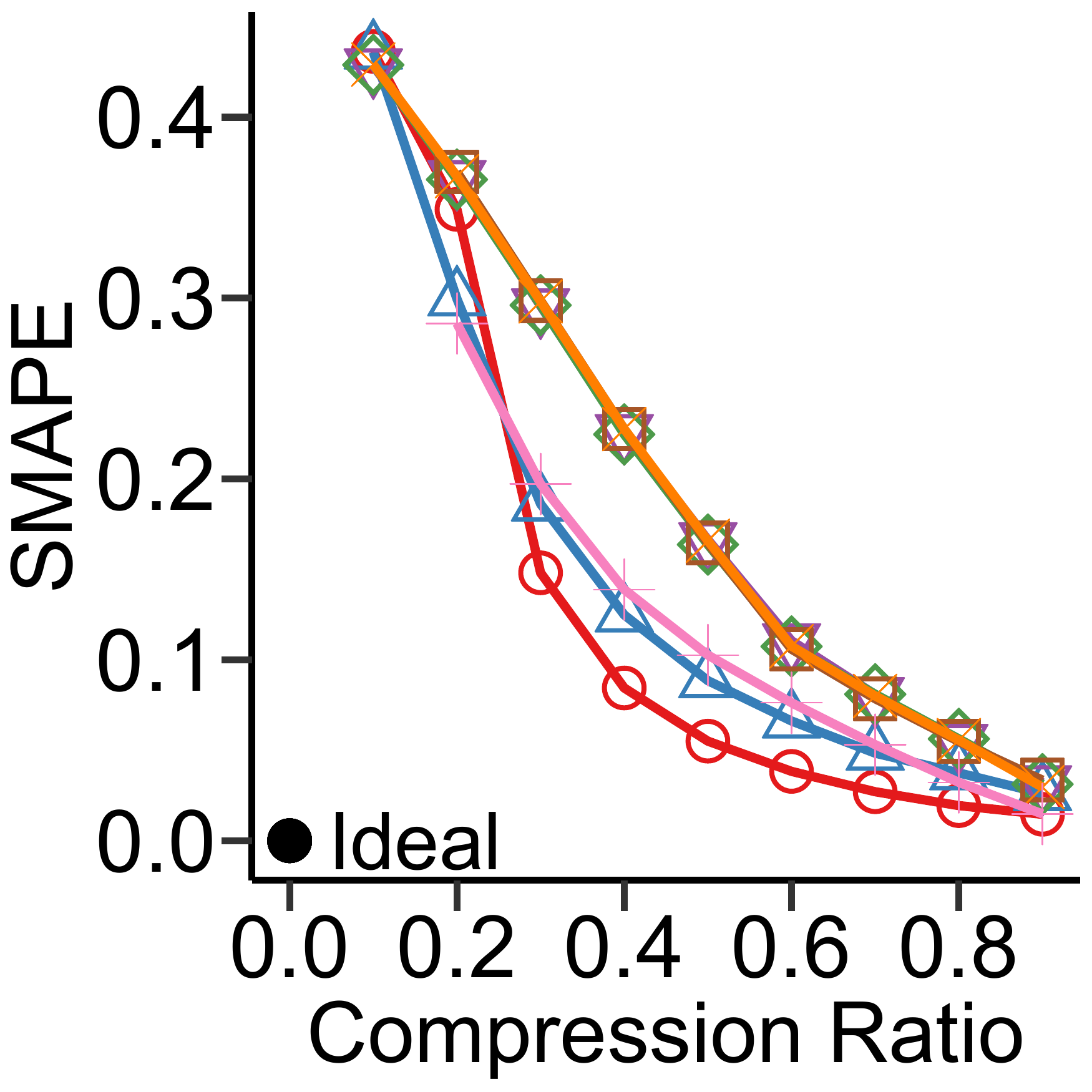}
	} 
	\subfigure[Skitter (\HOP)]{
		\includegraphics[width=0.145\textwidth]{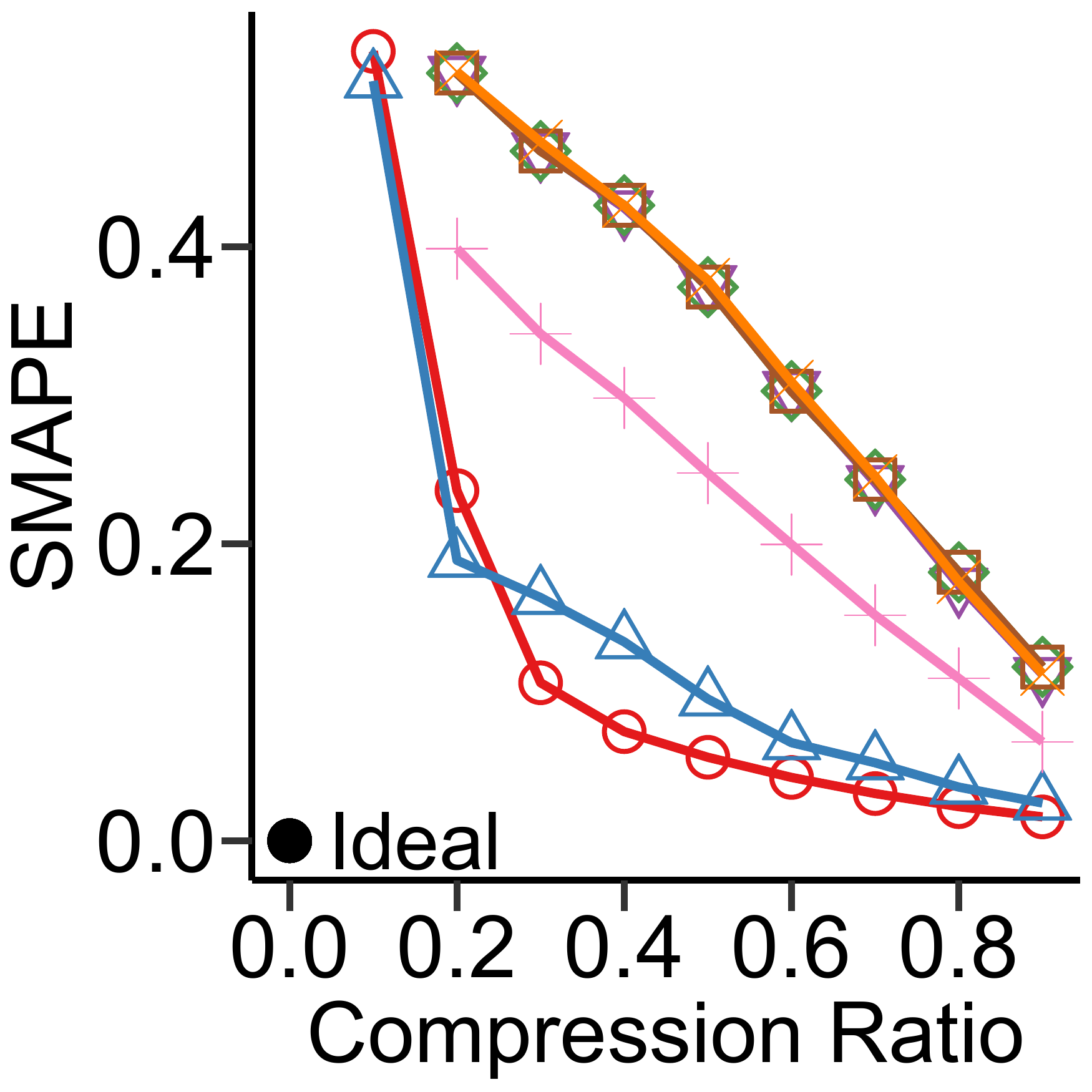}
	} 
	\subfigure[Wikipedia (\HOP)]{
		\includegraphics[width=0.145\textwidth]{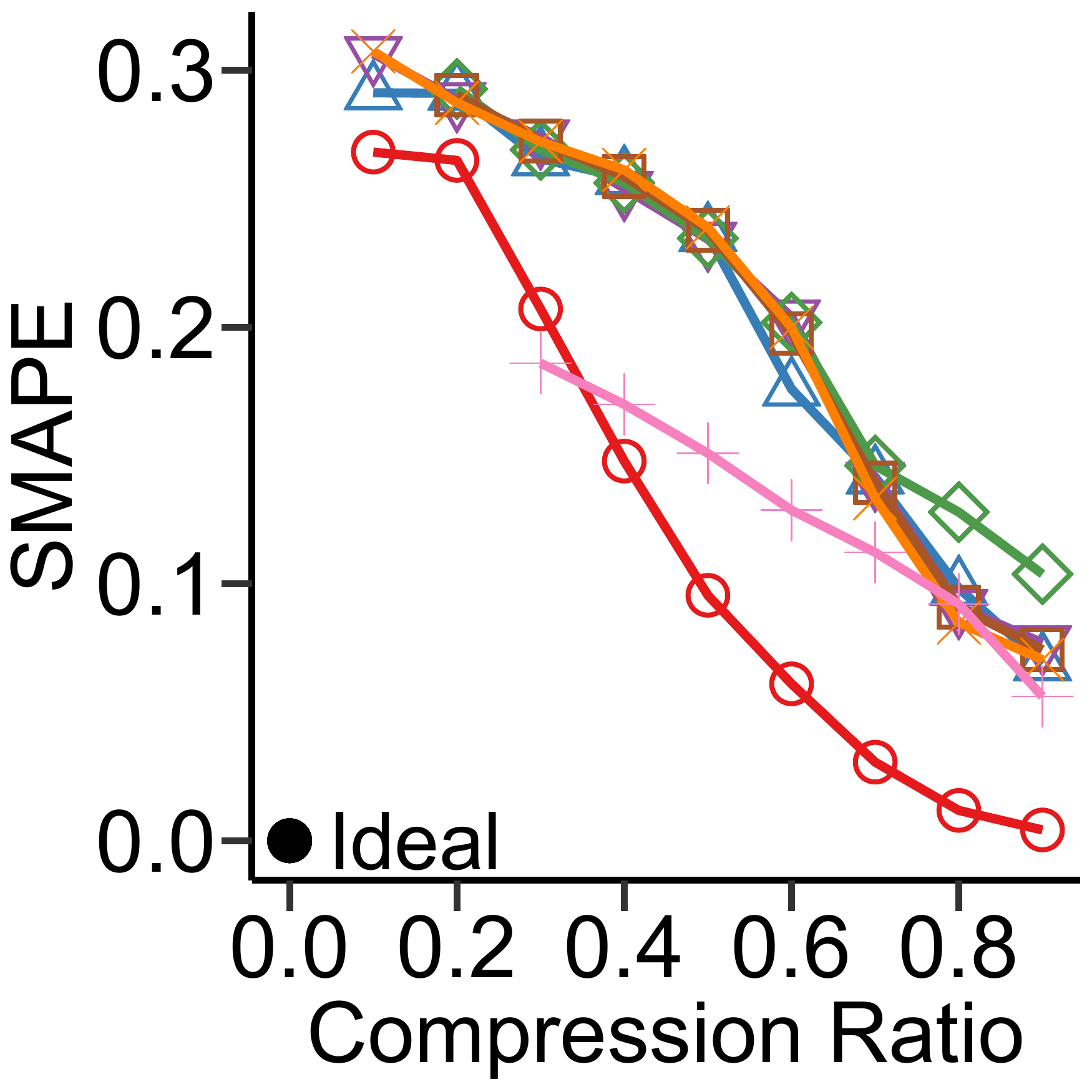}
	}
	\textbf{\small Spearman Correlation Coefficients (the higher, the better):} \hfill \ \ \\
		\subfigure[LastFM-Asia (\RWR)]{
		\includegraphics[width=0.145\textwidth]{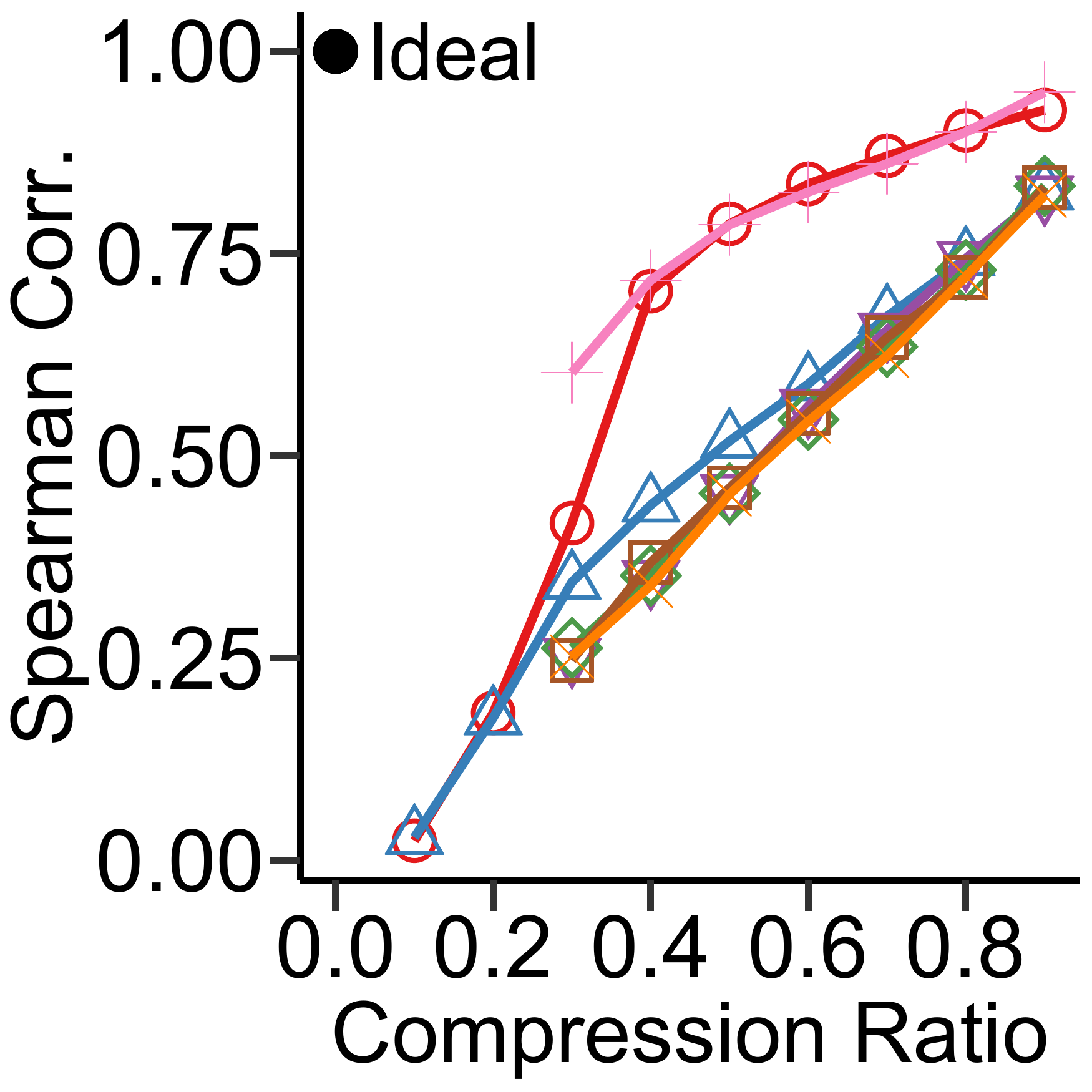}
	}
	\subfigure[Caida (\RWR)]{
		\includegraphics[width=0.145\textwidth]{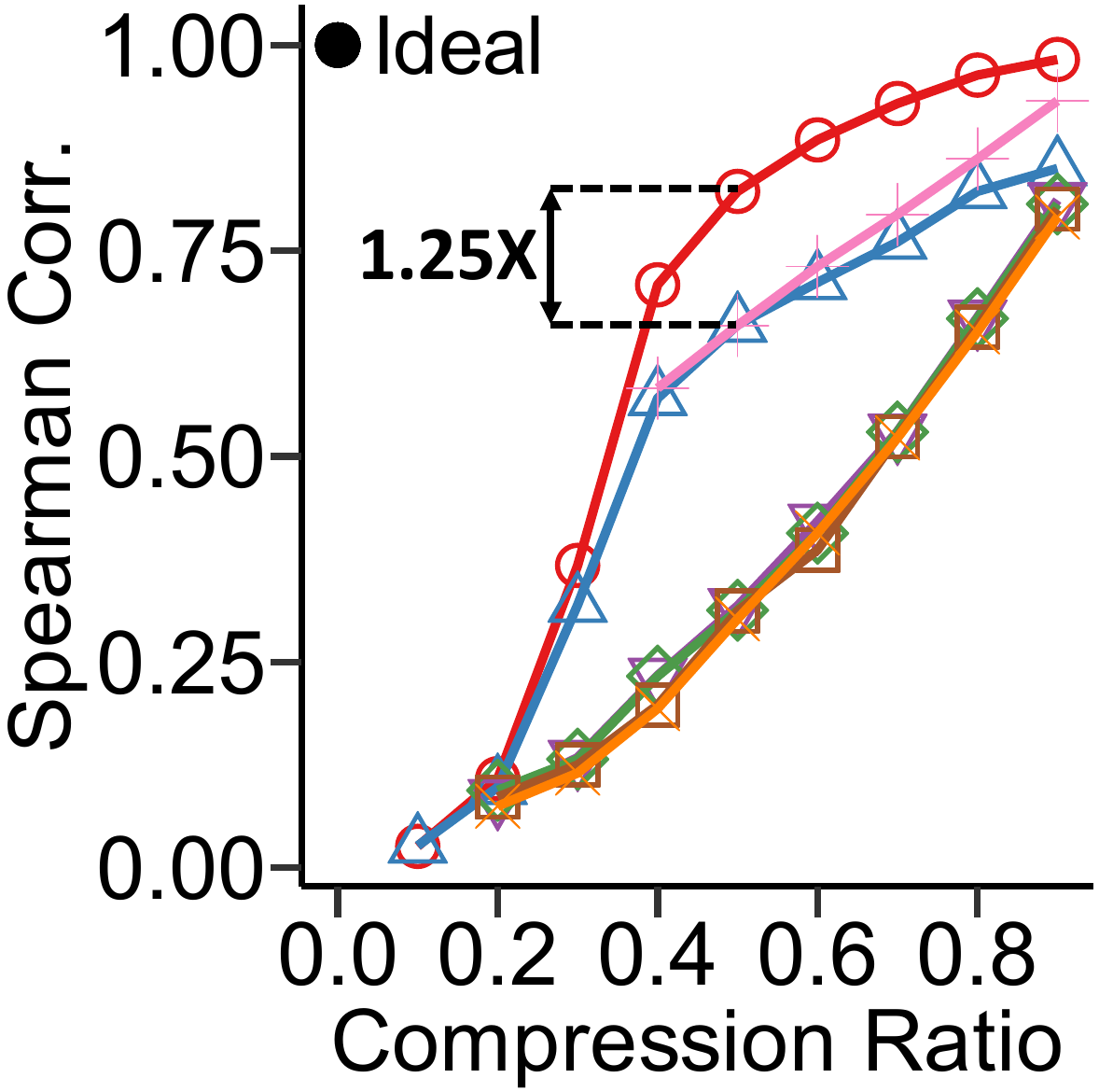}
	}
	\subfigure[DBLP (\RWR)]{
		\includegraphics[width=0.145\textwidth]{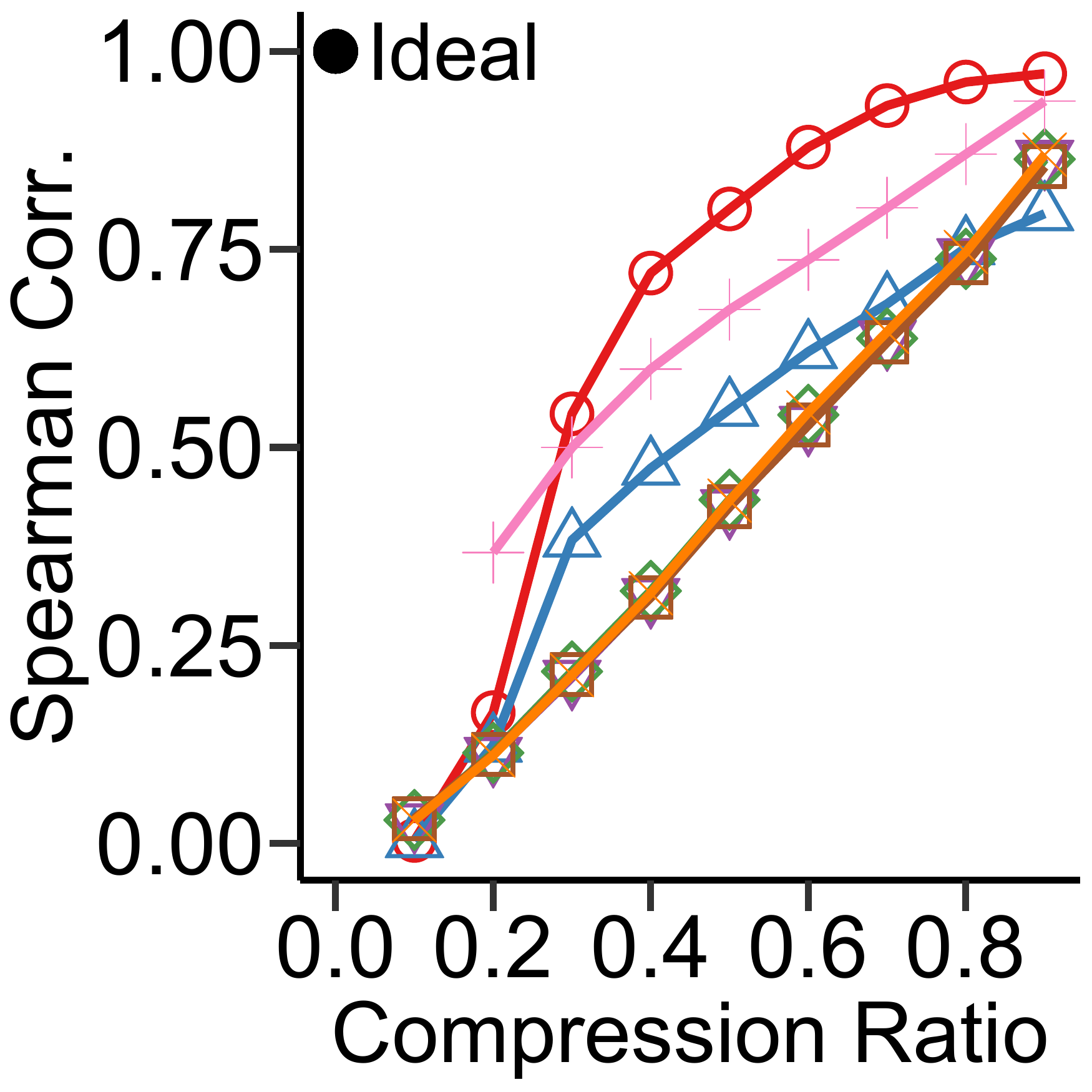}
	}
	\subfigure[Amazon0601 (\RWR)]{
		\includegraphics[width=0.145\textwidth]{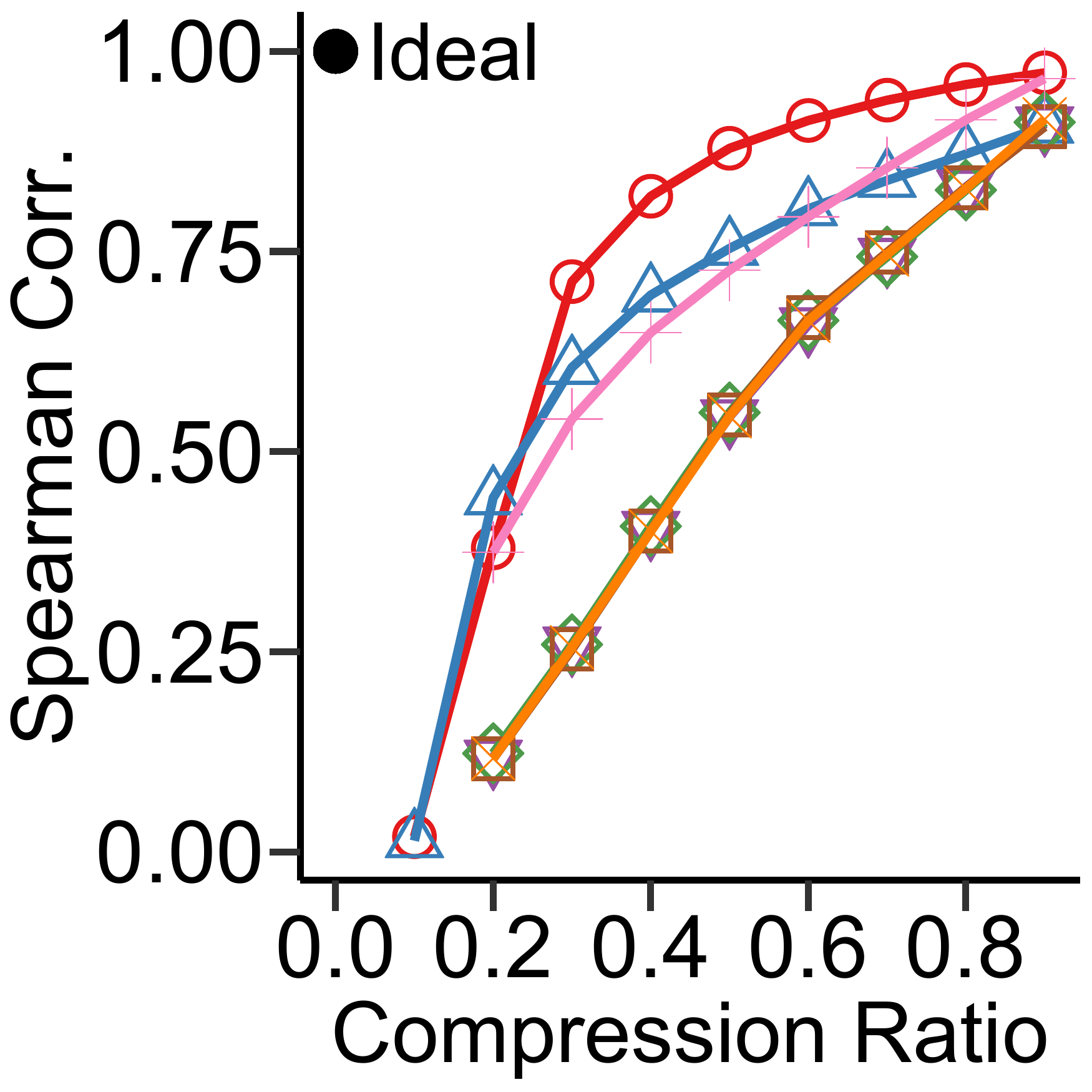}
	}
	\subfigure[Skitter (\RWR)]{
		\includegraphics[width=0.145\textwidth]{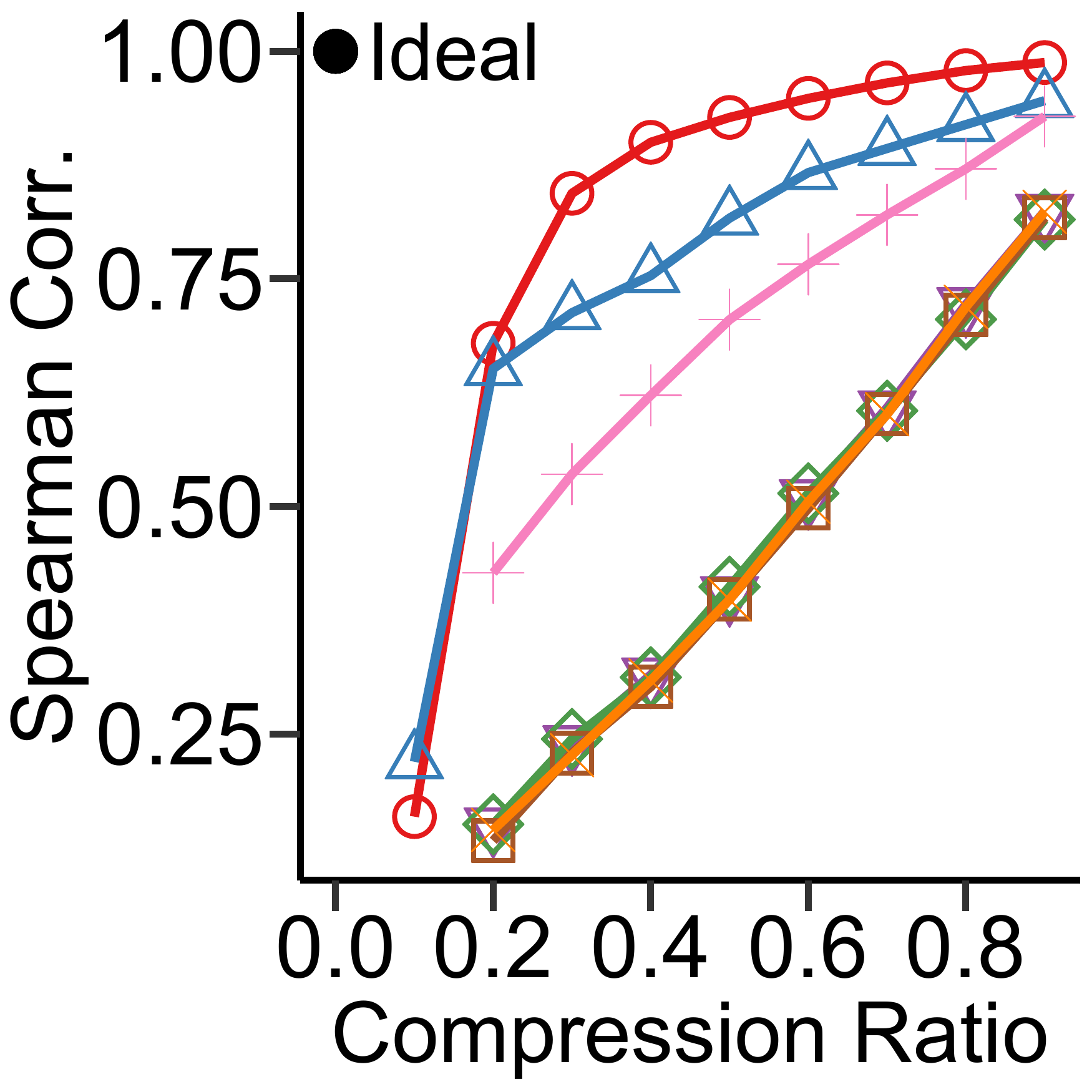}
	} 
	\subfigure[Wikipedia (\RWR)]{
		\includegraphics[width=0.145\textwidth]{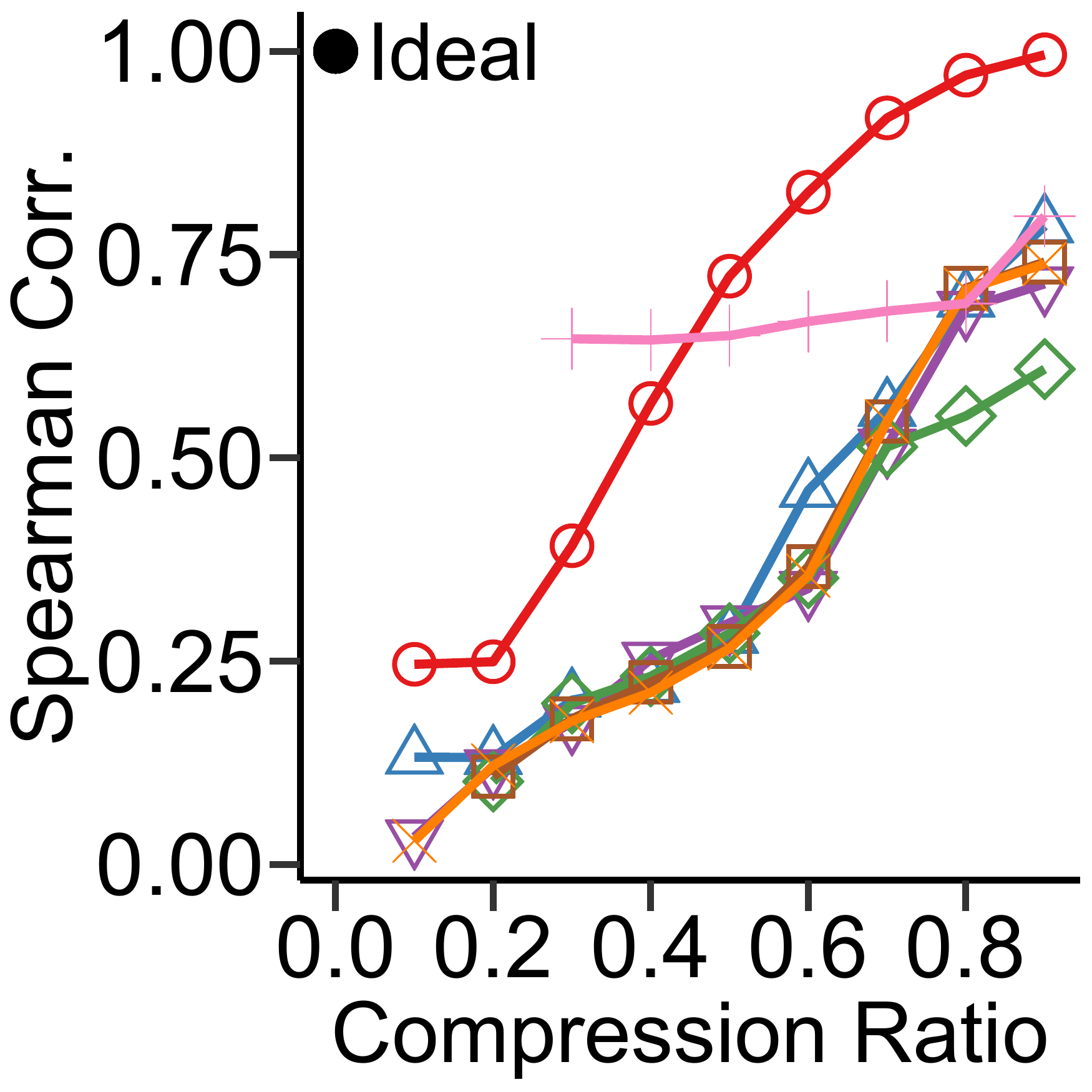}
	}
	\vspace{-3mm}
	\\
    \subfigure[LastFM-Asia (\HOP)]{
		\includegraphics[width=0.145\textwidth]{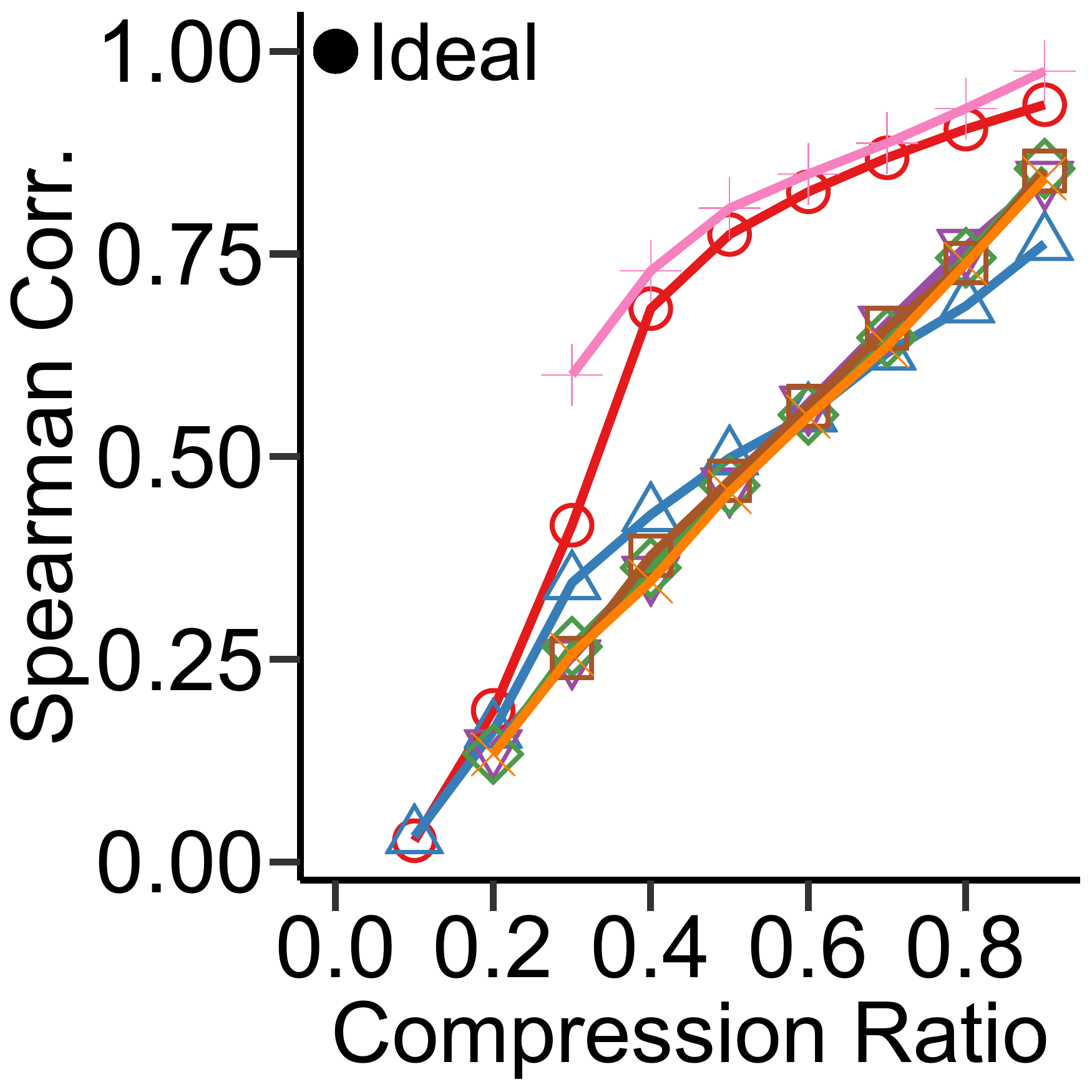}
	}
	\subfigure[Caida (\HOP)]{
		\includegraphics[width=0.145\textwidth]{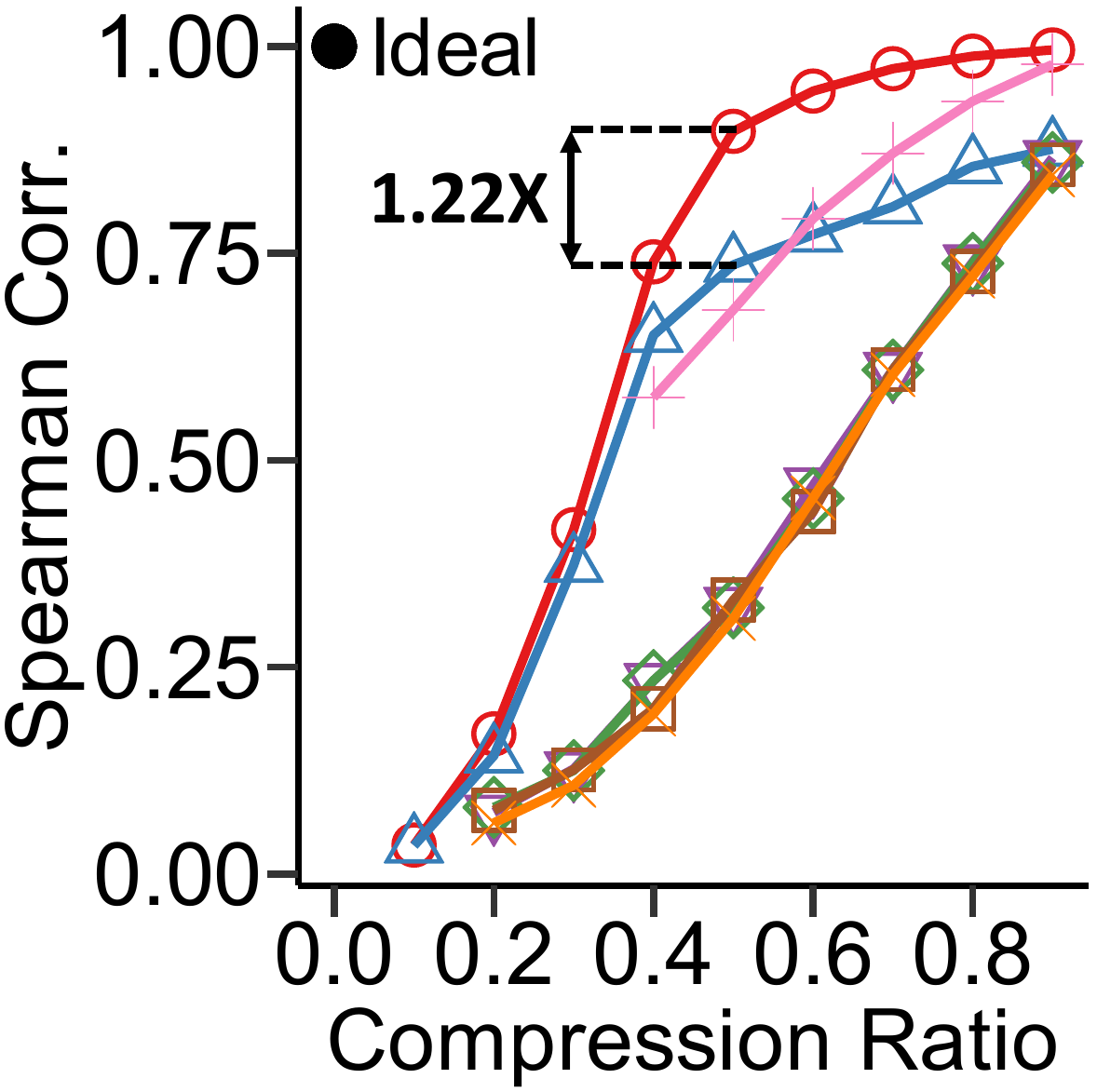}
	}
	\subfigure[DBLP (\HOP)]{
		\includegraphics[width=0.145\textwidth]{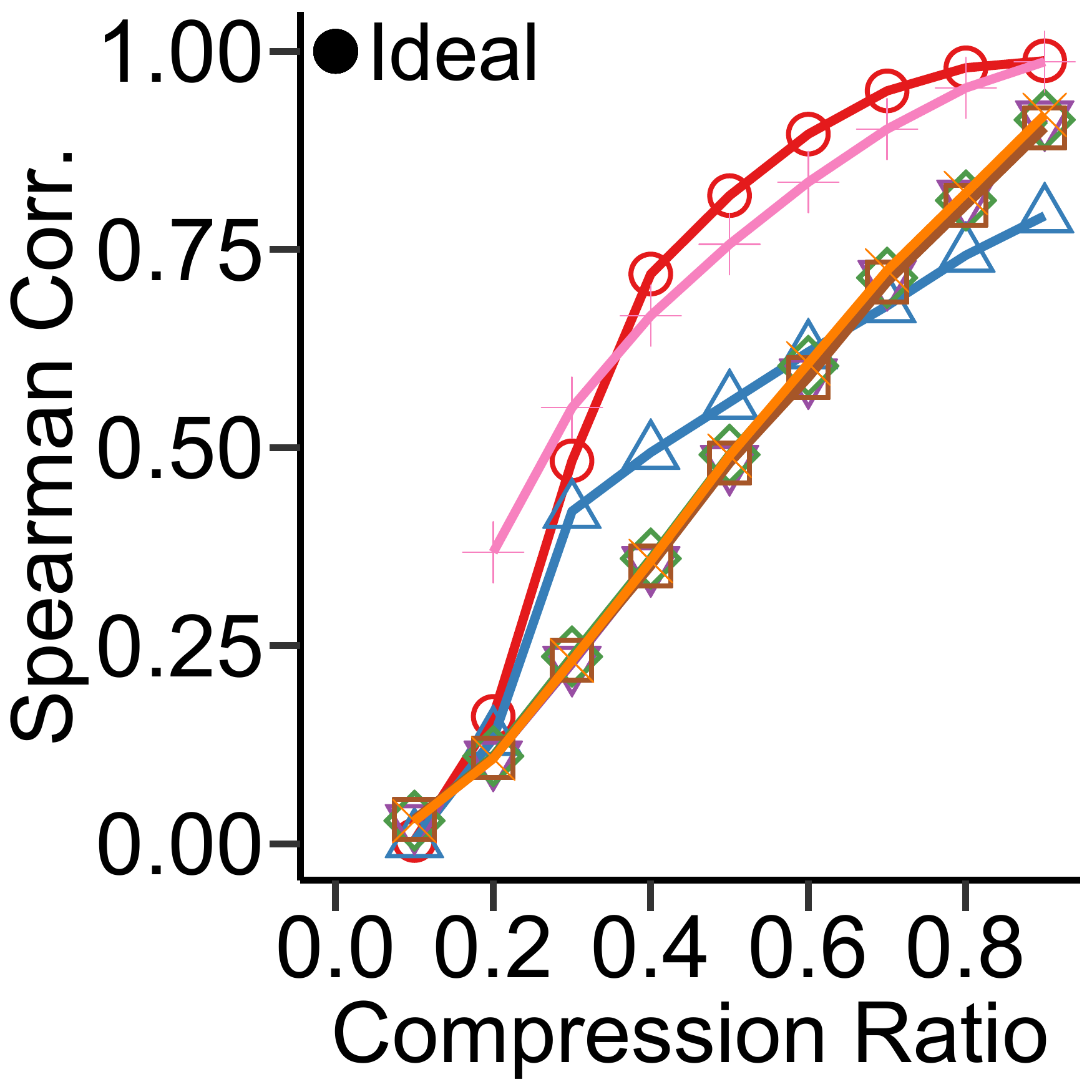}
	} 
	\subfigure[Amazon0601 (\HOP)]{
		\includegraphics[width=0.145\textwidth]{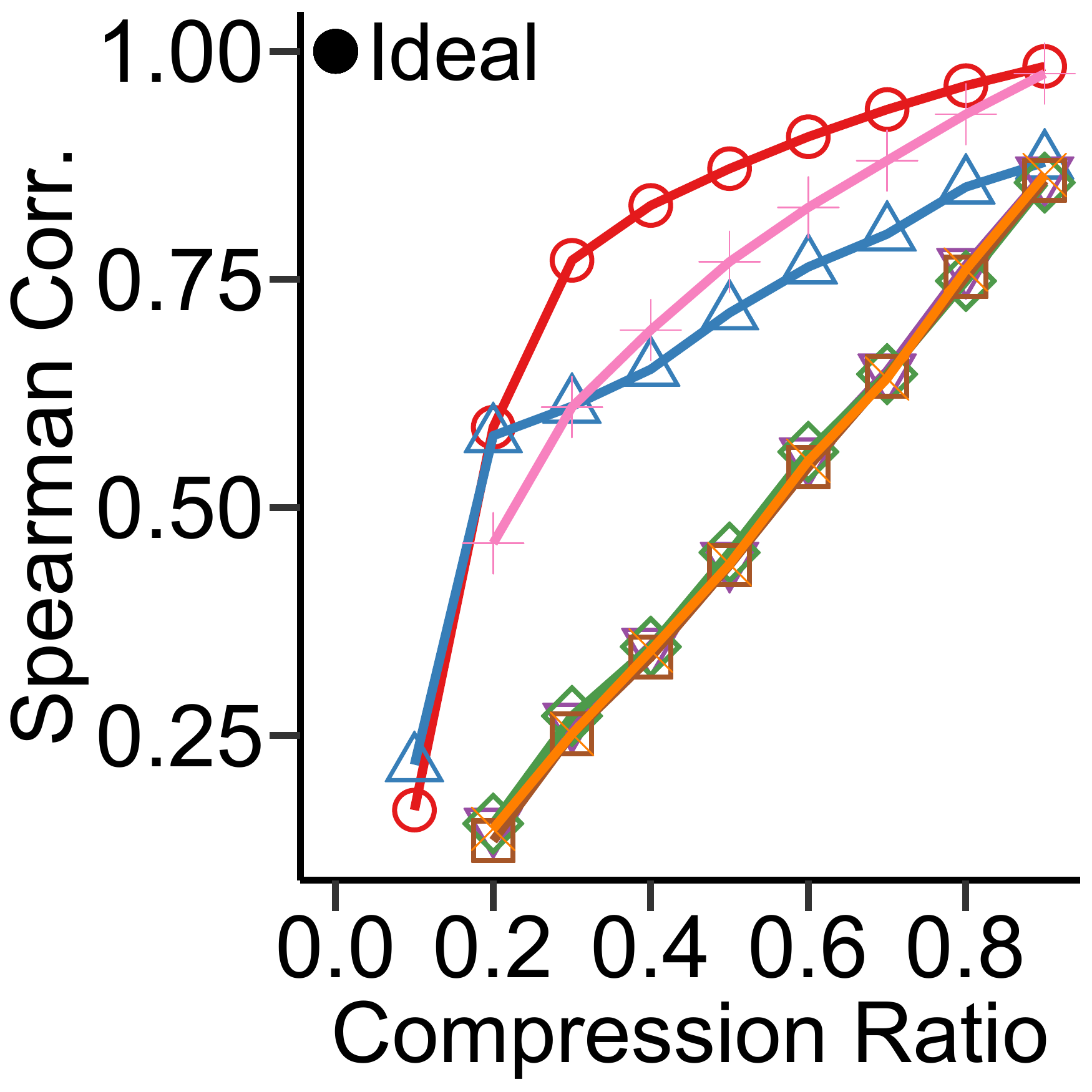}
	} 
	\subfigure[Skitter (\HOP)]{
		\includegraphics[width=0.145\textwidth]{ExperimentResult/PSMulti/HOP/SK_SpearmanCorr.pdf}
	} 
	\subfigure[Wikipedia (\HOP)]{
		\includegraphics[width=0.145\textwidth]{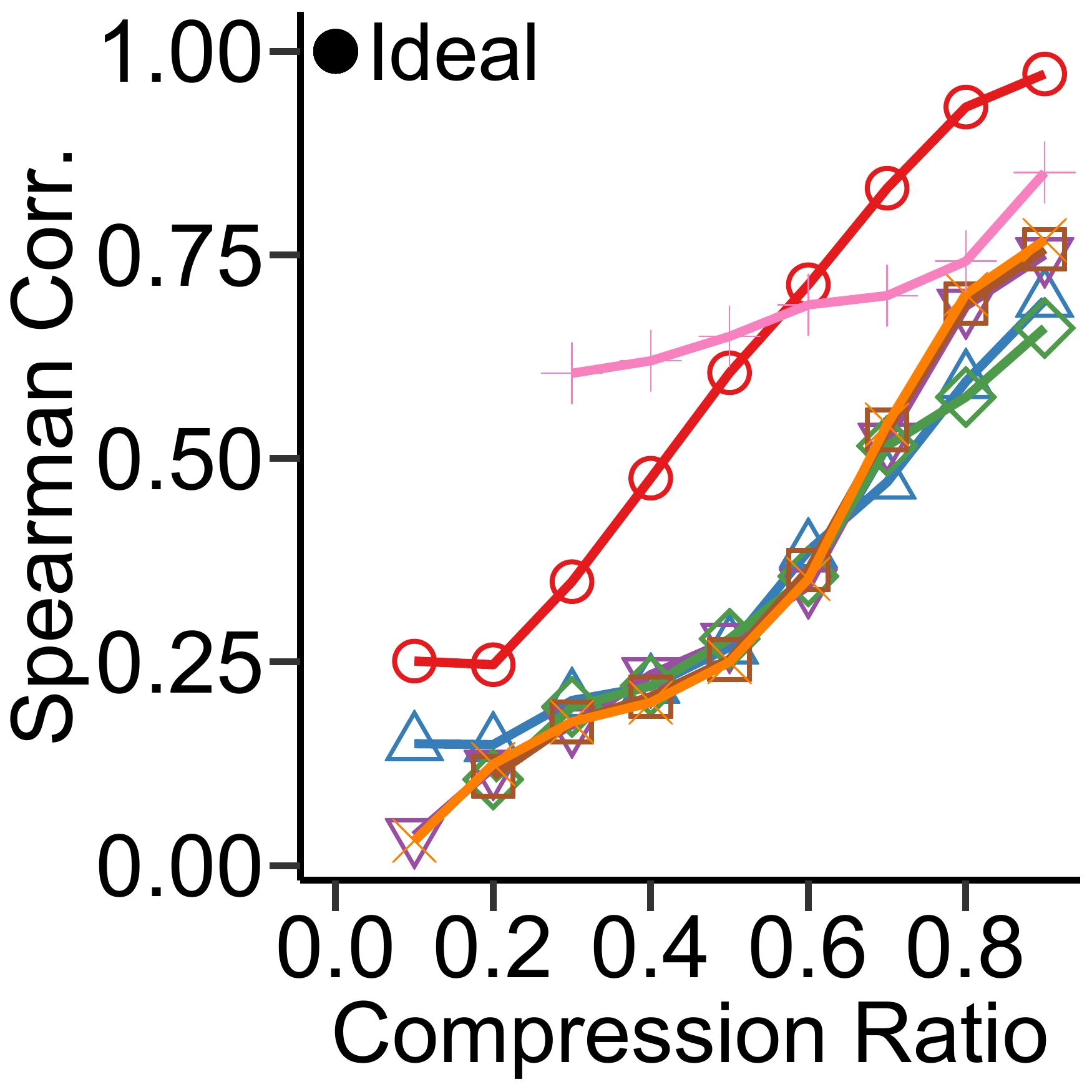}
	}
    \\
	\vspace{-1.5mm}
    \caption{\label{fig:Application}
    \underline{\smash{\textbf{\OurModel is useful for ``communication-free'' distributed multi-query processing.}}}
    Queries are answered more accurately from distributed personalized summary graphs (\method) than from non-personalized summary graphs (\SSumM) or distributed subgraphs (the others). The degree of personalization $\alpha$ is fixed to $1.25$. In some datasets, compression rates cannot be lowered further due to  imbalance among graph partitions.
    }
\end{figure*}

\smallsection{Effect of $\beta$.}
As seen in Fig.~\ref{fig:varMetricperAlphaRandomTNS}, in the majority of cases, \textbf{$\beta=0.1$ resulted in the best accuracy}, while the accuracy was not sensitive to $\beta$ as long as $\beta$ was not too close to $0$ or $1$.

\subsection{Q5. Application of \method (Fig.~\ref{fig:Application})}
\label{sec:q:application}

We demonstrate that \method can be useful for distributed multi-query answering. 
We assumed eight machines and applied \method and five graph partitioning algorithms \cite{ugander2013balanced, blondel2008fast, kabiljo2017social} to communication-free distributed multi-query processing, as described in Sect.~\ref{sec:application}. 
We randomly chose $100$ query nodes in the Wikipedia dataset and $500$ query nodes in the others dataset; and the averaged results are reported.
\textbf{Recall that summary graphs are not personalized only for query nodes.} Their target node sets together cover all nodes.

As seen in Fig.~\ref{fig:Application}, in almost all settings, \textbf{RWR and HOP queries were answered significantly more accurately from personalized summary graphs obtained by \method than from subgraphs obtained by graph partitioning.}
We obtained similar results for PHP queries, as reported in the online appendix\cite{appendixurl}.
For example, in the Caida dataset, answers were up to $2.52\times$ and $1.25\times$ more accurate in terms of \SMAPE and \Spearman (see Sect.~\ref{sec:experimetalsetting}), respectively, when the compression rate was $0.5$.
A shortcoming of our approach is that processing queries took longer on summary graphs than on subgraphs, which are uncompressed (see Fig.~\ref{fig:executionandquery}).

\section{Related Works}
\label{sec:relatedworks}

\smallsection{Graph Summarization:} 
While the term ``graph summarization'' refers to a specific way of compressing graphs \cite{lefevre2010grass,riondato2017graph,lee2020ssumm} in this work, it has been used also for a wide range of related concepts, as surveyed in \cite{liu2018graph}.
The most similar one is a lossless graph-compression technique \cite{navlakha2008graph,shin2019sweg,khan2015set,ko2020incremental,lee2022slugger} where the input graph is encoded ``losslessly'' by a summary graph, positive edge corrections, negative edge corrections, and optionally a hierarchy of nodes. 
Recently, LDME~\cite{yong2021efficient} reduces the search space of supernode pairs to be merged and accelerates the computation of saving and the update of encoding. Fan et al.~\cite{fan2021making} proposed a lossless graph contraction scheme that can be adapted for several types of graph queries (e.g., traingle counting and shortest distance), while \OurModel focuses on the neighborhood query, which is the key building block of many graph algorithms, as discussed in Appendix~\ref{sec:appendix:query}.

Among recent lossy graph summarization methods, UDS~\cite{kumar2018utility} uses memoization for summarizing a graph into the given number of supernodes while ensuring the proposed utility function does not drop below a certain threshold. T-BUDs~\cite{hajiabadi2021graph} use the minimum spanning tree of the two-hop graph, outperforming UDS in terms of speed and memory efficiency.
GLIMPSE~\cite{safavi2019personalized} summarizes a knowledge graph to capture facts preferred by a single user, who is not necessarily a node in the graph, based on the user's past queries.
The summary is a `subgraph' that captures local information and discards the others.

\smallsection{Graph Partitioning:}
Graph partitioning is to decompose a graph into subgraphs by partitioning nodes into groups for certain goals (e.g., to minimize normalized cuts).
Many approaches, including label propagation \cite{ugander2013balanced,bae2020label,awadelkarim2020prioritized},
local search \cite{kabiljo2017social}, and eigen decomposition \cite{newman2013spectral}, have been developed for
VLSI circuit placement \cite{augeri2004new},
sparse matrix factorization \cite{karypis1995analysis},
and storage sharding \cite{golab2014distributed,curino2010schism,kabiljo2017social}, etc.

\smallsection{Applications: Distributed Query Processing System:}
Distributed graph query processing systems can be divided into two types depending on whether multiple queries can be executed concurrently.
Some systems  \cite{kang2009pegasus, malewicz2010pregel, salihoglu2013gps, low2012distributed, wang2013asynchronous} are designed to handle one specific job at a time over the entire graph (e.g., PageRank and graph partitioning). They follow a vertex or edge-centric \cite{hong2014simplifying} scatter-gather model with batch processing.
The others \cite{shao2013trinity, hong2012green, sarwat2013horton+}, where a graph storage and a query processor exist on each machine, facilitate graph traversal over small areas in the graph. However, if  each query spans over nodes on the boundary of graph partitions, a large amount of inter-machine communication is inevitable, which eventually slows down the query processing time. 
Several multi-queriable distributed SPARQL engines \cite{huang2011scalable, schatzle2016s2rdf} were developed to handle queries on RDF graphs. 
Mayer et al. \cite{mayer2018q} proposed Q-graph for multi-query graph analysis that considers user-centric graph applications. 
EdgeFrame \cite{fuchs2020edgeframe}, a graph-specialized Spark DataFrame, caches the edge structure of a graph in compressed form on all workers in the cluster, which circumvents the inherent communication bottlenecks of worst-case optimal join (WCOJ) \cite{ngo2018worst, olteanu2015size, koutris2016worst} on distributed graphs.
Different from the previous studies,
we consider completely removing inter-machine communications, at the expense of exactness, using personalized summary graphs, as an application of \method to distributed multi-query processing (see Sects.~\ref{sec:application} and \ref{sec:q:application}).

\smallsection{Graph Visualization:} Our work is also related to graph visualization where the objective is to provide a pictorial representation of the nodes and edges where users focus the most. Rafiei \cite{rafiei2005effectively} and Ellis et al. \cite{ellis2006plot} use sampling and magnification to focus on a specific part of a graph. 
GMine \cite{rodrigues2006gmine} obtains hierarchical communities and offers multi-resolution graph exploration. It extracts a subgraph of interest based on the initial set of target nodes.
While these tools focus on small parts of graph for visualizations, \method summarizes the entire graph with for compression and query processing.

\section{Conclusion and Future Directions}
\label{sec:conclusion}

In this paper, we introduce the problem of finding summary graphs personalized to given target nodes (Problem~\ref{prob:personalizedgs}). 
We formulate the problem as an optimization problem, and we propose \OurModel, a linear-time algorithm (Theorem~\ref{thm:linearScalability}) that successfully summarizes a graph with one billion edges (Fig.~\ref{fig:random-scalability}).
Through extensive experiments using six real-world graphs and three types of node-similarity queries, we show that \OurModel provides summary graphs from which queries on target nodes are answered significantly more accurately than from non-personalized summary graphs obtained by state-of-the-art graph summarization methods (Fig.~\ref{fig:PSSingleRandomTNS}).
We also demonstrate the effectiveness of  \OurModel for communication-free distributed multi-query answering. Specifically, we show that queries are answered more accurately  from distributed personalized summary graphs than from distributed subgraphs (Fig.~\ref{fig:Application}).
The source code and the data are available at \cite{appendixurl} for reproducibility.
We leave the analysis of the hardness of Problem~\ref{prob:personalizedgs} and the design of theoretically sound algorithms for future work.

{\small  \smallsection{Acknowledgements:}
This work was supproted by National Research Foundation of Korea (NRF) grand funded by the Korea government (MSIT) (No. NRF-2020R1C1C1008296) and Institute of Information \& Communications Technology Planning \& Evaluation (IITP) grant funded by the Korea government (MSIT) (No.2019-0-00075, Artificial Intelligence Graduate School Program (KAIST)).}

\addcontentsline{toc}{section}{Appendices}
\renewcommand{\thesubsection}{\Alph{subsection}}

\begin{algorithm}[t]
    \small
    \DontPrintSemicolon
    \caption{\textbf{getNeighbors($\SummaryGraph$, $q$)}}
    \label{alg:NeighborQuery}
    \KwInput{(1) summary graph $\SummaryGraph = (S, P)$, (2) query node $q$\\
    }
    \KwOutput{approximate neighbor $\hat{N}_{q}$ of $q$ in $\ReconstructedGraph$}
    \SetAlgoLined
    $\hat{N}_{q} \leftarrow \emptyset$ \\
    $\bar{N}_{S_q} \leftarrow $ neighbors of $S_{q}$ in $\SummaryGraph$ (i.e., $\forall A$ where $\{A,S_q\} \in P$)\\ 
    % $V_{u}\leftarrow$ the supernode in $S$ where $u$ $\in V_{u}$\\
    \For{$\mathbf{each}$ $A \in \bar{N}_{S_q}$}{
        $\hat{N}_{q} \leftarrow \hat{N}_{q} \cup A$
    }
    $\hat{N}_{q} \leftarrow \hat{N}_{q} \setminus \{q\}$\\
    \textbf{return} $\hat{N}_{q}$ 
\end{algorithm}

\begin{algorithm}[t]
    \small
    \DontPrintSemicolon
    \caption{\label{alg:hops} Number of Hops (HOPS) on $\SummaryGraph$}
    \KwInput{(1) summary graph $\SummaryGraph = (S, P)$, (2) query node $q$\\
    }
    \KwOutput{distance vector $d\in \mathbb{R}^{|V|}$}
    $Q \leftarrow \emptyset$; $Q.$\normalfont{insert(}$q$\normalfont{);}\\
    $d\leftarrow -\textbf{1}$; $d_{q}\leftarrow 0$ \Comment*[r]{\textbf{1} = one vector of size |V|}
    \While{$Q \neq \emptyset$}{
        $u \leftarrow$ $Q.$\normalfont{pop()} \\
        \For{$\mathbf{each}$ $v\in$ \normalfont{\textbf{getNeighbors(}$\SummaryGraph$, $u$\normalfont{)}}}{
            \If{$d_{v} = -1$ }{            
                 $Q.$\normalfont{insert(}$v$\normalfont{)};
                 $d_{v}\leftarrow d_{u}+1$\\
            }
        }
    }
    \textbf{return} $d$ 
\end{algorithm}

\begin{algorithm}[t!]
    \small
    \caption{Random Walk with Restart (RWR) on $\SummaryGraph$}
    \label{alg:RWR}
    \DontPrintSemicolon
    \KwInput{(1) summary graph $\SummaryGraph = (S,P)$, \\
    \hspace{10mm}(2) random walk probability $p$, (3) query node $q$\\
    }
    \KwOutput{RWR score vector $r^{new}\in \mathbb{R}^{|V|}$}
    $V\leftarrow \bigcup_{A\in S}A$ \\
    $r^{old}\leftarrow \textbf{0}$ \Comment*[r]{\textbf{0} = zero vector of size |V|}
    $r^{new} \leftarrow \frac{1}{|V|}\cdot \textbf{1}$ \Comment*[r]{\textbf{1} = one vector of size |V|}
    \While{$r^{new}\neq r^{old}$}{
        $r^{old} \leftarrow r^{new}$; $r^{new} \leftarrow \textbf{0}$ \\
        \For{$\mathbf{each}\ u \in V$}{
            $\hat{N}_{u} \leftarrow$ \textbf{getNeighbors($\SummaryGraph$, $u$)} \\
               $r_{v}^{new} \leftarrow r_{v}^{new} + \frac{1}{|\hat{N}_{u}|}r_{u}^{old}, \forall v\in \hat{N}_{u}$
        }
        $r^{new} \leftarrow p\cdot r^{new}$ \\
        $r^{new}_{q} \leftarrow (1-p\cdot \sum_{v\in V} r_{v}^{new})$ \nonumber \\
    }
    \textbf{return} $r^{new}$ 
\end{algorithm}

\appendix
\subsection{Query Answering}
\label{sec:appendix:query}

As described in Alg.~\ref{alg:NeighborQuery},
the approximate neighbors of a given node $u\in V$ can be retrieved directly from a summary graph $\SummaryGraph$.
That is, the \textit{neighborhood query} can be answered efficiently on $\SummaryGraph$, without restoring the entire graph.
A wide range of graph algorithms (e.g., BFS, DFS, Dijkstra's, and PageRank) access graphs only through neighborhood queries, and thus also can be executed directly on $\SummaryGraph$.
Alg.~\ref{alg:hops} and Alg.~\ref{alg:RWR} describe how to answer \RWR and \HOP queries  on $\SummaryGraph$. Answers of  \PHP queries, which are used in Sect.~\ref{sec:experiments}, can be computed from those of \RWR queries (see \cite{wu2014fast} for details).
In Sect.~\ref{sec:experiments}, on weighted summary graphs,
\RWR and \HOP queries were processed considering superedge weights.

\bibliographystyle{IEEEtran}
\bibliography{reference.bib}

\begin{thebibliography}{10}
\providecommand{\url}[1]{#1}
\csname url@samestyle\endcsname
\providecommand{\newblock}{\relax}
\providecommand{\bibinfo}[2]{#2}
\providecommand{\BIBentrySTDinterwordspacing}{\spaceskip=0pt\relax}
\providecommand{\BIBentryALTinterwordstretchfactor}{4}
\providecommand{\BIBentryALTinterwordspacing}{\spaceskip=\fontdimen2\font plus
\BIBentryALTinterwordstretchfactor\fontdimen3\font minus
  \fontdimen4\font\relax}
\providecommand{\BIBforeignlanguage}[2]{{%
\expandafter\ifx\csname l@#1\endcsname\relax
\typeout{** WARNING: IEEEtran.bst: No hyphenation pattern has been}%
\typeout{** loaded for the language `#1'. Using the pattern for}%
\typeout{** the default language instead.}%
\else
\language=\csname l@#1\endcsname
\fi
#2}}
\providecommand{\BIBdecl}{\relax}
\BIBdecl

\bibitem{boldi2004webgraph}
P.~Boldi and S.~Vigna, ``The webgraph framework i: compression techniques,'' in
  \emph{WWW}, 2004.

\bibitem{page1999pagerank}
L.~Page, S.~Brin, R.~Motwani, and T.~Winograd, ``The pagerank citation ranking:
  Bringing order to the web,'' Stanford InfoLab, Tech. Rep., 1999.

\bibitem{dhulipala2016compressing}
L.~Dhulipala, I.~Kabiljo, B.~Karrer, G.~Ottaviano, S.~Pupyrev, and A.~Shalita,
  ``Compressing graphs and indexes with recursive graph bisection,'' in
  \emph{KDD}, 2016.

\bibitem{shin2019sweg}
K.~Shin, A.~Ghoting, M.~Kim, and H.~Raghavan, ``Sweg: Lossless and lossy
  summarization of web-scale graphs,'' in \emph{WWW}, 2019.

\bibitem{ramasco2004self}
J.~J. Ramasco, S.~N. Dorogovtsev, and R.~Pastor-Satorras, ``Self-organization
  of collaboration networks,'' \emph{Physical review E}, vol.~70, no.~3, p.
  036106, 2004.

\bibitem{leskovec2007dynamics}
J.~Leskovec, L.~A. Adamic, and B.~A. Huberman, ``The dynamics of viral
  marketing,'' \emph{TWEB}, vol.~1, no.~1, p.~5, 2007.

\bibitem{lee2020ssumm}
K.~Lee, H.~Jo, J.~Ko, S.~Lim, and K.~Shin, ``Ssumm: Sparse summarization of
  massive graphs,'' in \emph{KDD}, 2020.

\bibitem{khan2015set}
K.~U. Khan, W.~Nawaz, and Y.-K. Lee, ``Set-based approximate approach for
  lossless graph summarization,'' \emph{Computing}, vol.~97, no.~12, pp.
  1185--1207, 2015.

\bibitem{beg2018scalable}
M.~A. Beg, M.~Ahmad, A.~Zaman, and I.~Khan, ``Scalable approximation algorithm
  for graph summarization,'' in \emph{PAKDD}, 2018.

\bibitem{riondato2017graph}
M.~Riondato, D.~Garc{\'\i}a-Soriano, and F.~Bonchi, ``Graph summarization with
  quality guarantees,'' \emph{DMKD}, vol.~31, no.~2, pp. 314--349, 2017.

\bibitem{lefevre2010grass}
K.~LeFevre and E.~Terzi, ``Grass: Graph structure summarization,'' in
  \emph{SDM}, 2010.

\bibitem{ko2020incremental}
J.~Ko, Y.~Kook, and K.~Shin, ``Incremental lossless graph summarization,'' in
  \emph{KDD}, 2020.

\bibitem{chierichetti2009compressing}
F.~Chierichetti, R.~Kumar, S.~Lattanzi, M.~Mitzenmacher, A.~Panconesi, and
  P.~Raghavan, ``On compressing social networks,'' in \emph{KDD}, 2009.

\bibitem{buehrer2008scalable}
G.~Buehrer and K.~Chellapilla, ``A scalable pattern mining approach to web
  graph compression with communities,'' in \emph{WSDM}, 2008.

\bibitem{fan2012query}
W.~Fan, J.~Li, X.~Wang, and Y.~Wu, ``Query preserving graph compression,'' in
  \emph{SIGMOD}, 2012.

\bibitem{henecka2015lossy}
W.~Henecka and M.~Roughan, ``Lossy compression of dynamic, weighted graphs,''
  in \emph{FiCloud}, 2015.

\bibitem{tsalouchidou2018scalable}
I.~Tsalouchidou, F.~Bonchi, G.~D.~F. Morales, and R.~Baeza-Yates, ``Scalable
  dynamic graph summarization,'' \emph{TKDE}, vol.~32, no.~2, pp. 360--373,
  2018.

\bibitem{liu2018graph}
Y.~Liu, T.~Safavi, A.~Dighe, and D.~Koutra, ``Graph summarization methods and
  applications: A survey,'' \emph{CSUR}, vol.~51, no.~3, pp. 1--34, 2018.

\bibitem{tobler1970computer}
W.~R. Tobler, ``A computer movie simulating urban growth in the detroit
  region,'' \emph{Economic Geography}, vol.~46, no. sup1, pp. 234--240, 1970.

\bibitem{shao2013trinity}
B.~Shao, H.~Wang, and Y.~Li, ``Trinity: A distributed graph engine on a memory
  cloud,'' in \emph{SIGMOD}, 2013.

\bibitem{sarwat2013horton+}
M.~Sarwat, S.~Elnikety, Y.~He, and M.~F. Mokbel, ``Horton+ a distributed system
  for processing declarative reachability queries over partitioned graphs,''
  \emph{PVLDB}, vol.~6, no.~14, pp. 1918--1929, 2013.

\bibitem{ngo2018worst}
H.~Q. Ngo, E.~Porat, C.~R{\'e}, and A.~Rudra, ``Worst-case optimal join
  algorithms,'' \emph{JACM}, vol.~65, no.~3, pp. 1--40, 2018.

\bibitem{khan2018smart}
A.~Khan, G.~Segovia, and D.~Kossmann, ``On smart query routing: for distributed
  graph querying with decoupled storage,'' in \emph{ATC}, 2018.

\bibitem{appendixurl}
``Supplementary materials: Appendix, code, datasets,''
  \url{https://github.com/ShinhwanKang/ICDE22-PeGaSus}, 2021.

\bibitem{grunwald2007minimum}
P.~D. Gr{\"u}nwald, \emph{The minimum description length principle}.\hskip 1em
  plus 0.5em minus 0.4em\relax MIT press, 2007.

\bibitem{broder2000min}
A.~Z. Broder, M.~Charikar, A.~M. Frieze, and M.~Mitzenmacher, ``Min-wise
  independent permutations,'' \emph{Journal of Computer and System Sciences},
  vol.~60, no.~3, pp. 630--659, 2000.

\bibitem{blum1973time}
M.~Blum, R.~W. Floyd, V.~R. Pratt, R.~L. Rivest, R.~E. Tarjan \emph{et~al.},
  ``Time bounds for selection,'' \emph{Journal of Computer and System
  Sciences}, vol.~7, no.~4, pp. 448--461, 1973.

\bibitem{blondel2008fast}
V.~D. Blondel, J.-L. Guillaume, R.~Lambiotte, and E.~Lefebvre, ``Fast unfolding
  of communities in large networks,'' \emph{JSTAT}, vol. 2008, no.~10, p.
  P10008, 2008.

\bibitem{lancichinetti2009community}
A.~Lancichinetti and S.~Fortunato, ``Community detection algorithms: a
  comparative analysis,'' \emph{Physical review E}, vol.~80, no.~5, p. 056117,
  2009.

\bibitem{hager2013exact}
W.~W. Hager, D.~T. Phan, and H.~Zhang, ``An exact algorithm for graph
  partitioning,'' \emph{Mathematical Programming}, vol. 137, no.~1, pp.
  531--556, 2013.

\bibitem{hager1999graph}
W.~W. Hager and Y.~Krylyuk, ``Graph partitioning and continuous quadratic
  programming,'' \emph{SIAM Journal on Discrete Mathematics}, vol.~12, no.~4,
  pp. 500--523, 1999.

\bibitem{felner2005finding}
A.~Felner, ``Finding optimal solutions to the graph partitioning problem with
  heuristic search,'' \emph{Annals of Mathematics and Artificial Intelligence},
  vol.~45, no.~3, pp. 293--322, 2005.

\bibitem{brunetta1997branch}
L.~Brunetta, M.~Conforti, and G.~Rinaldi, ``A branch-and-cut algorithm for the
  equicut problem,'' \emph{Mathematical Programming}, vol.~78, no.~2, pp.
  243--263, 1997.

\bibitem{donath1972algorithms}
W.~E. Donath and A.~J. Hoffman, ``Algorithms for partitioning of graphs and
  computer logic based on eigenvectors of connection matrices,'' \emph{IBM
  Technical Disclosure Bulletin}, vol.~15, no.~3, pp. 938--944, 1972.

\bibitem{donath2003lower}
------, ``Lower bounds for the partitioning of graphs,'' in \emph{Selected
  Papers Of Alan J Hoffman: With Commentary}.\hskip 1em plus 0.5em minus
  0.4em\relax World Scientific, 2003, pp. 437--442.

\bibitem{feather}
B.~Rozemberczki and R.~Sarkar, ``{Characteristic Functions on Graphs: Birds of
  a Feather, from Statistical Descriptors to Parametric Models},'' in
  \emph{CIKM}, 2020.

\bibitem{leskovec2005graphs}
J.~Leskovec, J.~Kleinberg, and C.~Faloutsos, ``Graphs over time: densification
  laws, shrinking diameters and possible explanations,'' in \emph{KDD}, 2005.

\bibitem{yang2015defining}
J.~Yang and J.~Leskovec, ``Defining and evaluating network communities based on
  ground-truth,'' \emph{KAIS}, vol.~42, no.~1, pp. 181--213, 2015.

\bibitem{kunegis2013konect}
J.~Kunegis, ``Konect: the koblenz network collection,'' in \emph{WWW}, 2013.

\bibitem{barabasi1999emergence}
A.-L. Barab{\'a}si and R.~Albert, ``Emergence of scaling in random networks,''
  \emph{Science}, vol. 286, no. 5439, pp. 509--512, 1999.

\bibitem{ugander2013balanced}
J.~Ugander and L.~Backstrom, ``Balanced label propagation for partitioning
  massive graphs,'' in \emph{WSDM}, 2013.

\bibitem{kabiljo2017social}
I.~Kabiljo, B.~Karrer, M.~Pundir, S.~Pupyrev, A.~Shalita, A.~Presta, and
  Y.~Akhremtsev, ``Social hash partitioner: a scalable distributed hypergraph
  partitioner,'' \emph{arXiv preprint arXiv:1707.06665}, 2017.

\bibitem{awadelkarim2020prioritized}
A.~Awadelkarim and J.~Ugander, ``Prioritized restreaming algorithms for
  balanced graph partitioning,'' in \emph{KDD}, 2020.

\bibitem{tong2008random}
H.~Tong, C.~Faloutsos, and J.-Y. Pan, ``Random walk with restart: fast
  solutions and applications,'' \emph{KAIS}, vol.~14, no.~3, pp. 327--346,
  2008.

\bibitem{zhang2012evaluating}
C.~Zhang, L.~Shou, K.~Chen, G.~Chen, and Y.~Bei, ``Evaluating geo-social
  influence in location-based social networks,'' in \emph{CIKM}, 2012.

\bibitem{guan2011assessing}
Z.~Guan, J.~Wu, Q.~Zhang, A.~Singh, and X.~Yan, ``Assessing and ranking
  structural correlations in graphs,'' in \emph{SIGMOD}, 2011.

\bibitem{goodwin1999asymmetry}
P.~Goodwin and R.~Lawton, ``On the asymmetry of the symmetric mape,''
  \emph{International journal of forecasting}, vol.~15, no.~4, pp. 405--408,
  1999.

\bibitem{spearman1961proof}
C.~Spearman, ``The proof and measurement of association between two things.''
  1961.

\bibitem{watts1998collective}
D.~J. Watts and S.~H. Strogatz, ``Collective dynamics of
  ‘small-world’networks,'' \emph{nature}, vol. 393, no. 6684, pp. 440--442,
  1998.

\bibitem{navlakha2008graph}
S.~Navlakha, R.~Rastogi, and N.~Shrivastava, ``Graph summarization with bounded
  error,'' in \emph{SIGMOD}, 2008.

\bibitem{lee2022slugger}
K.~Lee, J.~Ko, and K.~Shin, ``Slugger: Hierarchical summarization of massive
  graphs,'' in \emph{ICDE}, 2022.

\bibitem{yong2021efficient}
Q.~Yong, M.~Hajiabadi, V.~Srinivasan, and A.~Thomo, ``Efficient graph
  summarization using weighted lsh at billion-scale,'' in \emph{SIGMOD}, 2021.

\bibitem{fan2021making}
W.~Fan, Y.~Li, M.~Liu, and C.~Lu, ``Making graphs compact by lossless
  contraction,'' in \emph{SIGMOD}, 2021.

\bibitem{kumar2018utility}
K.~A. Kumar and P.~Efstathopoulos, ``Utility-driven graph summarization,''
  \emph{PVLDB}, vol.~12, no.~4, 2018.

\bibitem{hajiabadi2021graph}
M.~Hajiabadi, J.~Singh, V.~Srinivasan, and A.~Thomo, ``Graph summarization with
  controlled utility loss,'' in \emph{KDD}, 2021.

\bibitem{safavi2019personalized}
T.~Safavi, C.~Belth, L.~Faber, D.~Mottin, E.~M{\"u}ller, and D.~Koutra,
  ``Personalized knowledge graph summarization: From the cloud to your
  pocket,'' in \emph{ICDM}, 2019.

\bibitem{bae2020label}
M.~Bae, M.~Jeong, and S.~Oh, ``Label propagation-based parallel graph
  partitioning for large-scale graph data,'' \emph{IEEE Access}, vol.~8, pp.
  72\,801--72\,813, 2020.

\bibitem{newman2013spectral}
M.~E. Newman, ``Spectral methods for community detection and graph
  partitioning,'' \emph{Physical Review E}, vol.~88, no.~4, p. 042822, 2013.

\bibitem{augeri2004new}
C.~J. Augeri and H.~H. Ali, ``New graph-based algorithms for partitioning vlsi
  circuits,'' in \emph{ISCAS}, 2004.

\bibitem{karypis1995analysis}
G.~Karypis and V.~Kumar, ``Analysis of multilevel graph partitioning,'' in
  \emph{Supercomputing}, 1995.

\bibitem{golab2014distributed}
L.~Golab, M.~Hadjieleftheriou, H.~Karloff, and B.~Saha, ``Distributed data
  placement to minimize communication costs via graph partitioning,'' in
  \emph{SSDBM}, 2014.

\bibitem{curino2010schism}
C.~Curino, E.~Jones, Y.~Zhang, and S.~Madden, ``Schism: a workload-driven
  approach to database replication and partitioning,'' \emph{PVLDB}, vol.~3,
  no. 1-2, pp. 48--57, 2010.

\bibitem{kang2009pegasus}
U.~Kang, C.~E. Tsourakakis, and C.~Faloutsos, ``Pegasus: A peta-scale graph
  mining system implementation and observations,'' in \emph{ICDM}, 2009.

\bibitem{malewicz2010pregel}
G.~Malewicz, M.~H. Austern, A.~J. Bik, J.~C. Dehnert, I.~Horn, N.~Leiser, and
  G.~Czajkowski, ``Pregel: a system for large-scale graph processing,'' in
  \emph{SIGMOD}, 2010.

\bibitem{salihoglu2013gps}
S.~Salihoglu and J.~Widom, ``Gps: A graph processing system,'' in \emph{SSDBM},
  2013.

\bibitem{low2012distributed}
Y.~Low, J.~Gonzalez, A.~Kyrola, D.~Bickson, C.~Guestrin, and J.~M. Hellerstein,
  ``Distributed graphlab: A framework for machine learning and data mining in
  the cloud,'' \emph{PVLDB}, vol.~5, no.~8, 2012.

\bibitem{wang2013asynchronous}
G.~Wang, W.~Xie, A.~J. Demers, and J.~Gehrke, ``Asynchronous large-scale graph
  processing made easy,'' in \emph{CIDR}, 2013.

\bibitem{hong2014simplifying}
S.~Hong, S.~Salihoglu, J.~Widom, and K.~Olukotun, ``Simplifying scalable graph
  processing with a domain-specific language,'' in \emph{CGO}, 2014.

\bibitem{hong2012green}
S.~Hong, H.~Chafi, E.~Sedlar, and K.~Olukotun, ``Green-marl: a dsl for easy and
  efficient graph analysis,'' in \emph{ASPLOS}, 2012.

\bibitem{huang2011scalable}
J.~Huang, D.~J. Abadi, and K.~Ren, ``Scalable sparql querying of large rdf
  graphs,'' \emph{PVLDB}, vol.~4, no.~11, pp. 1123--1134, 2011.

\bibitem{schatzle2016s2rdf}
A.~Sch{\"a}tzle, M.~Przyjaciel-Zablocki, S.~Skilevic, and G.~Lausen, ``S2rdf:
  Rdf querying with sparql on spark,'' \emph{PVLDB}, vol.~9, no.~10, 2016.

\bibitem{mayer2018q}
C.~Mayer, R.~Mayer, J.~Grunert, K.~Rothermel, and M.~A. Tariq, ``Q-graph:
  preserving query locality in multi-query graph processing,'' in
  \emph{GRADES-NDA}, 2018.

\bibitem{fuchs2020edgeframe}
P.~Fuchs, P.~Boncz, and B.~Ghit, ``Edgeframe: Worst-case optimal joins for
  graph-pattern matching in spark,'' in \emph{GRADES-NDA}, 2020.

\bibitem{olteanu2015size}
D.~Olteanu and J.~Z{\'a}vodn{\`y}, ``Size bounds for factorised representations
  of query results,'' \emph{TODS}, vol.~40, no.~1, pp. 1--44, 2015.

\bibitem{koutris2016worst}
P.~Koutris, P.~Beame, and D.~Suciu, ``Worst-case optimal algorithms for
  parallel query processing,'' in \emph{ICDT}, 2016.

\bibitem{rafiei2005effectively}
D.~Rafiei, ``Effectively visualizing large networks through sampling,'' in
  \emph{VIS}, 2005.

\bibitem{ellis2006plot}
G.~Ellis and A.~Dix, ``The plot, the clutter, the sampling and its lens:
  occlusion measures for automatic clutter reduction,'' in \emph{AVI}, 2006.

\bibitem{rodrigues2006gmine}
J.~F. Rodrigues~Jr, H.~Tong, A.~J. Traina, C.~Faloutsos, and J.~Leskovec,
  ``Gmine: a system for scalable, interactive graph visualization and mining,''
  in \emph{VLDB}, 2006.

\bibitem{wu2014fast}
Y.~Wu, R.~Jin, and X.~Zhang, ``Fast and unified local search for random walk
  based k-nearest-neighbor query in large graphs,'' in \emph{SIGMOD}, 2014.

\end{thebibliography}

\end{document}